\documentclass{./sty/ametsocV6.1}

\def\scriptO{{{\it O}\kern -.42em {\it `}\kern + .20em}}
\def\RR{{{\rm l}\kern - .15em {\rm R} }}
\def\PP{{{\rm l}\kern - .15em {\rm P} }}
\def\L2{{{\sf L}^2}}
\def\H1{{{\sf H}^1}}
\def\PN2{{\PP_{N}-\PP_{N-2}}}

\def\complex{{{\rm C} \kern - .53em {\rm l} \kern + .38em}}

\def\a1{{ | \lambda_{\min} |}}

\def\l1{{   \lambda_{\min}  }}

\def\bu0{{\underline {\bf 0}}}

\def\bu{{\bf u}}

\def\bS{{\bf s}}

\def\u0{{\underline 0}}

\newcommand{\pp}[2]{\frac{\partial #1}{\partial #2} }

\usepackage{subfig}
\usepackage{wrapfig}
\usepackage{enumitem}

\nolinenumbers

\nolinenumbers
\title{Modeling Turbulence in the Atmospheric Boundary Layer
       with Spectral Element and Finite Volume Methods}

 \authors{Ananias Tomboulides,\aff{c}   
  Matthew Churchfield,\aff{b}
  Paul Fischer,\aff{d,e}
  \\
  Michael Sprague,\aff{b}
  and
  Misun Min\correspondingauthor{Email address: mmin@mcs.anl.gov (Misun Min)},\aff{a}
 }
 \affiliation{\aff{a}{First Affiliation}\\
 \aff{b}{Second Affiliation}\\
 \aff{c}{Third Affiliation}\\
 \aff{d}{Fourth Affiliation}
 }
 
 \affiliation{\aff{a}{Mathematics and Computer Science, Argonne National Laboratory, Lemont, IL, USA}\\
 \aff{b}{Computational Science Center, National Renewable Energy Laboratory, Golden, CO, USA}\\
 \aff{c}{Mechanical Engineering, Aristotle University of Thessaloniki, Tessaloniki, Greece}\\
 \aff{d}{Computer Science, University of Illinois Urbana-Champaign, Urbana, IL, USA}\\
 \aff{e}{Mechanical Science \& Engineering, University of Illinois Urbana-Champaign, Urbana, IL, USA}
\nolinenumbers
}

\nolinenumbers
\abstract{
We present large-eddy-simulation (LES) modeling approaches for the simulation of atmospheric boundary
layer turbulence that are of direct relevance to wind energy production.
In this paper, we study a GABLS benchmark problem using high-order spectral element code Nek5000/RS
and a block-structured second-order finite-volume code AMR-Wind
which are supported under the DOE's Exascale Computing Project (ECP)
Center for Efficient Exascale Discretizations (CEED) and ExaWind projects, respectively,
targeting application simulations on various acceleration-device based exascale computing platforms.
As for Nek5000/RS we demonstrate our newly developed subgrid-scale (SGS) models based on mean-field eddy viscosity (MFEV),
high-pass filter (HPF), and Smagorinsky (SMG) with traction boundary conditions. For the traction boundary conditions,
a novel analytical approach is presented that solves for the surface friction velocity and surface kinematic temperature flux.
For AMR-Wind, standard SMG is used and discussed in detail the traction boundary conditions for convergence.
We provide low-order statistics, convergence and turbulent structure analysis.
Verification and convergence studies were performed for both codes at various resolutions and it was found that
Nek5000/RS demonstrates convergence with resolution for all ABL bulk parameters, including boundary layer and low level jet (LLJ) height.
Extensive comparisons are presented with simulation data from the literature.
}

\nolinenumbers

\begin{document}

\maketitle

%
%
%
%
%

%

\section{Introduction}

Accurate simulations of the atmospheric boundary layer (ABL) are central to
engineering design questions related to wind farms, buildings, and urban
canyons.   In this paper, we explore turbulence modeling for the ABL
in the context of two general-purpose codes that are capable of supporting the
complex geometries required of engineering design codes, namely the spectral
element code, Nek5000/RS,~\cite{nek5000,nekrs}\footnote{Nek5000/RS represents
the two codes, Nek5000 and NekRS where NekRS is a GPU variant of Nek5000.} 
and the finite-volume code, AMR-Wind,~\cite{Sprague-etal:2020,Sharma-etal:2024}.   
These codes were developed as part of the U.S. Department of
Energy's Exascale Computing Project and are designed to run on both CPU- and
GPU-based platforms.  Extensive performance studies for the GEWEX (Global
Energy and Water Cycle Experiment) Atmospheric Boundary Layer Study
(GABLS)~\cite{beare2006intercomparison} are presented in~\cite{min22a}.  Here,
we explore the impacts of discretizations, subgrid-scale LES models, and wall
models on mean and rms velocity and temperatures profiles on turbulence
morphology, and on the energy spectra.  We present inter-code comparisons and
comparisons with results in the literature, including the pseudospectral
results of \cite{Sullivan-etal:2008}.

ABL flows feature turbulent mixing, vertical diffusion, vertical and horizontal
heat exchanges, and Coriolis effects due to planetary rotation and curvature,
with the additional complexity from the regional-scale weather patterns and terrain. 
Significant studies have been applied to ABL flows~\cite{Moeng:1984, 
Sullivan-etal:2008, Churchfield-Moriarty:2020}.
We focus on high-order numerical computation of stably stratified ABL flows 
using large eddy simulation (LES).
The governing equations, incompressible Navier--Stokes (NS),
are solved in filtered form such that the larger, 
energy-containing eddies are directly resolved, and the remaining SGS 
turbulence is modeled. 
The stably stratified atmospheric boundary layer is a key component of 
Earth-system modeling, as well as of large-scale weather, climate, and 
ocean models~\cite{Fernando2010}, \cite{Mahrt2014}, \cite{Large1994}, 
\cite{McWilliams_2004}, \cite{Cuxart2006}, \cite{Svensson2009}, \cite{Holtslag2013},
\cite{Heisel2023}. 

The ABL community has set up a sequence of benchmark problems,
the GEWEX (Global Energy and Water Cycle Experiment) Atmospheric Boundary Layer Study
(GABLS)~\cite{beare2006intercomparison},
to quantify the effects of numerical modeling and discretization choices. 
These benchmarks represent the atmospheric boundary layer in regional and large-scale.
Atmospheric models are considered important benchmarks for improving modeling approaches
for the study of wind energy, climate, and weather on all scales
\cite{churchfield2017}.

 \begin{wrapfigure}{r}{0.4\textwidth}
  \centering
  \includegraphics[width=0.38\textwidth]{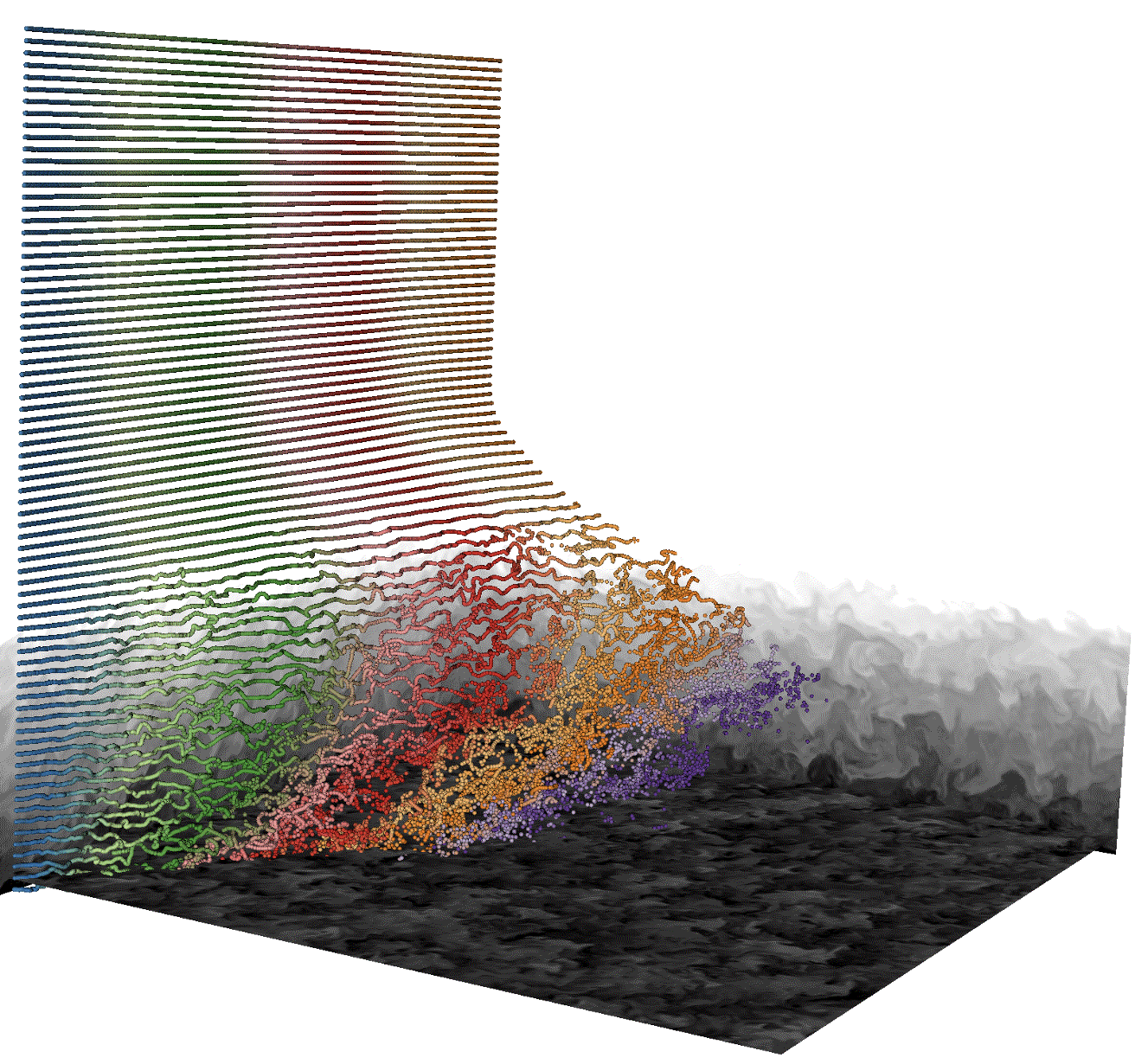}
  \caption{\label{abl_flow}
    NekRS simulation for the atmospheric boundary layer flows with particle tracer
    (Simulation by Lidquist~\cite{lindquist21}.}
 \end{wrapfigure}
In this paper, we consider the GABLS1 benchmark, illustrated in
Fig.~\ref{abl_flow},  which is a well-documented stably-stratified flow
problem.  The studies are conducted using the Argonne-developed
open-source Navier--Stokes (NS) solver, Nek5000/RS, which is based on
high-order spectral element (SE) discretizations~\cite{dfm02}.  NekRS~\cite{nekrs} is a GPU-accelerated
version of Nek5000~\cite{nek5000} developed under the ECP CEED
project, targeting application simulations on various acceleration-device 
based exascale computing platforms~\cite{nekrs,min-sc22}. 

In our earlier reports~\cite{ANL22_report}, ~\cite{ANL23_report}, we demonstrated 
our newly developed SGS models based on the work of~\cite{sullivan94}, \cite{Moeng2015} 
that involve the use of a mean-field eddy viscosity (MFEV)
in conjunction with either a high-pass filter (HPF) method, an algebraic Smagorinsky method,
or the solution of an SGS turbulent kinetic energy equation (TKE) for the isotropic small scale motion. 
The model fidelity and scaling performance of Nek5000/RS on DOE's leadership computing platforms 
in comparison with that of
AMR-Wind, a block-structured second-order finite-volume code with adaptive-mesh-refinement 
capabilities, were discussed in~\cite{min-ijhpca24}.
Here we focus on model fidelity of Nek5000/RS in comparison to that of AMR-Wind as well as issues related to numerical convergence.


This paper is organized as follows. 
Section 2 presents the governing equations for our LES modeling approach. 
Section 3 briefly describes the GABLS benchmark problem, Section 4 discusses our SGS models, and Section 4 presents verification and convergence studies.
Section 5 presents the way the traction boundary conditions are obtained, and Section 6 contains the results obtained with the newly implemented SGS models.
In Section 7 we present some of the conclusions from our study.

\section{Large Eddy Simulation Model}
For the atmospheric LES, we consider the governing equations consisting of 
the incompressible Navier--Stokes (NS) and  potential temperature equations 
in nondimensional form, solved in a {\em spatially filtered} resolved-scale 
formulation defined as 
\index{Navier-Stokes}
\begin{eqnarray}
    \label{eq:ns}
    \pp{\bar u_i}{t} + \bar u_j\frac{\partial {\bar u_i}}{\partial x_j}
     &=& -\frac{1}{\bar \rho}\frac{\partial {\bar p}}{\partial x_i}
         -\pp{\tau_{ij}}{x_j} + f_i - \frac{\theta^{\prime}}{\theta_0} g_i, \\
    \label{eq:inc}
    \frac{\partial {\bar u_j}}{\partial x_j} &=& 0, \\
    \label{eq:energy}
    \pp{\bar \theta}{t} + {\bar u_j}\frac{\partial {\bar \theta}}{\partial x_j}
     &=& -\pp{\tau_{\theta j}}{x_j}\,,
\end{eqnarray}
where 
$\bar u_i$ is the $i$th component of the resolved-scale velocity vector, $\bar \rho$ is
the density,  $\bar p$ is the pressure, $g_i$ is the gravity acceleration vector,
and $\bar \theta$ is the potential temperature in the resolved scale.
The scalar $\theta^{\prime}/\theta_0$ in the buoyancy force is defined by 
\begin{eqnarray}
 \frac{\theta^{\prime}}{\theta_0} = \frac{\bar{\theta}-\theta_0}{\theta_0},
\end{eqnarray}
where $\theta_0$ is the reference potential temperature. $f_i$ represents
the Coriolis acceleration defined by
\begin{eqnarray}
 f_{i} = -2 \epsilon_{i3k} \Omega \bar{u}_k,
\end{eqnarray}
where $\epsilon_{ijk}$ is the alternating unit tensor and $\Omega$ is the
planetary rotation rate vector at the point of interest on the planet (which is
dependent on latitude), and $j=3$ corresponds to the vertical direction.  


\noindent
In addition, $\tau_{ij}$ and $\tau_{\theta j}$ are the stress tensors in the momentum
and energy equations including SGS modeling terms defined as
\begin{equation}
 \tau_{ij} = -\frac{2}{Re} S_{i j} + \tau^{sgs}_{ij} =
   -\frac{1}{Re} \left( \frac{\partial{\bar u_i}}{\partial x_j} +
   \frac{\partial {\bar u_j}}{\partial x_i} \right) +
   \tau^{sgs}_{ij}, 
\label{eqn.tauij}
\end{equation}
\noindent
and
\begin{equation}
   \tau_{\theta j} = -\frac{1}{Pe} \frac{\partial{\bar \theta}}{\partial x_j} +
   \tau^{sgs}_{\theta j}, 
\end{equation}
where $Re$ is the Reynolds number, $Pe$ is the Peclet number,
$S_{ij}$ is the resolved-scale strain-rate tensor, and
$\tau^{sgs}_{ij}$ and $\tau^{sgs}_{\theta j}$ are the SGS stress tensors.

\vspace{-0.5em}
\section{ABL GABLS Benchmark}
We consider the GABLS benchmark problem~\cite{beare06} which is a stable ABL where
the ground temperature is cooler than the air temperature and continues to cool over the 
duration of the simulation. 
We define the domain as $\Omega = L_x \times L_y \times L_z =$ 400 m $\times$ 400 m $\times$ 400 m,
with the streamwise direction $x$, the spanwise direction $y$, and  the vertical direction $z$.
We initialize our simulations at time $t=0$ with a constant velocity in the
streamwise direction equal to the geostrophic wind speed of $U=8$ m/s. We define
the initial potential temperature by 265 K in $0\le z\le 100$ m and linearly increase at
a rate of 0.01 K/m in the range of 100 m $\le z\le 400$ m.  The reference potential
temperature is 263.5 K.  The Reynolds number is $Re= UL_b/\nu$,
where $L_b=100$ m is the thickness of the initial thermal boundary layer and $\nu$ 
is the molecular viscosity, and it is $\approx 50M$, which precludes direct numerical 
simulation (DNS) wherein all turbulent scales are resolved.
We add an initial perturbation to the temperature with an amplitude of 0.1 K
on the potential temperature field for $0 \le z\le 50$ m.

Periodic boundary conditions (BCs) are used in the streamwise and spanwise directions.
At the top boundary, ($z=400$ m), a stress-free, rigid lid is applied for momentum, and
the heat flux for the energy equation is set consistent with the 0.01 K/m temperature
gradient initially prescribed in the upper region of the flow.
At the bottom boundary, we perform simulations with impenetrable traction BCs
for the velocity, where the specified shear stress comes from Monin-Obukhov
similarity theory~\cite{monin1954}. For the energy equation, a heat flux is
applied that is derived from the same theory and a specified potential
temperature difference between the flow at a height, $z_1$, and the surface.
The surface temperature is from the GABLS specification following the rule
$\theta_b(t) = 265 - 0.25 t$, where $t$ is in hours.  Because the boundary
conditions are periodic (lateral), or the mass flow rate through the boundaries
is zero (top and bottom),  pressure boundary conditions are not needed.

\vspace{-0.5em}
\section{SGS Modeling Approaches in Nek5000/NekRS}

We have extended the range of SGS modeling approaches in Nek5000/NekRS based on 
the MFEV approach of~\cite{sullivan94} and have implemented three
different ways to include small0scale isotropic motion as described in our earlier ANL reports,
~\cite{ANL22_report} and~\cite{ANL23_report}. Our SGS modeling approaches investigated are
summarized below,

\begin{table}[h]
\centering
\begin{tabular}{|l|l|l|}
\hline
 Model Name & SGS Anisotropic  & SGS Isotropic   \\
\hline
 MFEV/HPF   &  MFEV &  HPF   \\
 MFEV/SMG   &  MFEV &  SMG   \\
 MFEV/SGS-TKE   &  MFEV &  SGS-TKE   \\
\hline
\end{tabular}
\end{table}
\noindent
where HPF refers to the high pass filter of~\cite{Stolz_Schlatter_2005}, 
SMG to the algebraic Smagorinsky model~\cite{Smagorinsky:1963}, and TKE to the solution of a SGS turbulent 
kinetic energy equation~\cite{sullivan94}. 

We consider traction boundary conditions (BCs) along the lower wall in all simulations discussed here,
in which the normal velocity component is set to zero and traction BCs are specified for the two
horizontal velocity components; in addition, heat flux BCs are specified for the potential
temperature, based on the Monin-0Obukhov log law~\cite{monin1954}.


We implemented the traction BCs for the horizontal velocity components in the context 
of the log law, following the approach of ~\cite{Grotjans_Menter1998} 
and~\cite{Kuzmin_2007}, which is suitable for finite element methods based on a weighted residual formulation.
The traction BCs imposed on the tangential velocity are based on the horizontally averaged slip velocity that develops 
at the boundary or at a specified sampling $z-$location from the lower wall.

The SGS stress tensors $\tau^{sgs}_{ij}$ and $\tau^{sgs}_{\theta j}$ 
are expressed in terms of a non-isotropic part, $\left\langle\tau^{sgs}_{i j}\right\rangle$, and an isotropic part,
$\tau^{\prime}_{i j} $. Thus, the sub-grid-scale dissipation is based on a 
non-isotropic MFEV $\nu_T$, obtained by the horizontally averaged mean strain rate, and an isotropic, 
fluctuating part $\nu_t$.
The law of the wall is effected through the use of the MFEV concept, and the approach originally  
by~\cite{SCHUMANN1975376} is used to convert the horizontally averaged traction to local values based on the local slip velocity 
in each of the horizontal directions. The SGS model of~\cite{sullivan94} for the momentum is based on the following expression:
\begin{equation}
 \tau^{sgs}_{i j}=\left\langle\tau^{sgs}_{i j}\right\rangle + \tau^{\prime}_{i j} 
 = -2 \nu_{T}\left\langle S_{i j}\right\rangle - 2 \gamma \nu_t S_{i j}. 
\label{eqn.tijdef2}
\end{equation}

For the energy equation, the definition of $\tau_{\theta j}$ is
\begin{equation}
\quad \tau_{\theta j}= \left\langle\tau^{sgs}_{\theta z}\right\rangle + \tau^{\prime}_{\theta j} = -\nu_{\Theta} \frac{\partial\langle \theta \rangle}{\partial z} - \nu_\theta \frac{\partial \theta}{\partial x_j},
\label{eqn.tauthetadef}
\end{equation}
where $\left\langle \ \ \right\rangle$ denotes averaging over the
homogeneous directions and $\nu_T$ is an average eddy viscosity, which is expressed in
terms of mean flow quantities. In Eq.~(\ref{eqn.tijdef2}), $\gamma$ is
an ``isotropy factor,'' which accounts for variability in the SGS constants due to
anisotropy of the mean flow. 
When the fluctuating (isotropic) part of 
turbulent motion is taken into account through the use of the fluctuating strain rate, $\nu_t$ in Eq.~(\ref{eqn.tijdef2}) 
is nonzero and the full stress tensor has to be taken into account. The diffusivities $\nu_\Theta$ and $\nu_\theta$ in~(\ref{eqn.tauthetadef})
are given by $\nu_\Theta = \nu_T/Pr_t$ and $\nu_\theta = \gamma \nu_t/Pr_t$, where $Pr_t$ is either $1$ or $1/3$(~\cite{sullivan94}). 
Thus, the momentum and potential temperature equations are given by
\begin{eqnarray}
 \label{eq:ns3}
  \frac{\partial \bar{u}_{i}}{\partial t}
  +\bar{u}_{j}\frac{\partial \bar{u}_{i}}{\partial x_{j}}
  & = & -\frac{\partial \bar{p}}{\partial x_{i}}-\pp{\tau_{ij}}{x_j}
        -2 \epsilon_{i3k} \Omega \bar{u}_k 
    +  (1-\delta_{i3}) \frac{\partial }{\partial z}\nu_{T} 
   \frac{\partial\langle \bar{u}_i\rangle}{\partial z} - \frac{\theta^{\prime}}{\theta_0} g_i \quad  \\ 
  & = & -\frac{\partial \bar{p}}{\partial x_{i}}
        + \frac{\partial}{\partial x_j} \left(\frac{1}{\operatorname{Re}} +\gamma \nu_t \right) 
          2S^n_{ij} - 2 \epsilon_{i3k} \Omega \bar{u}_k 
    +   (1-\delta_{i3}) 
      \frac{\partial }{\partial z}\nu_{T} \frac{\partial\langle \bar{u}_i \rangle}{\partial z} - \frac{\theta^{\prime}}{\theta_0} g_i, \quad
\\
  \frac{\partial \bar{\theta}}{\partial t}
  + \bar{u}_{j}\frac{\partial \bar{\theta}}{\partial x_{j}}
    \label{eq:energy_filt3}
  & = & \frac{\partial }{\partial x_{j}} \left(\frac{1}{\operatorname{Pe}}
   +\frac{\gamma \nu_t}{Pr_t} \right) \frac{\partial \bar{\theta}}{\partial x_{j} }
   + \frac{\partial }{\partial z}\frac{\nu_{T}}{Pr_t} \frac{\partial\langle \bar{\theta} \rangle}{\partial z}.
\end{eqnarray}


The expression for the MFEV $\nu_T$ is derived so that the law-of-the-wall behavior
can be recovered in the absence of any resolved turbulence, as explained below.
Following~\cite{sullivan94}, we impose a ``constant flux'', traction-type boundary condition at $z=z_{1}$, 
which states that the sum of the SGS and resolved momentum fluxes be equal to the surface stress, i.e.,
\begin{equation}
\left[\left\langle\tau^{sgs}_{u w}\right\rangle^{2}+\left\langle\tau^{sgs}_{v w}\right\rangle^{2}\right]^{1 / 2}
 +\left[\langle u w\rangle^{2}+\langle v w\rangle^{2}\right]^{1 / 2}=u_{\tau}^{2} .
\label{eqn.stress}
\end{equation}
As described in~\cite{ANL22_report}, this traction boundary condition in Nek is imposed at the
first grid point in the vertical direction, which is assumed to be a point inside the log layer at a location $z=z_1$,
where the boundary condition for the vertical velocity component is defined to be zero.
For this reason the second term in Eq.~(\ref{eqn.stress}) corresponding to the resolved momentum fluxes is
identically equal to zero.
In~\cite{sullivan94}, a predictive relationship for the MFEV at the first grid point 
$z_1$, $\nu_T^\star=\nu_T(z_1)$ is obtained by invoking the approximation
that the fluctuating components of strain are neglected compared with the mean strain so
that only the horizontally averaged SGS stress in Eq.~(\ref{eqn.tijdef2}) is retained.
This leads to
\begin{equation}
\begin{aligned}
\left\langle\tau^{sgs}_{u w}\right\rangle &=-\nu_{T} \frac{\partial\langle u\rangle}{\partial z}, \\ 
\left\langle\tau^{sgs}_{v w}\right\rangle &=-\nu_{T} \frac{\partial\langle v\rangle}{\partial z}.
\label{eqn.tijsim}
\end{aligned}
\end{equation}

A model for MFEV at any height, which is consistent with this idea is as follows:
\begin{equation}
\nu_{T}=\nu_{T}^{\star} \frac{\kappa z_{1}}{u_{\tau} \phi_{m}\left(z_{1}\right)} \sqrt{2\left\langle S_{i j}\right\rangle\left\langle S_{i j}\right\rangle},
 \end{equation}
\noindent
where $u_{\tau}$ is the friction velocity, $\kappa$ the von Karman constant, and $\phi_{m}$ the Monin-Obukhov stability function 
for momentum. The expression for $\nu_T^\star$ is
\begin{equation}
\nu_{T}^{\star}=\frac{u_{\tau} \kappa z_{1}}{\phi_{m}\left(z_{1}\right)}.
\label{eqn.nuTsim}
\end{equation}

Equation~(\ref{eqn.nuTsim}) provides an adaptive method for estimating the MFEV
needed to force the computed wind speed derivative to match with similarity theory at $z = z_1$. 
In contrast to~\cite{sullivan94}, a similar correction was also applied to the SGS potential temperature
field, and $\tau^{sgs}_{\theta z}$ becomes 
\begin{equation}
\left\langle\tau^{sgs}_{\theta z}\right\rangle =-\frac{\nu_{T}}{Pr_t} \frac{\partial\langle \theta \rangle}{\partial z}.
\label{eqn.tautheta}
\end{equation}

In our approach, following~\cite{Grotjans_Menter1998} and~\cite{Kuzmin_2007}, as was described in the previous subsection, in our approach 
the boundary of the computational domain
is not located exactly at the wall but at a finite distance from the wall corresponding to a fixed value of $z_1^+=z/z_0$.
Strictly speaking, this implies that a boundary layer of width $z_1$ (corresponding to the specified value of $z_1^+$) should
be removed from the computational domain; however, it is assumed that this width is very small at high Reynolds numbers
and can be considered negligible, so that the equations can be solved in the whole domain with traction BCs prescribed
on the lower boundary. Since the choice of $z_1^+$ is rather arbitrary, we have found that values of $z_1^+$ up
to $10$ at the target $Re$ produce averaged results that do not differ significantly.

In all apporaches described below for the modeling of the small-scale isotropic motion, 
the SGS dissipation is effected through a non-isotropic MFEV obtained by the horizontally-averaged mean strain rate.
In the HPF approch, the isotropic, fluctuating, part of the SGS modeling is taken into account through the
use of a high-pass filter of~\cite{Stolz_Schlatter_2005}, which is not eddy-viscosity based and thus for this case 
$\nu_t$ in~(\ref{eqn.tijdef2}) is equal to zero. Instead, an additional term of the form $-\chi H_{N} * \bar{u}_{i}$ 
and $-\chi H_{N} * \bar{\theta}$ is added to the momentum and energy equations, respectively, as explained in~\cite{ANL22_report}.

In the SMG approach, the isotropic, fluctuating part is taken into account through an algebraic Smagorinsky 
(SMG) model based on the fluctuating strain rate. The expression used for the isotropic part of the eddy viscosity $\nu_t$ is 
\begin{equation}
\nu_t=\left(C_s \Delta\right)^2 \sqrt{2 S^{\prime}_{i j} S^{\prime}_{i j}},
\end{equation}
and $S^{\prime}_{i j}$ is given by
\[ S^{\prime}_{i j} = S_{i j}-\left\langle S_{i j}\right\rangle.\]
The Smagorinsky constant $C_s$ is written in terms of $C_k$ and $C_{\varepsilon}$ as
\begin{equation}
C_s=\left(C_k \sqrt{\frac{C_k}{C_\epsilon}}\right)^{1 / 2}.
\end{equation}
From~\cite{sullivan94}, the SGS constants are $C_k = 0.1$ and $C_{\varepsilon}=0.93$. 
The isotropy factor $\gamma$ is obtained from
\begin{equation}
\gamma=\frac{S^{\prime}}{S^{\prime}+\langle S\rangle},
\end{equation}
where
\begin{equation}
\langle S\rangle=\sqrt{2\left\langle S_{i j}\right\rangle\left\langle S_{i j}\right\rangle}.
\end{equation}
and
\begin{equation}
S^{\prime}=\sqrt{2 \left\langle S^{\prime}_{i j} S^{\prime}_{i j} \right\rangle}, 
\end{equation}

In the third approach, TKE, a transport equation is solved for the SGS turbulent kinetic energy equation,according to~\cite{sullivan94}: 

\begin{equation}
\left(\frac{\partial}{\partial t}+u_j \frac{\partial}{\partial x_j}\right) e =2 \gamma \nu_t S^{\prime}_{i j} S^{\prime}_{i j} + \frac{g}{\theta_0} \tau_{\theta z} -C_\varepsilon \frac{e^{3 / 2}}{L}+\frac{\partial}{\partial x_i} \left(\frac{1}{\operatorname{Re}} +2 \gamma \nu_t \right) \frac{\partial e}{\partial x_i},
\label{eqn.sgstke_equation}
\end{equation}
where the fluctuating eddy viscosity, $\nu_t$, is given by
\begin{equation}
\nu_t=C_k L e^{1 / 2}.
\label{eqn.nut_sgstke}
\end{equation}
%
%
The model length scale $L$ that appears in the SGS TKE equation~(\ref{eqn.sgstke_equation}) 
and in expression~(\ref{eqn.nut_sgstke}) 
for the fluctuating eddy viscosity is obtained as follows. For unstable stratification,
the length scale $L$ is defined by
\[L=L_{SMG}=\Delta,\]
while for stable stratification, $L$ is reduced as suggested by~\cite{deardorff80} and is obtained by the following expression:
\[L=L_{DRD}=\frac{0.76 e^{1 / 2}}{\left(\frac{g}{\theta_0} \frac{\partial \theta}{\partial z}\right)^{1 / 2}}.\]
The definition of $\tau_{\theta z}$ is
\[\quad \tau_{\theta z}=-\nu_\theta \frac{\partial \theta^{\prime}}{\partial z},\]
where
\[\theta^{\prime}=\theta-\left\langle \theta \right\rangle\]
The constant $C_{\varepsilon}$ is given from
\[C_{\varepsilon}=0.19+0.74 L / \Delta.\]
We have the SGS constant $C_k = 0.1$; $C_{\varepsilon}=0.93$ 
for the case $L=L_{SMG}=\Delta$, whereas it is variable 
for the case $L=L_{DRD}$. All other quantities such as 
$S^{\prime}_{i j}$, $\gamma$, and $\langle S\rangle$ 
are defined as in the second case SMG. 

\vspace{-0.5em}
\section{Traction Boundary Condition}
\label{s:tractionBC}

\def\bt{{\bf t}}
\def\btau{{\bf \tau}}
\def\bn{{\bf n}}
\def\bS{{\bf S}}

\noindent
Following~\cite{SCHUMANN1975376}, the exact form of the traction boundary condition corresponding to the local value of the 
momentum flux for each of the two horizontal components of the tangential velocity $\bu_t=\bu - \bn (\bn \cdot \bu)$, and
for the case where the normal to the boundary direction is aligned with the $z$-direction, is obtained as follows:
\begin{equation}
 \bt_{w} = \tau_w \frac{\bu_t}{|\bu_t|},
\end{equation}
where 
\begin{eqnarray}
\tau_w = u^2_{\tau}. 
\end{eqnarray}
The momentum flux at the boundary is based on a horizontally averaged value of the friction velocity $u_{\tau}$ obtained 
by using the law of the wall for rough walls which for the velocity and temperature is defined as
\begin{eqnarray}
u^+     & = & \frac{|\bu_t|}{u_\tau}=\frac{1}{\kappa}\left(\ln\frac{z_1}{z_0} + \beta_m \frac{z_1}{L}\right), \\
\theta^+ & = & \frac{|\Delta \theta|}{\theta_\tau}=\frac{1}{\kappa}\left(\ln\frac{z_1}{z_0} + \beta_h \frac{z_1}{L}\right),
\label{eqn.loglaw}
\end{eqnarray}
where $\kappa$ is the von Karman constant, $z_1$ the location of the lower wall and $z_0$ is a roughness related 
length scale; $\bu_t$ is the velocity vector parallel to the wall, and $\Delta \theta$ 
is the difference between the actual wall temperature located at $z=0$ and the temperature at $z_1^+$ which is the 
location of the lower boundary in the computation. The constants $\beta_m$ and $\beta_h$ are in general not equal
to each other and their values are usually taken to be $4.8$ and $7.2$, respectively.
The boundary conditions are implemented through the derivatives of the above quantities in the wall normal $z$ 
direction as follows:
\begin{eqnarray}
 \frac{\partial |\bu_t|}{\partial z} = \frac{u_\tau}{\kappa z_1} \phi_m
   \quad{\rm and}\quad
 \frac{\partial |\theta|}{\partial z} = \frac{\theta_\tau}{\kappa z_1} \phi_h,
 \end{eqnarray}
\noindent
where 
\begin{eqnarray}
 \phi_m \left(1 + \beta_m \frac{z_1}{L}\right)
   \quad{\rm and}\quad
 \phi_h \left(1 + \beta_h \frac{z_1}{L}\right).
 \end{eqnarray}
The value of $u_{\tau}$ is coupled with the corresponding value of $\theta_{\tau}$ through
the Monin-Obukhov lentgh $L$ defined as
\begin{equation}
\label{eqn.MOlength}
 L = u^2_{\tau} \left( \frac{\kappa g \theta_{\tau}}{\theta_0}\right)^{-1}.
\end{equation}

\noindent
The values of $u_{\tau}$ and $\theta_{\tau}$ are obtained by solving the two-equation 
system~(\ref{eqn.loglaw}); and its solution procedure is described below. 
Rewriting equations~(\ref{eqn.loglaw}) and using definition~(\ref{eqn.MOlength}), we obtain
the following system:

\begin{equation}
\begin{aligned}
&\left( \ln \frac{z_1}{z_{0}} \right) u_{\tau}^{2} - \kappa U u_{\tau} + \beta_{m} \kappa g z {\tilde{\theta}_{\tau}}= 0, \\ 
&\beta_h \kappa g z {\tilde{\theta}_{\tau}}^{2} + \left( \ln \frac{z_1}{z_{0}} \right) u_{\tau}^{2} {\tilde{\theta}_{\tau}} 
- \kappa \Delta \tilde{\theta} u_{\tau}^{2} = 0,
\label{eqn.system}
\end{aligned}
\end{equation}
where $U=|\left< \bu_t\right>|$ is the absolute value of the horizontally averaged slip velocity in the horizontal direction
and $\Delta \tilde{\theta} = | \left< \Delta \theta \right> | / \theta_0 $ is the nondimensional absolute value of the horizontally averaged 
temperature difference between the wall temperature located at $z=0$ and the computed temperature at $z_1$ which is the location 
of the lower boundary of the computational domain. Both $\left< \bu_t\right>$ and $\left< \Delta \theta \right>$ are obtained from the 
velocity and temperature solutions at the previous timestep during the simulation. 
In system~(\ref{eqn.system}) the only unknowns are $u_{\tau}$ and ${\tilde{\theta}_{\tau}} = \theta_{\tau}/\theta_0$ and the 
solution can be obtained analytically as

\begin{equation}
\tilde{\theta}_{\tau 1,2}=\frac{u_{\tau 1,2}}{\beta_m \kappa g z_1} \left(\kappa U - u_{\tau 1,2} \ln \frac{z_1}{z_0} \right),
\label{eqn.solthetau}
\end{equation}
and
\begin{equation}
u_{\tau 1,2}=\frac{1}{2} \frac{2\frac{\beta_{h}}{\beta_{m}}-1}{\frac{\beta_{h}}{\beta m}-1} 
             \frac{\kappa U}{\ln \frac{z_1}{z_0}} 
             \left[ 1 \pm \frac{\sqrt{\Delta}}{\left( 2 \frac{\beta_{h}}{\beta_{m}}-1\right)}\right],
\label{eqn.solutau}
\end{equation}
where $Ri$ is the bulk Richardson number with $Ri=g z_1 \Delta \tilde{\theta}/U^2$ and $\Delta = 1+4 Ri\left(\beta_{h}-\beta_{m}\right)$.
For the case that $\beta_m$ and $\beta_h$ are assumed to be equal, the solution simplifies to 

\begin{equation}
\begin{aligned}
&u_{\tau}=\frac{\kappa U}{\ln \frac{z_1}{z_0}}\left(1-\beta_{m} Ri\right), \\ 
&\tilde{\theta}_{\tau}=\frac{\kappa \Delta \tilde{\theta}}{\ln \frac{z_1}{z_0}} \left(1-\beta_{m} Ri\right), 
 \end{aligned} 
\end{equation}
which is the same solution as that obtained by~\cite{basu2008}. However, when the constants $\beta_m$ and $\beta_h$
are not equal and assuming that $\beta_h>\beta_m$, the general solution is given by~(\ref{eqn.solthetau}) and~(\ref{eqn.solutau}). 
We note that only one of the two solutions for $u_{\tau}$ in~(\ref{eqn.solutau}) leads to a positive value for 
$\theta_{\tau}$, and this is the smallest root $u_{\tau 2}$, which corresponds to the negative sign in front of $\Delta$. Thus, this is 
the solution used to specify $u_{\tau}$ and $\theta_{\tau}$ in our computations when traction BCs are used for the horizontal velocities.

\begin{figure*}[!t]
  \begin{center}
   \subfloat[]
   {
    \includegraphics[width=0.44\textwidth]{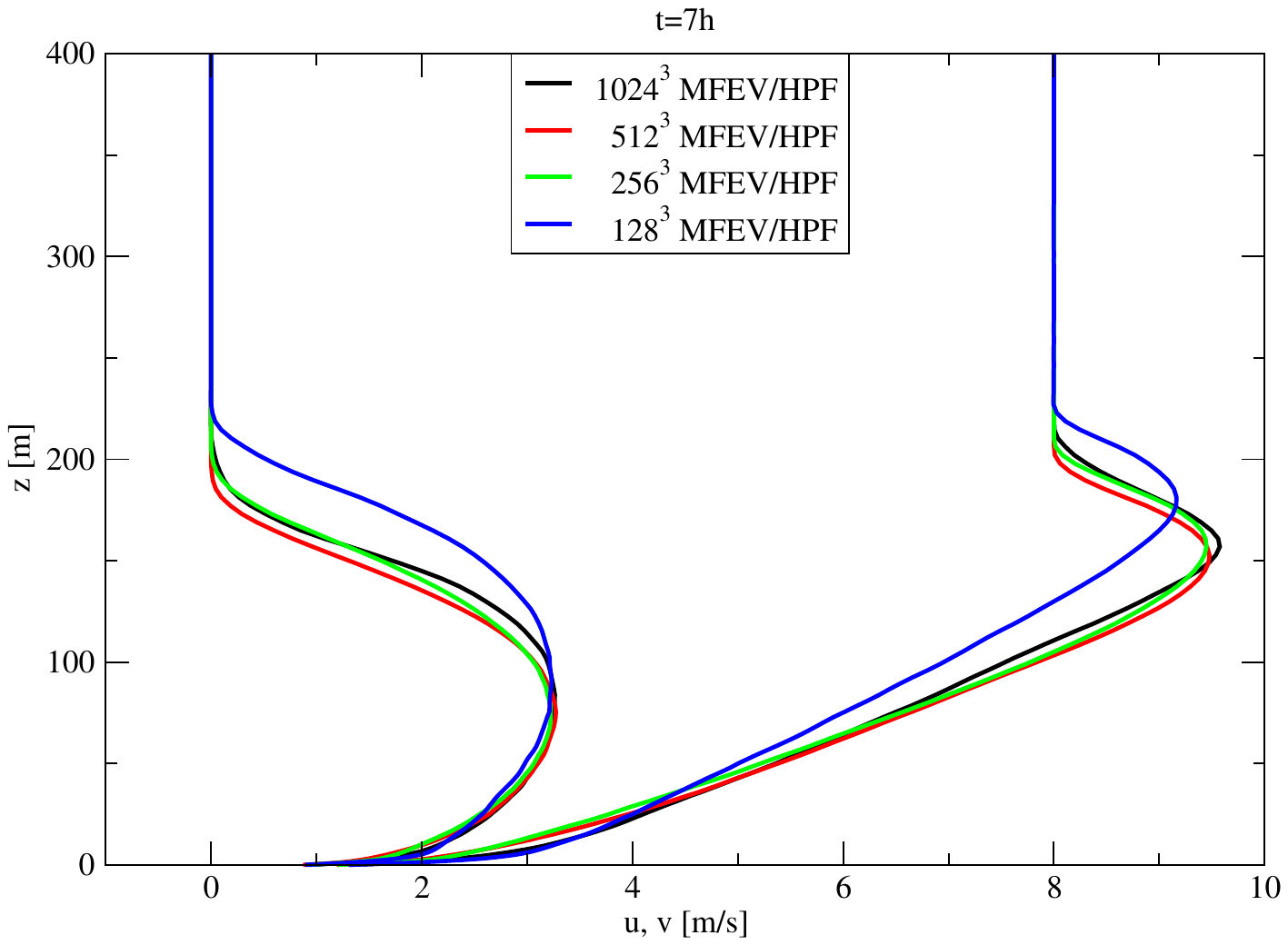}
   }
   \hspace{-1em}
   \subfloat[]
   {
    \includegraphics[width=0.44\textwidth]{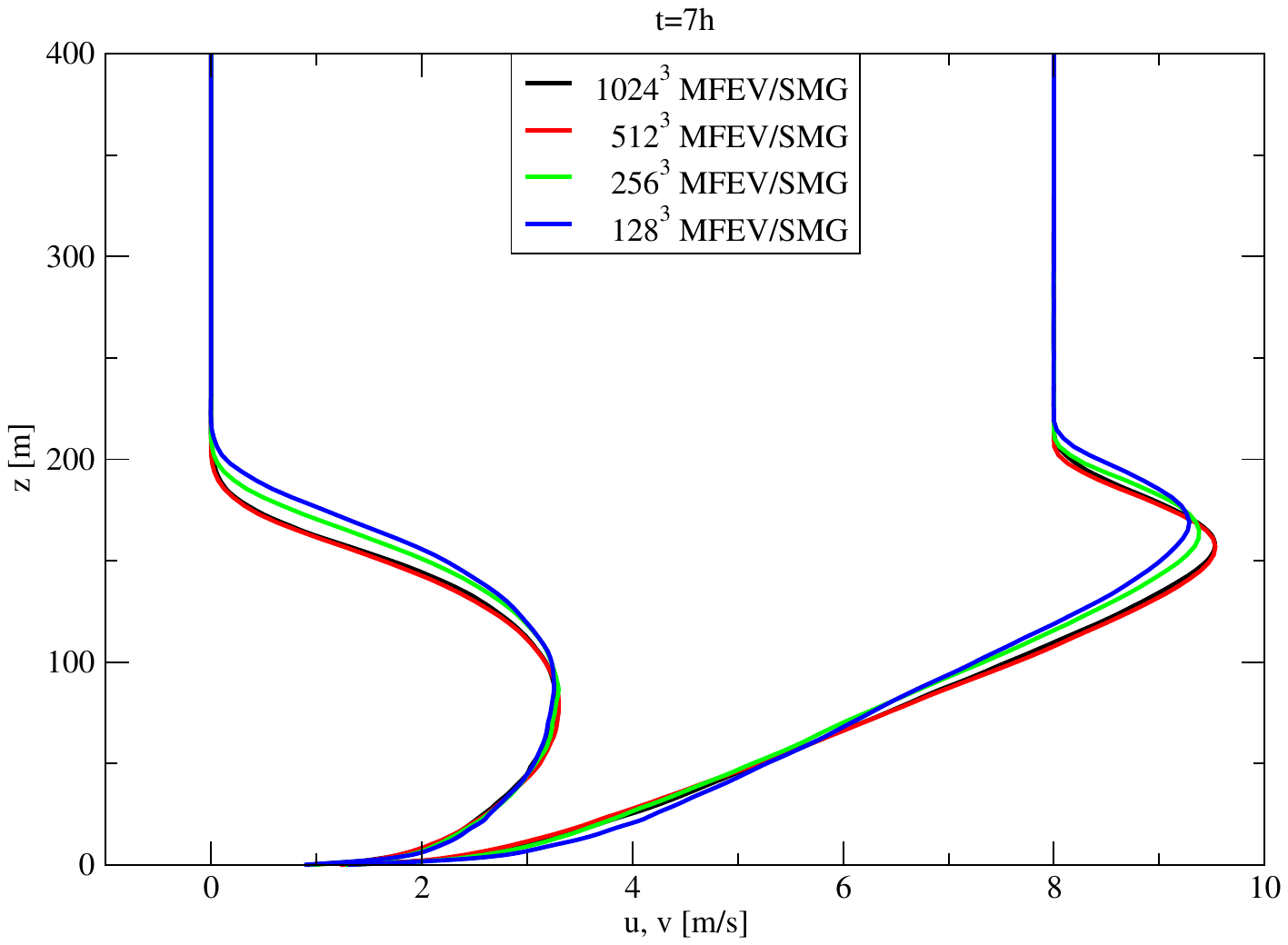}
   }
   \\
   \vspace{-1em}
   \subfloat[]
   {
    \includegraphics[width=0.44\textwidth]{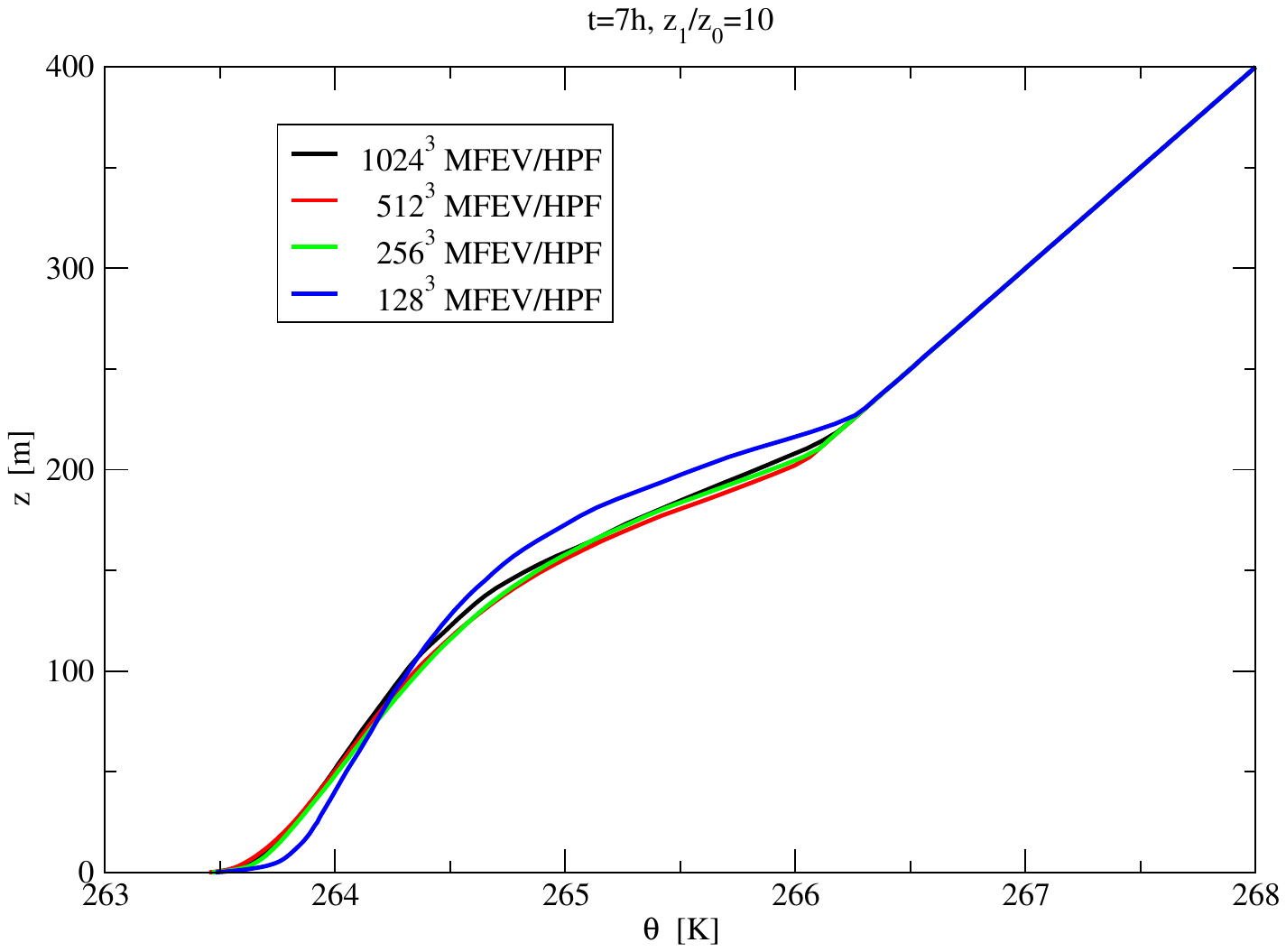}
   }
   \hspace{-1em}
   \subfloat[]
   {
    \includegraphics[width=0.44\textwidth]{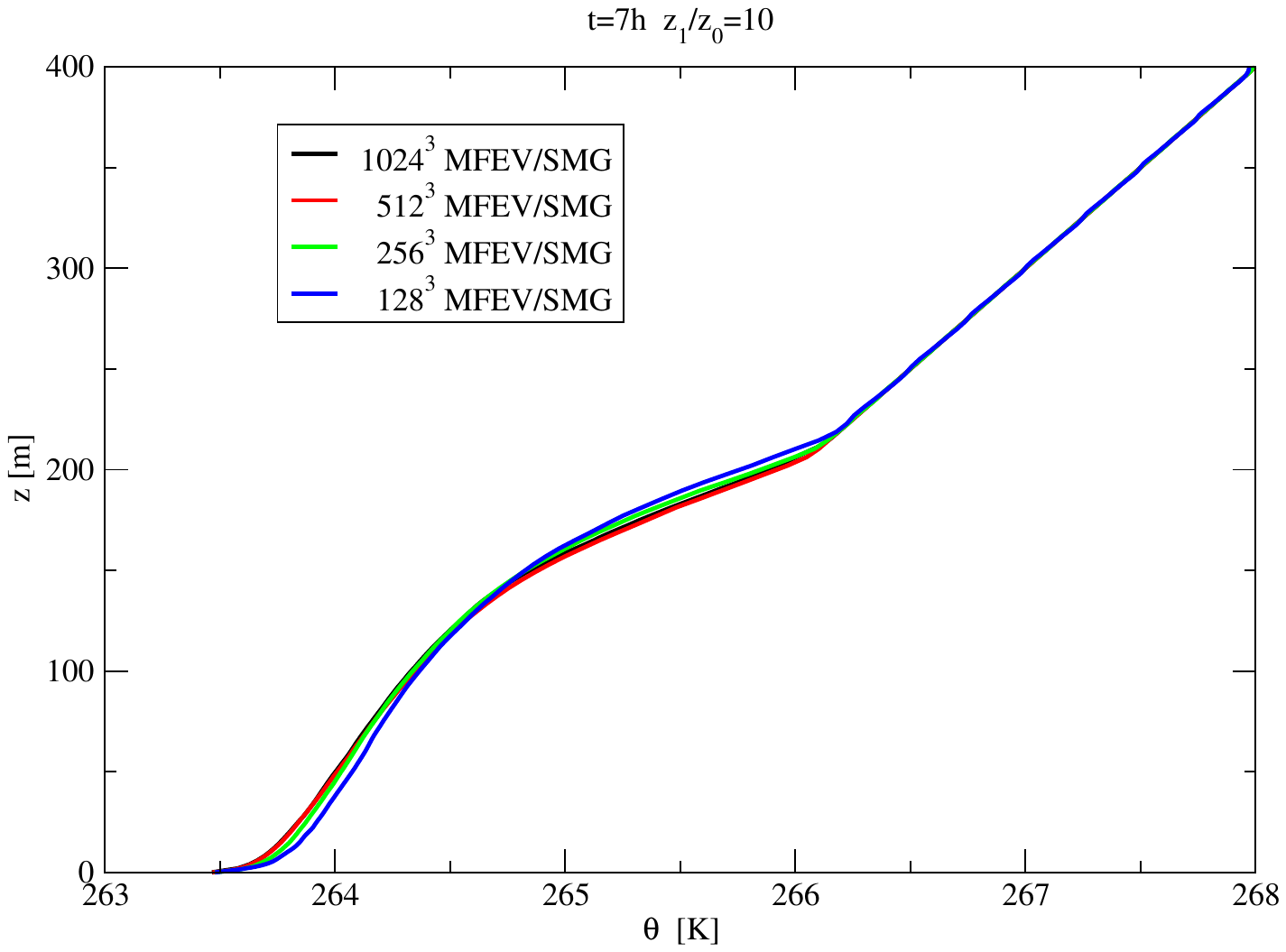}
    }
  \end{center}
  \caption{\label{fig.mfev_hpf_smg} 
   Nek5000/RS Convergence in horizontally averaged velocity 
   with (a) MEFV/HPF and (b) MFEV/SMG and potential temperature
   profiles at $t=7h$ with (c) MEFV/HPF and (d)  MFEV/SMG.}
\end{figure*}

\begin{figure*}
  \begin{center}
   \subfloat[]
   {
    \includegraphics[width=0.44\textwidth]{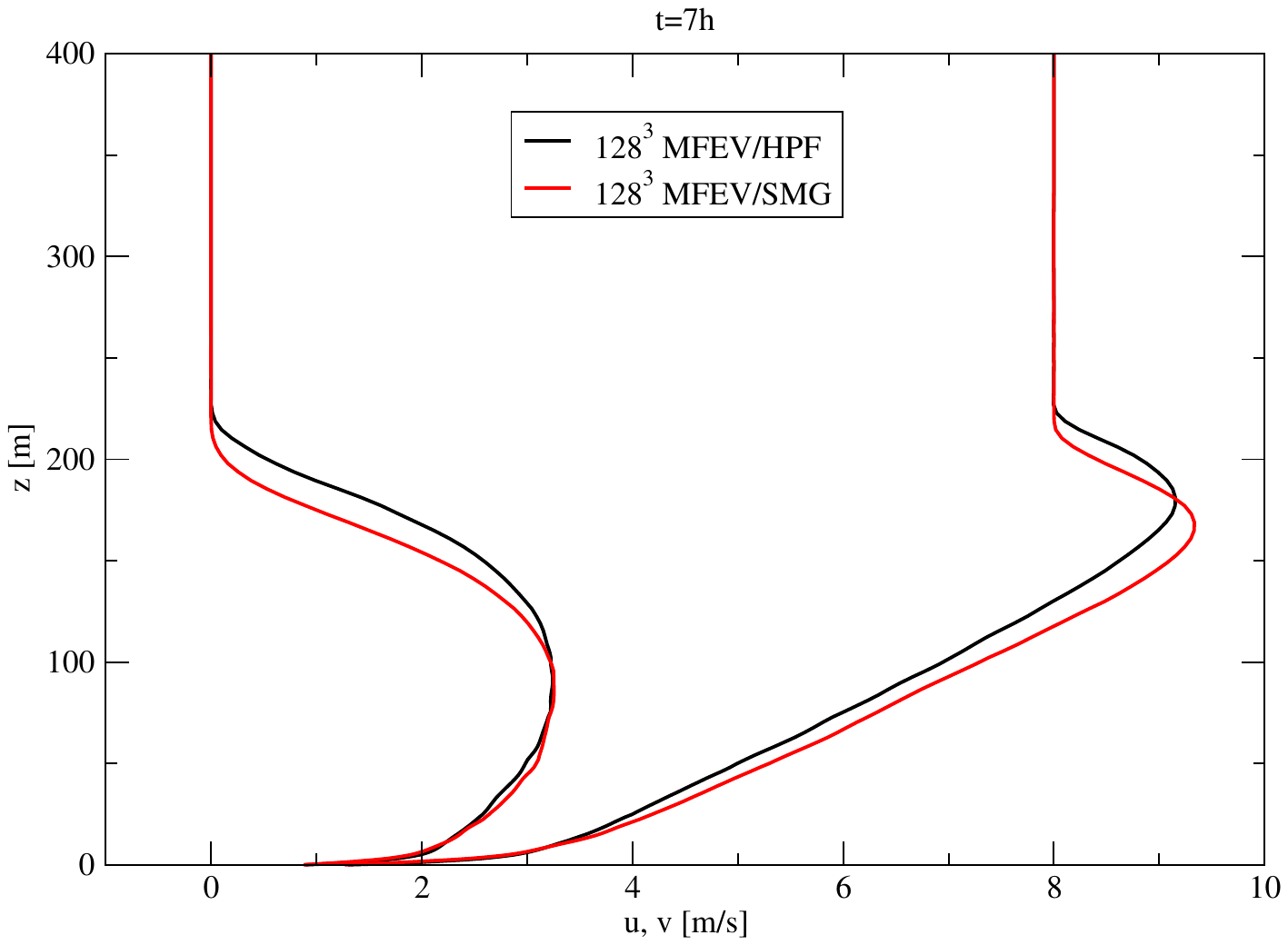}
   }
    \hspace{-1em}
   \subfloat[]
   {
    \includegraphics[width=0.44\textwidth]{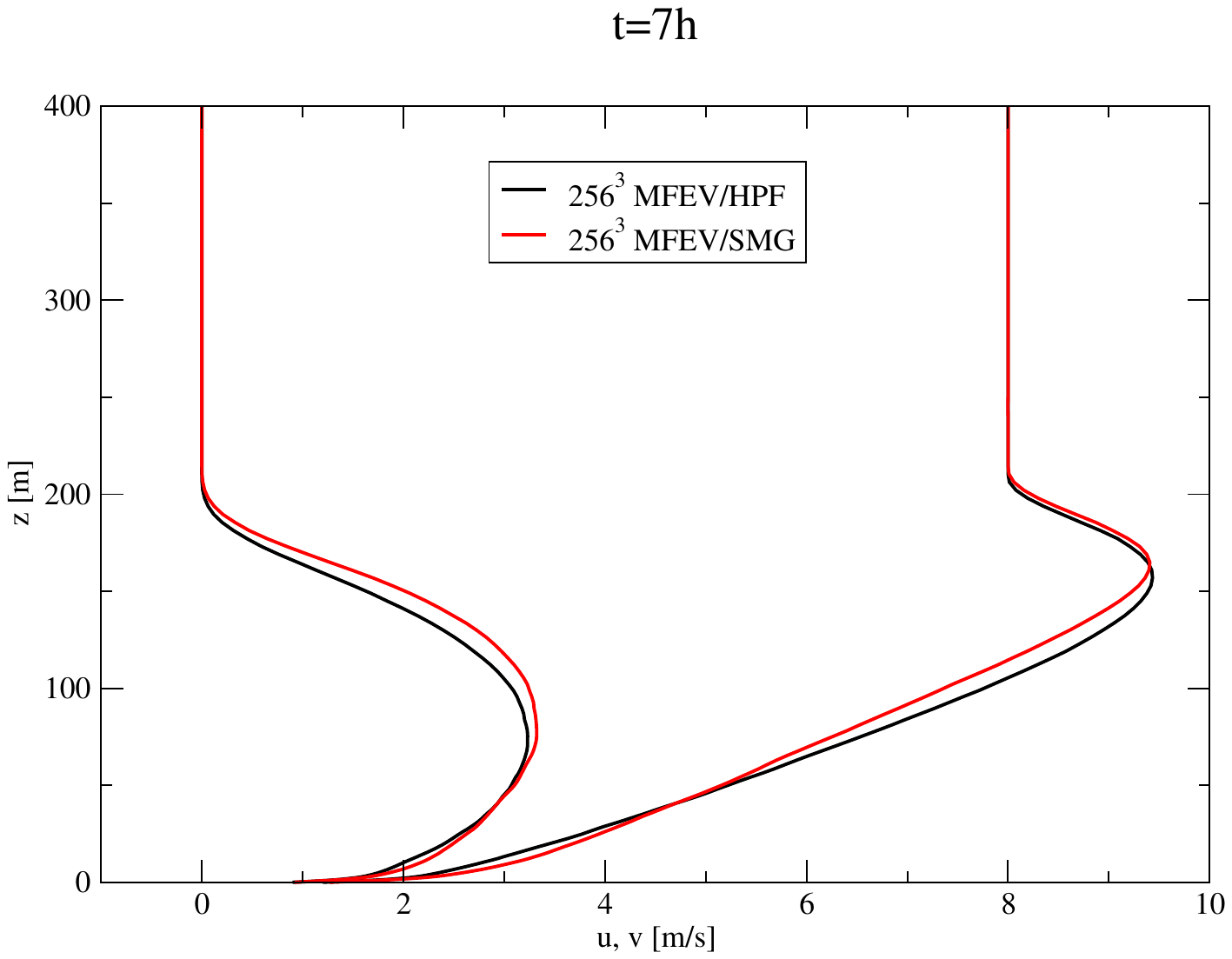}
   }
    \\
    \vspace{-1em}
   \subfloat[]
   {
    \includegraphics[width=0.44\textwidth]{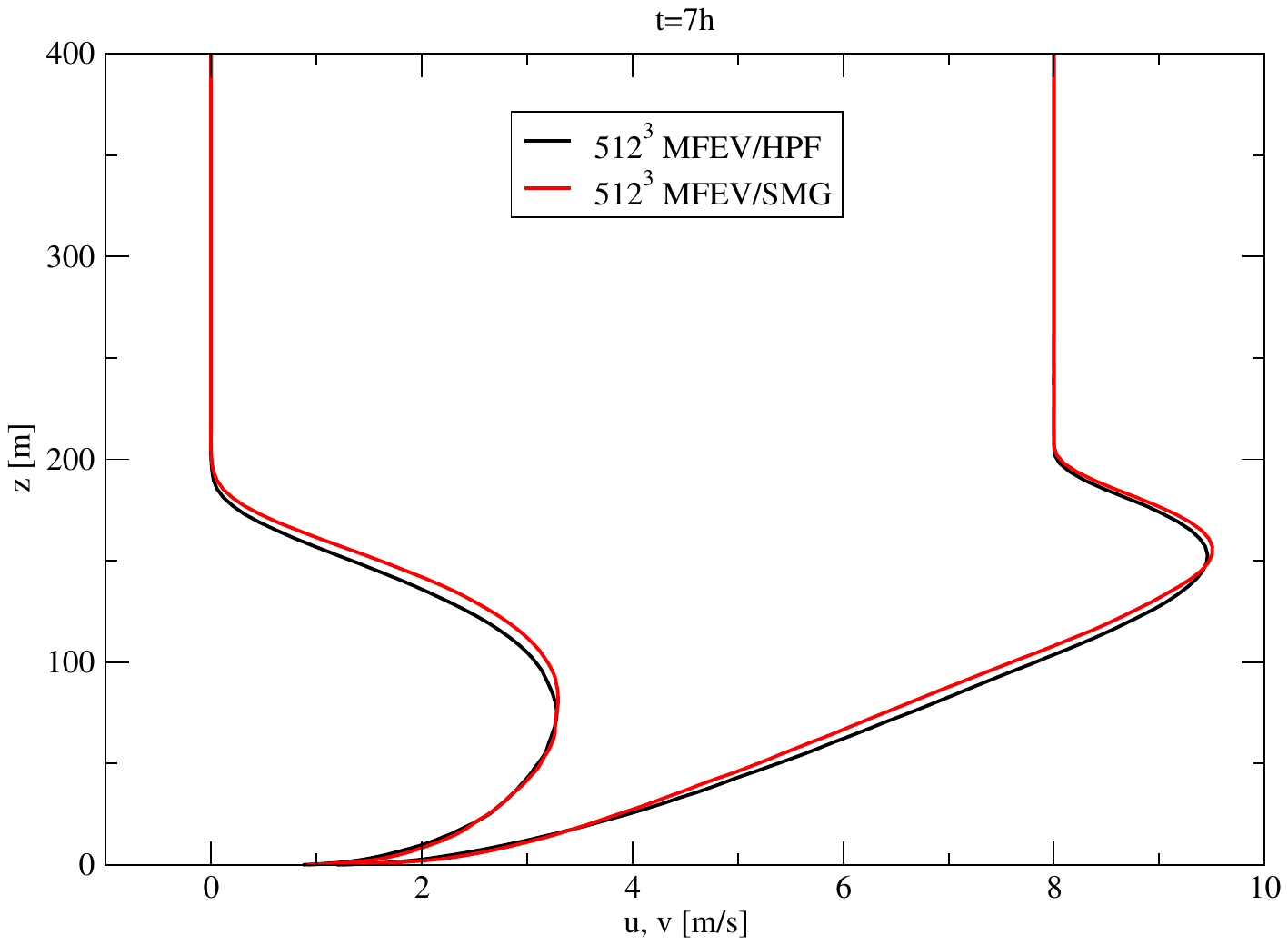}
   }
    \hspace{-1em}
   \subfloat[]
   {
    \includegraphics[width=0.44\textwidth]{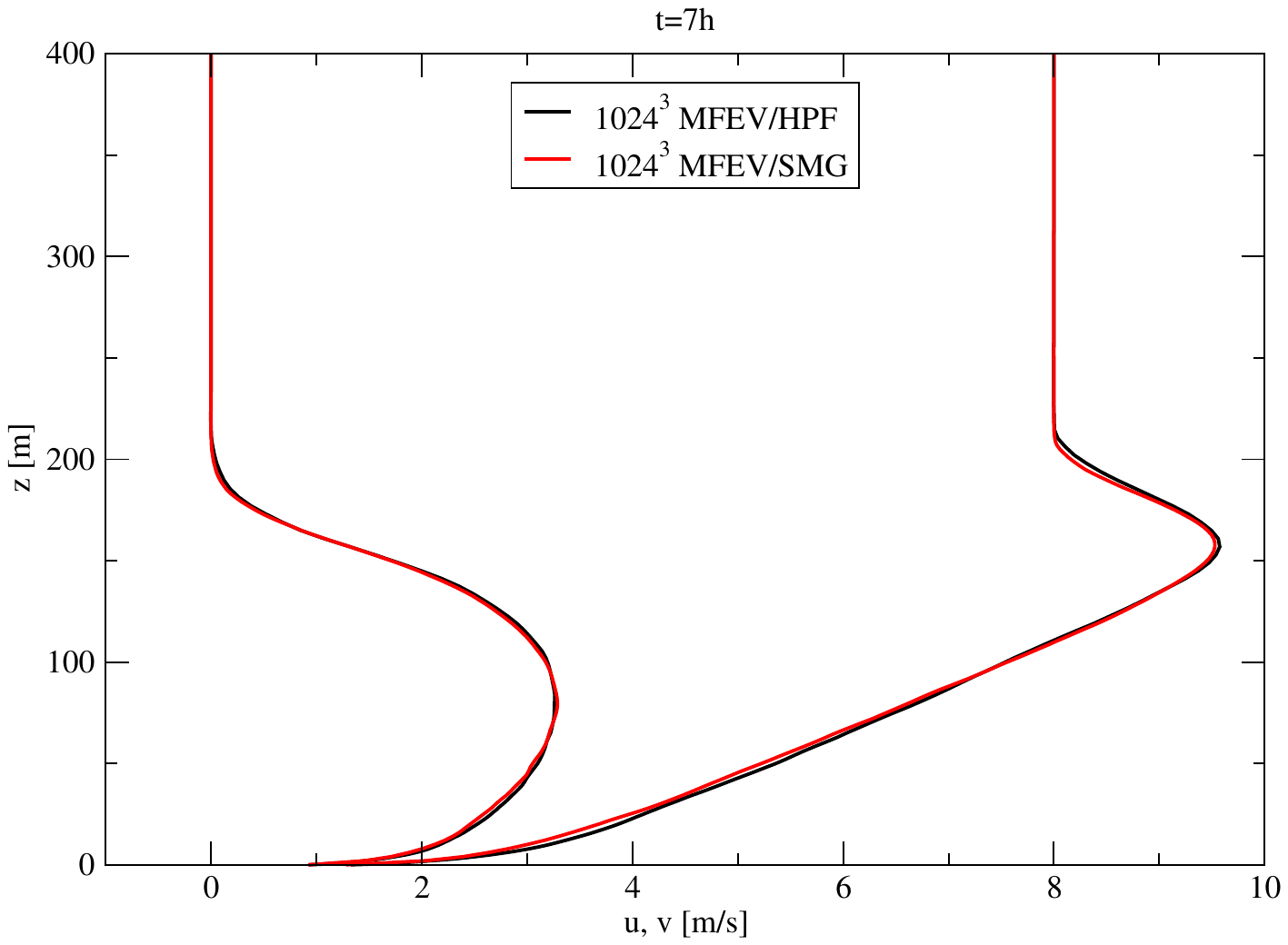}
   }
  \end{center}
  \caption{\label{fig.plot_MFEV_HPF_SMG} 
   Nek5000/RS Comparison between mean profiles obtained with MEFV/HPF and MFEV/SMG with 
   resolution (a) $n=128^3$, (b) $n=256^3$, (c) $n=512^3$, and (d) $n=1024^3$ .}
\end{figure*}
\begin{figure*}
  \begin{center}
   \subfloat[]
   {
    \includegraphics[width=0.4\textwidth]{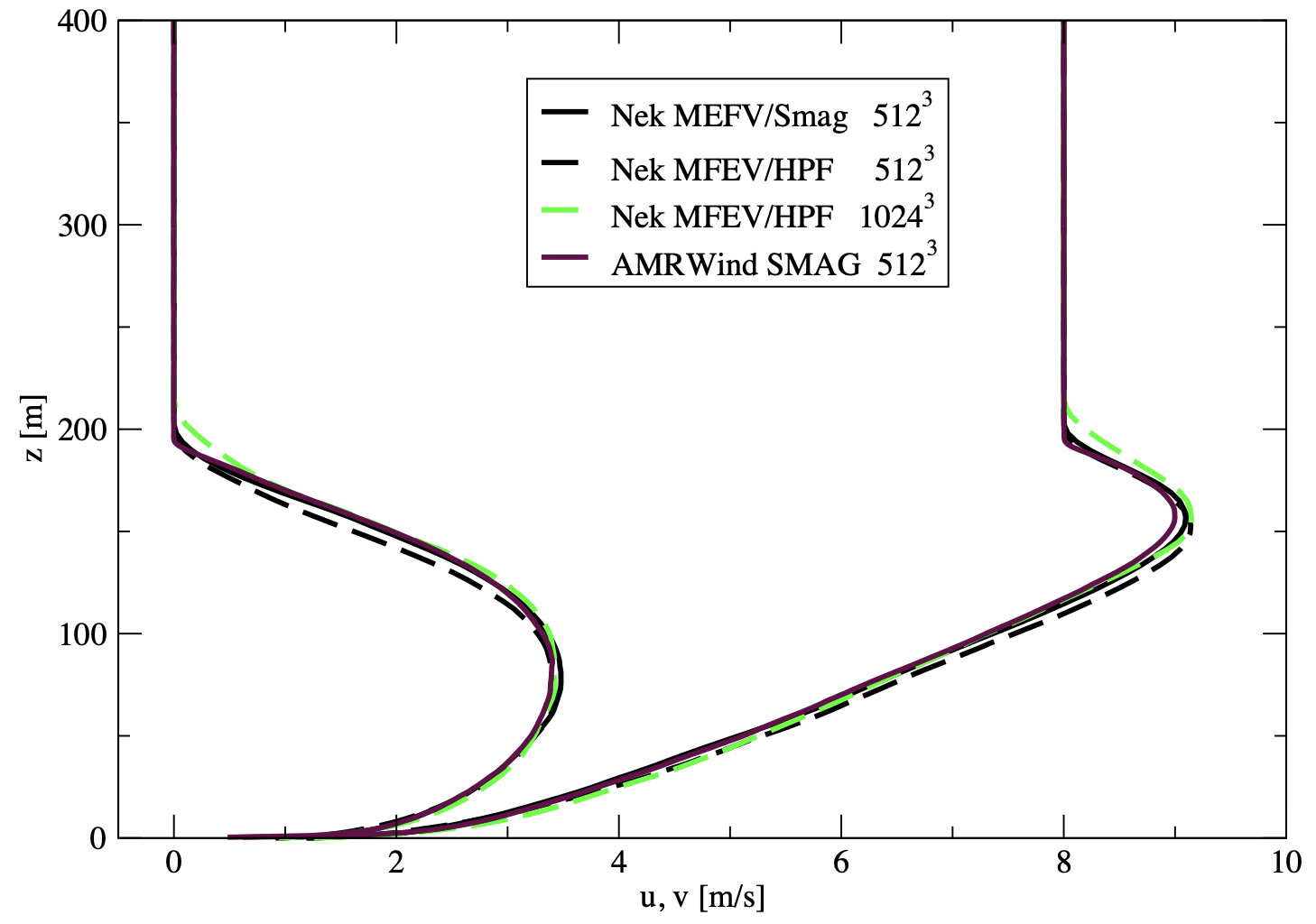}
   }
   \hspace{1em}
   \subfloat[]
   {
    \includegraphics[width=0.4\textwidth]{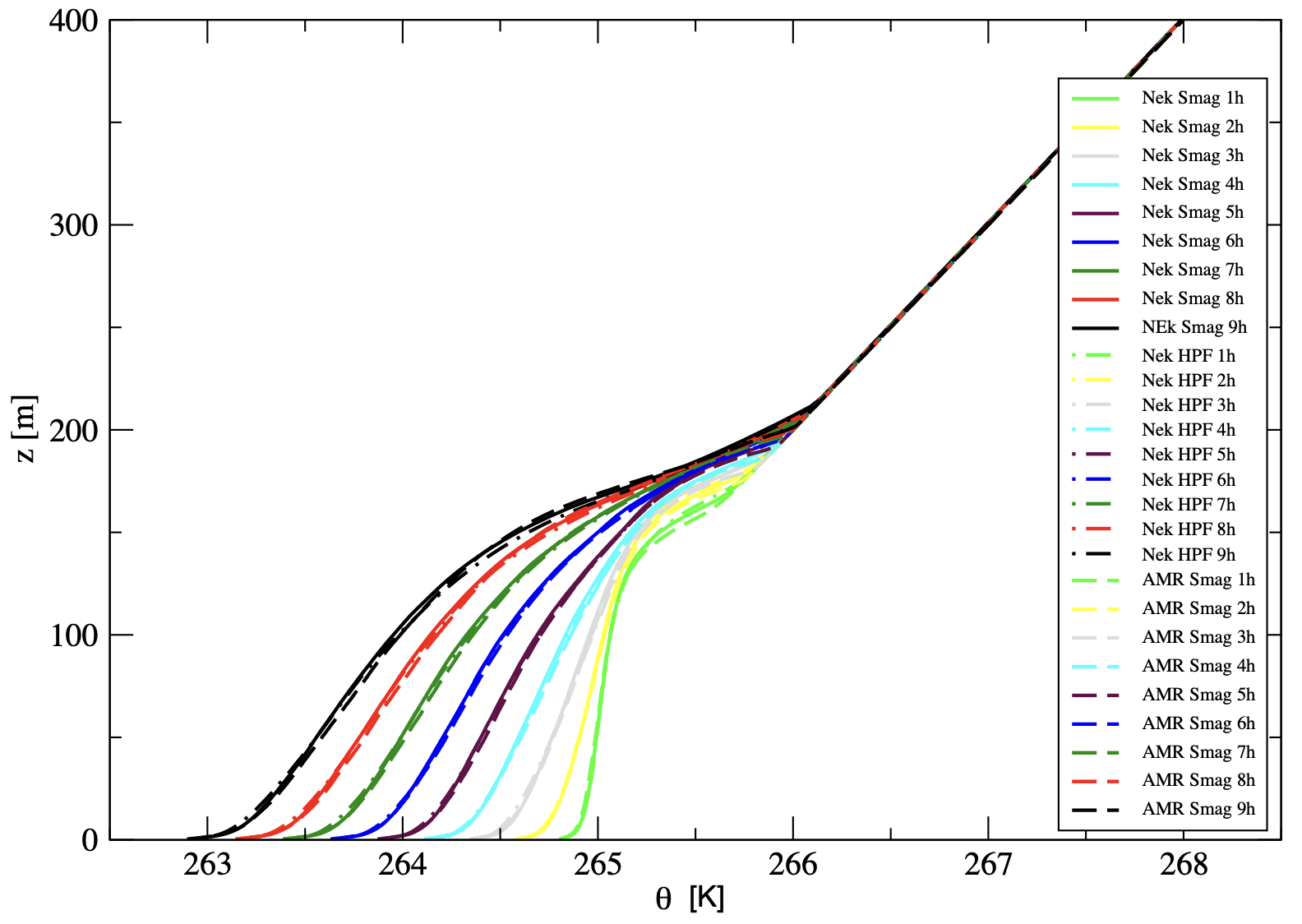}
   }
  \end{center}
  \caption{\label{fig.nek-amr-512}
    Nek5000/RS horizontally averaged (a) streamwise, spanwise velocities and 
    (b) potential temperature at $t=6h$ using MFEV/SMG and MFEV/HPF
    with traction boundary conditions, compard with AMR-Wind for $512^3$.}
\end{figure*}

\begin{figure*}
  \begin{center}
    \includegraphics[width=0.44\textwidth]{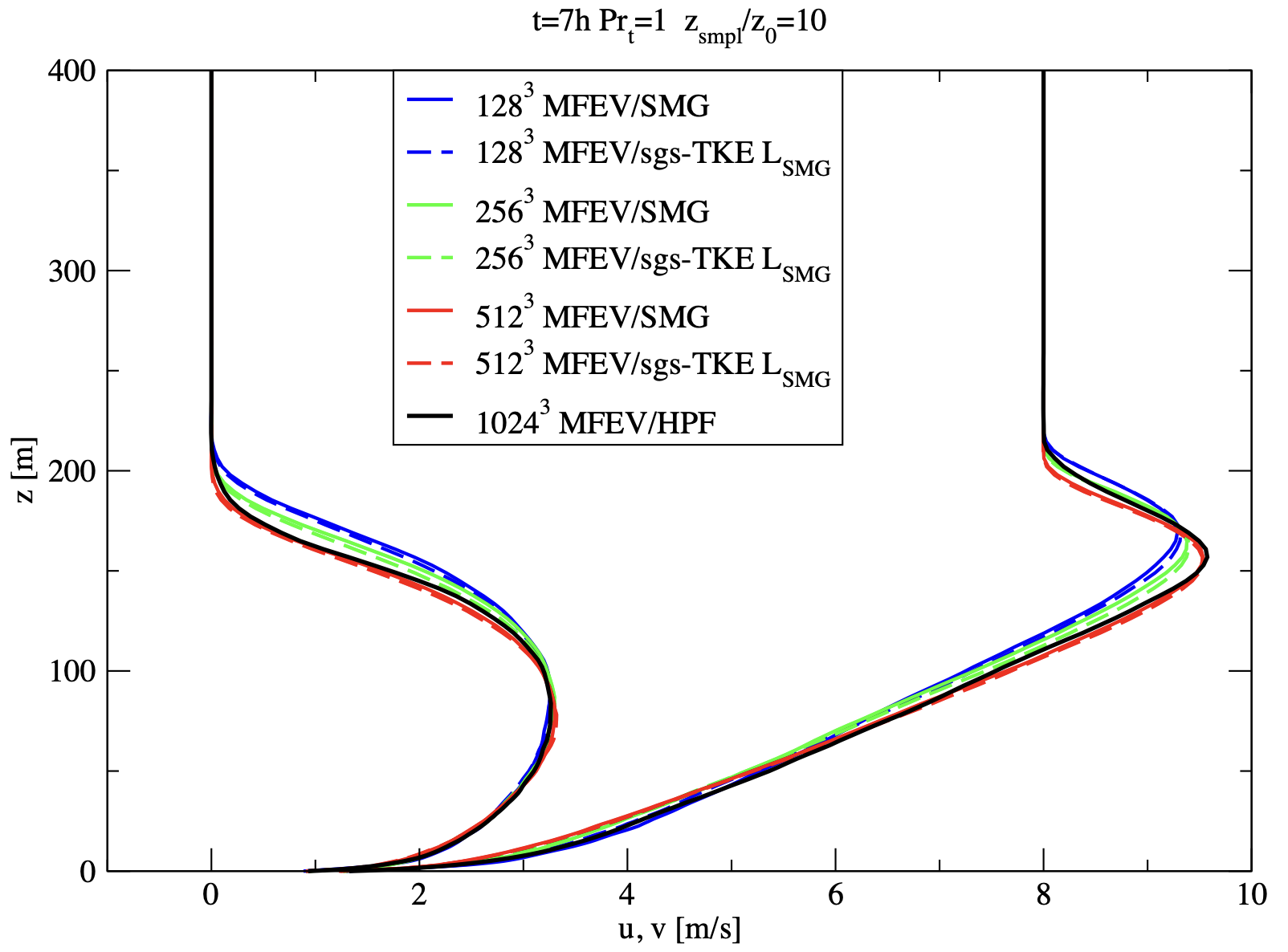}
  \end{center}
  \caption{\label{fig.nek-tke-smg} 
   Nek5000/RS Convergence for MEFV/SMG and MFEV/HPF.}
\end{figure*}

\begin{figure*}
  \begin{center}
   \subfloat[]
   {
    \includegraphics[width=0.44\textwidth]{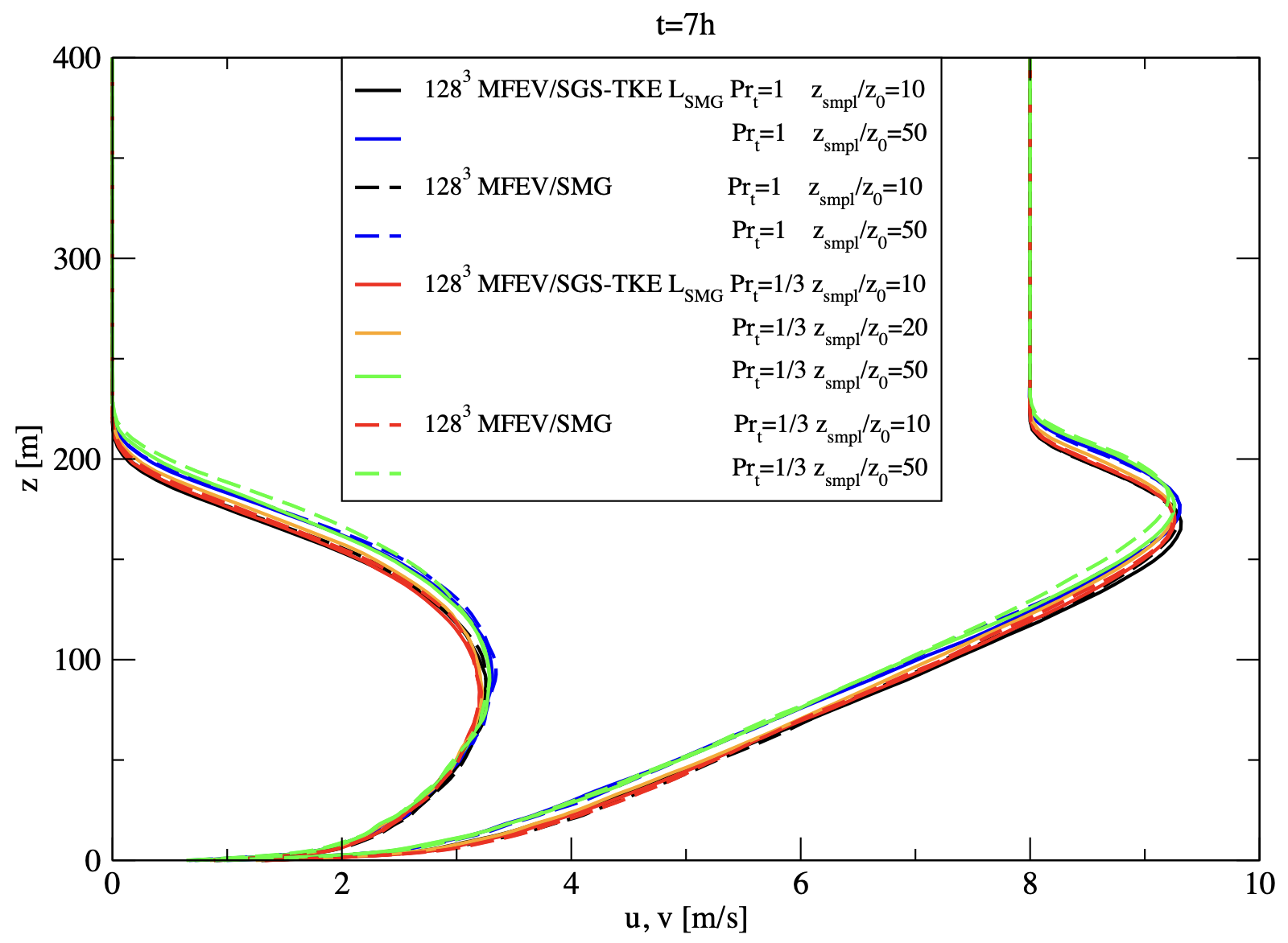}
   }
    \hspace{1em}
   \subfloat[]
   {
    \includegraphics[width=0.44\textwidth]{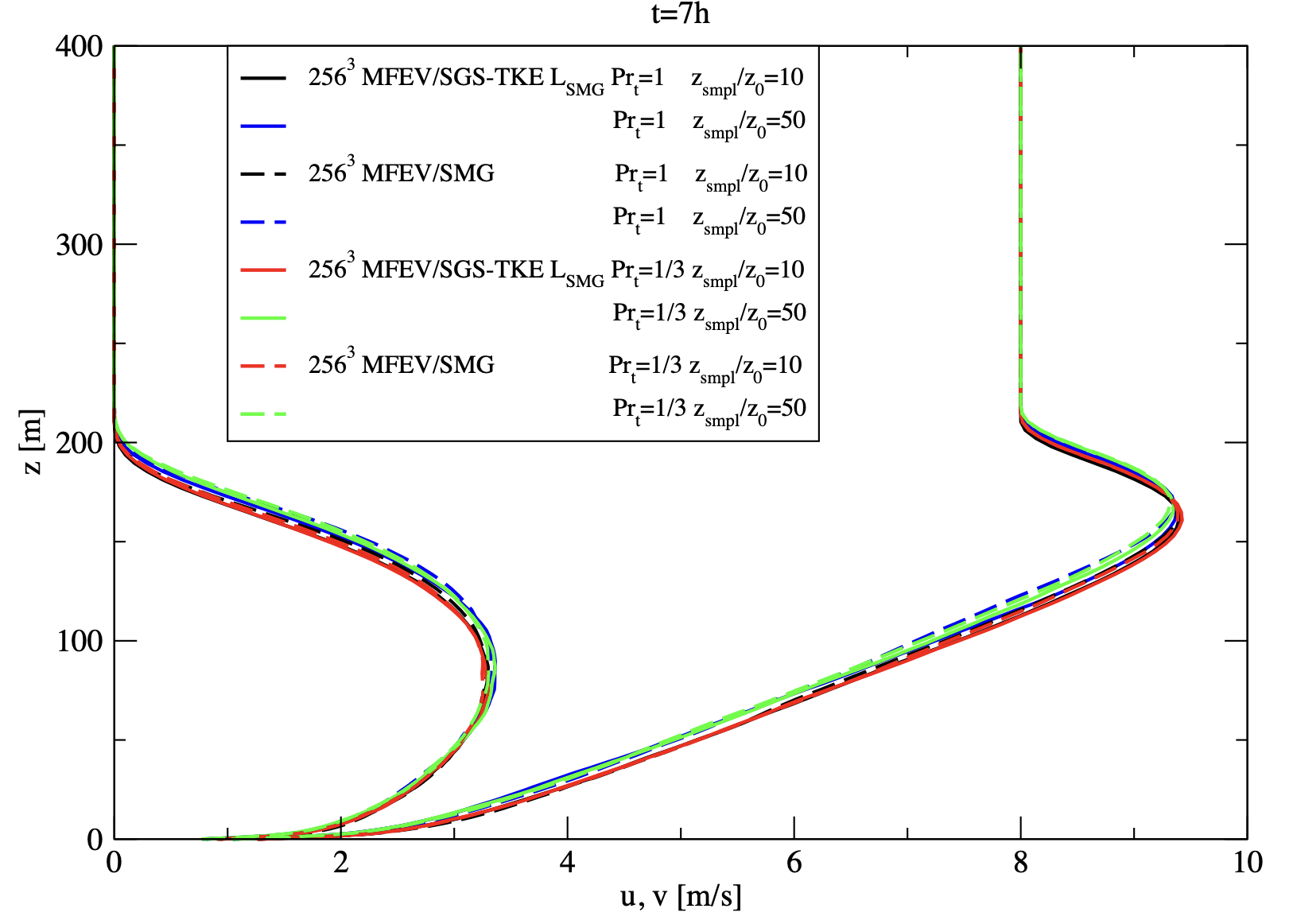}
   }
    \\
    \vspace{-1em}
   \subfloat[]
   {
    \includegraphics[width=0.44\textwidth]{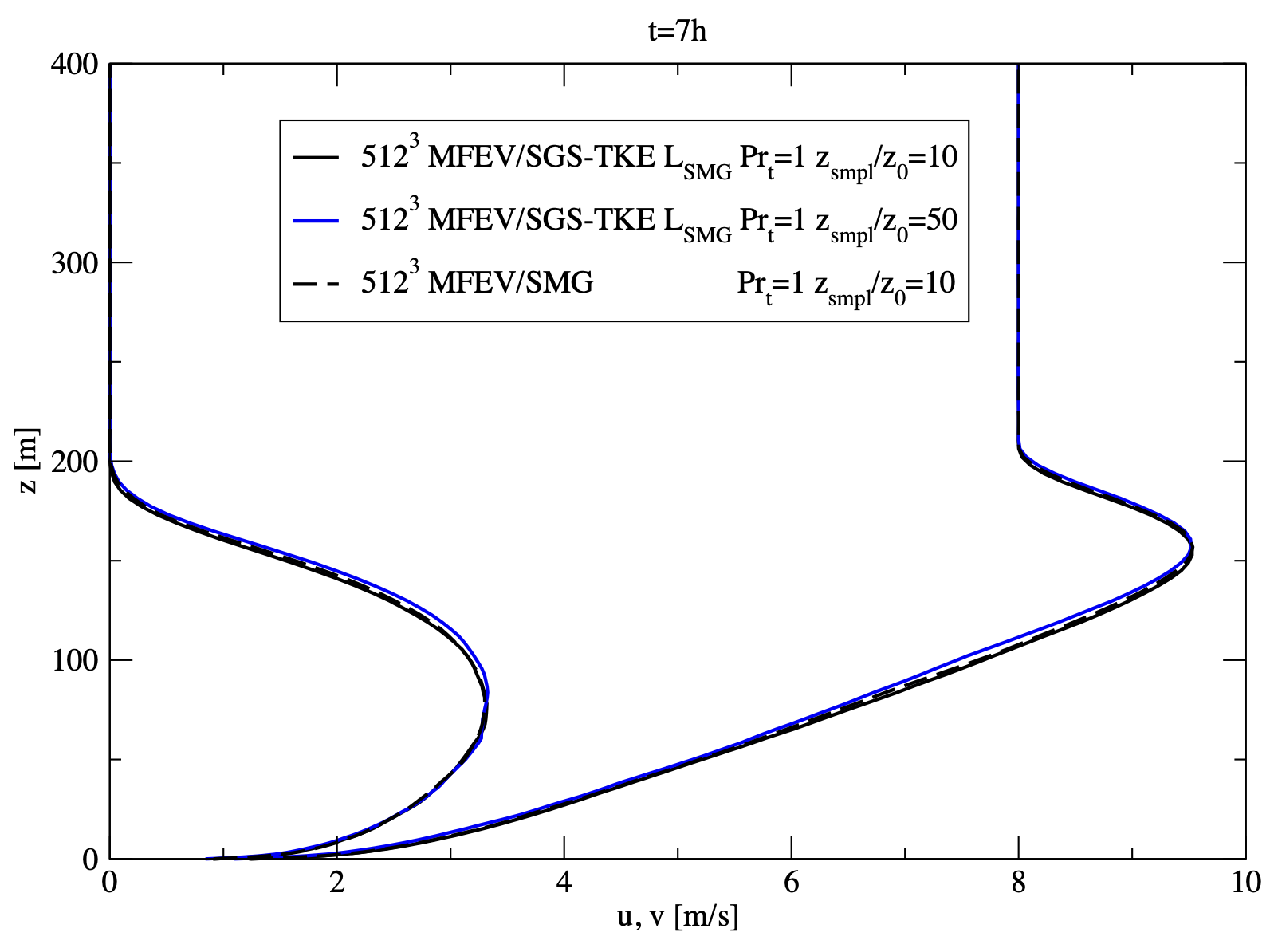}
   }
  \end{center}
  \caption{\label{fig.nek-pr1_z10_z50} Nek5000/RS, The effect of sampling location: 
   horizontally averaged streamwise and spanwise velocity profiles at $t=7h$ 
   using $z_{smpl}/z_0$ values between 10 and 50 for resolutions 
   (a) $n=128^3$, (b) $n=256^3$, and (c) $n=512^3$.}
\end{figure*}
\begin{figure*}[h]
  \begin{center}
   \subfloat[]
   {
    \includegraphics[width=0.44\textwidth]{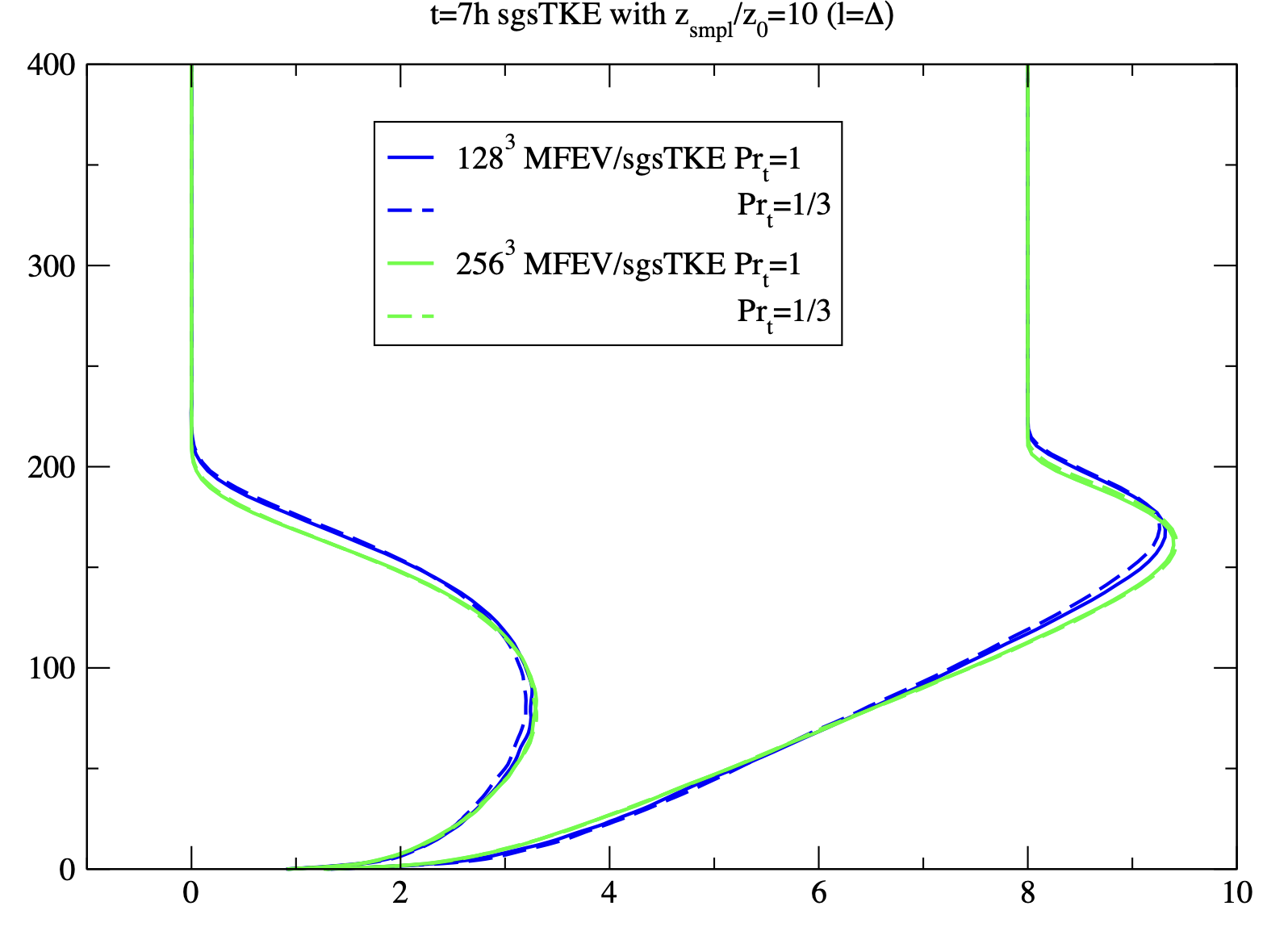}
   }
    \hspace{1em}
   \subfloat[]
   {
    \includegraphics[width=0.44\textwidth]{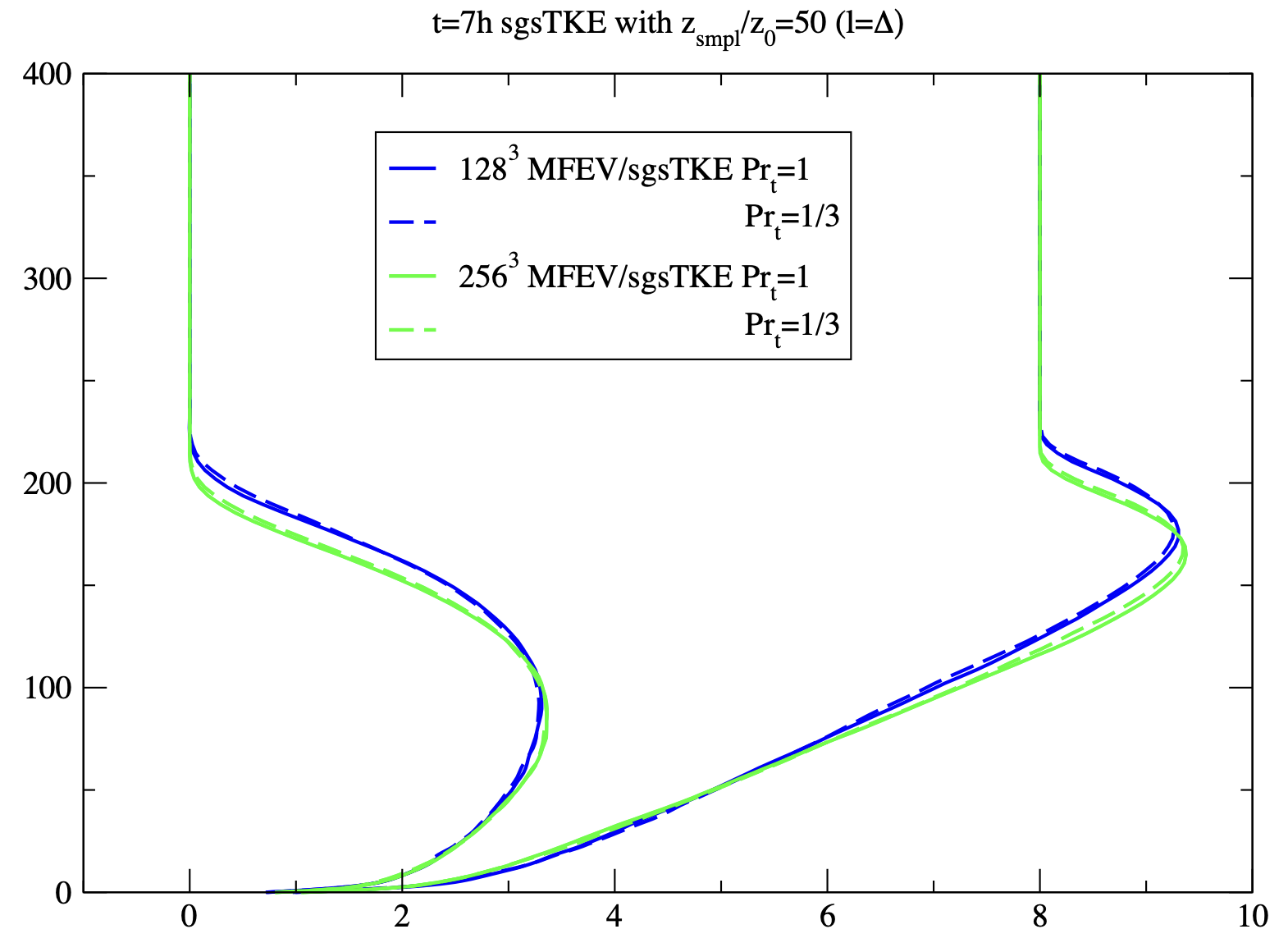}
   } 
  \end{center}
  \caption{\label{fig.nek-var-pr-conv-z10-z50} Nek5000/RS,
   The effect of $Pr_t$: horizontally averaged streamwise and spanwise velocity 
   profiles at $t=7h$ at resolutions
   $128^3$, and $256^3$, using (a) $z_{smpl}/z_0=10$ and (b) $z_{smpl}/z_0=50$, 
   with $Pr_t$ taking values $1$ and $1/3$.}
\end{figure*}
\begin{figure*}
  \begin{center}
   \subfloat[]
    {
    \includegraphics[width=0.44\textwidth]{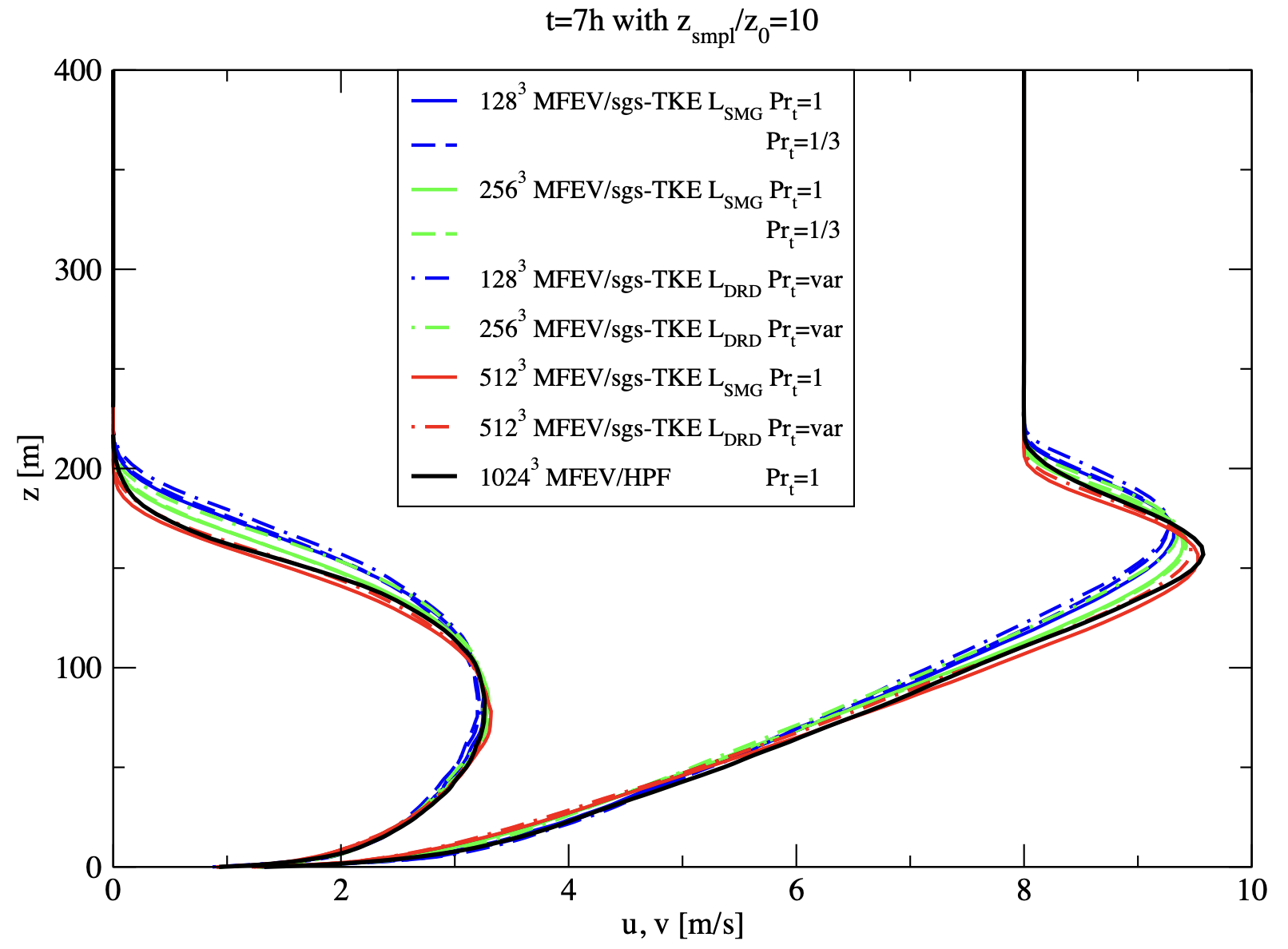}
    }
    \hspace{1em}
   \subfloat[]
    {
    \includegraphics[width=0.44\textwidth]{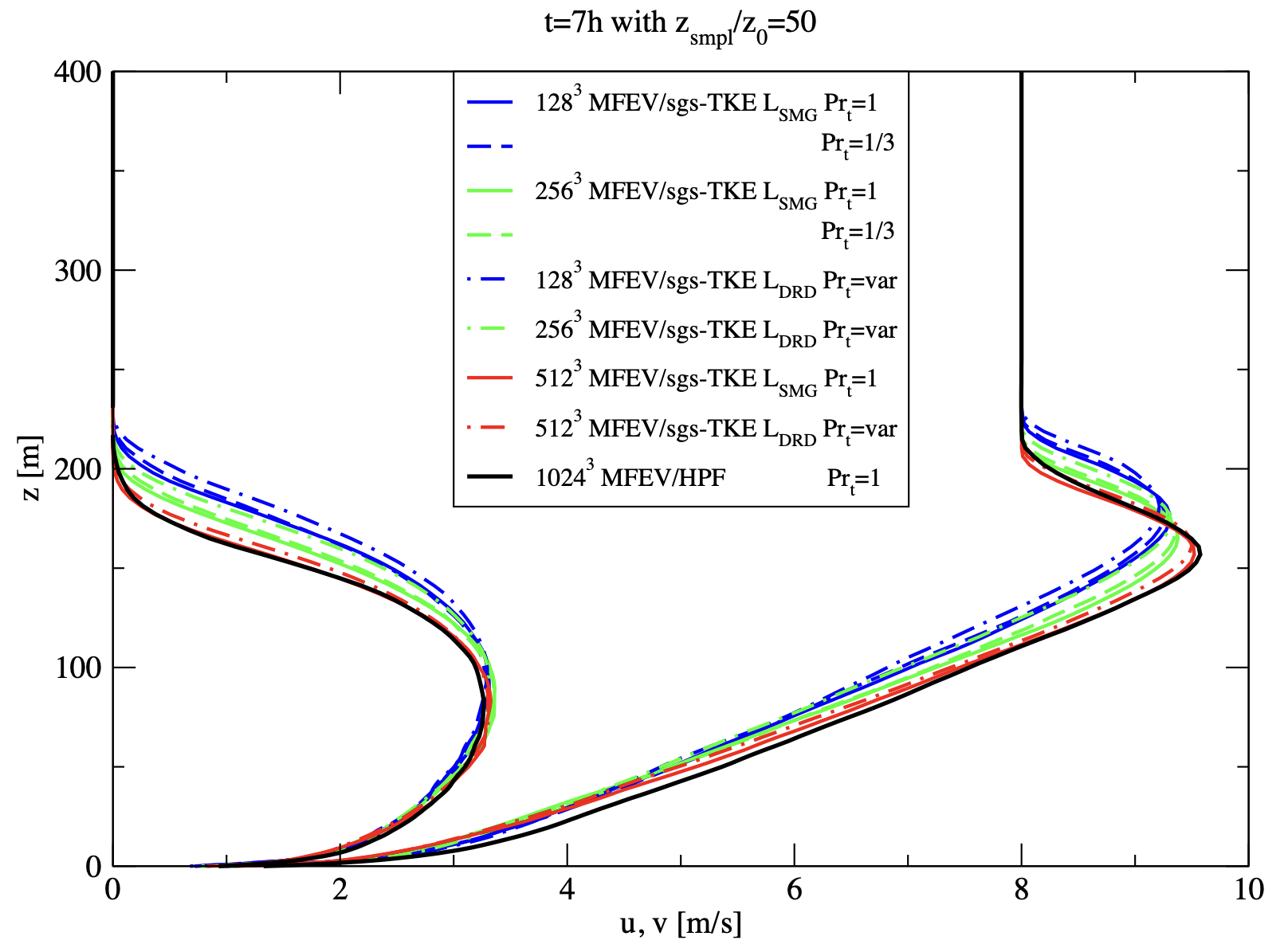}
    }
  \end{center}
  \caption{\label{fig.mean_sgstke_vs_deard_z10_z50}
    Nek5000/RS,
    Horizontally averaged streamwise and spanwise velocity profiles at $t=7h$ 
    for resolutions $n=128^3$, $n=256^3$, and $n=512^3$ using MFEV/SGS-TKE 
    and $L_{SMG}$ vs $L_{DRD}$ for (a) $z_{smpl}/z_0=10$ and (b) $z_{smpl}/z_0=50$.}
\end{figure*}
\begin{figure*}
  \begin{center}
   \subfloat[]
    {
    \includegraphics[width=0.44\textwidth]{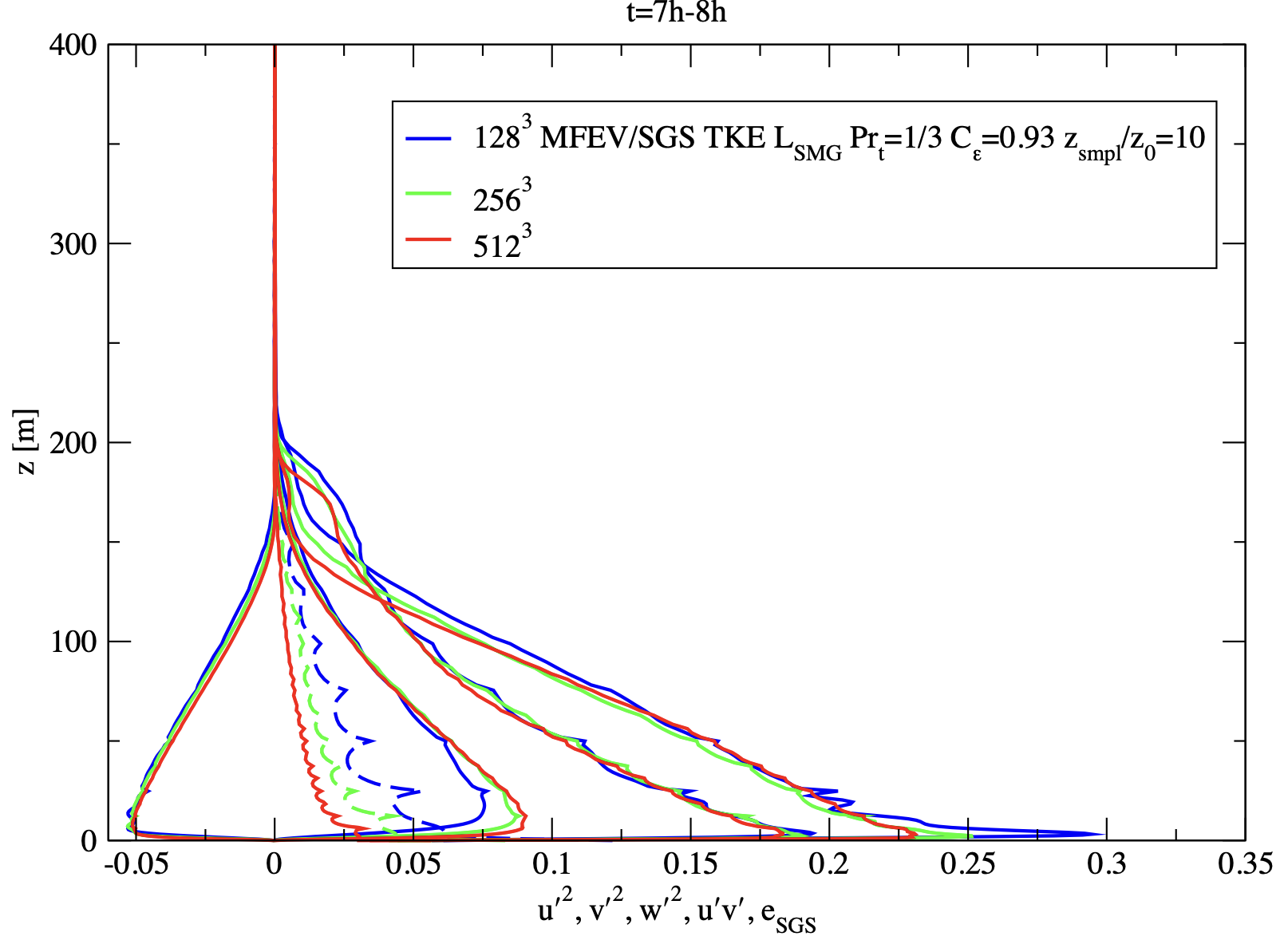}
    }
    \hspace{1em}
   \subfloat[]
    {
    \includegraphics[width=0.44\textwidth]{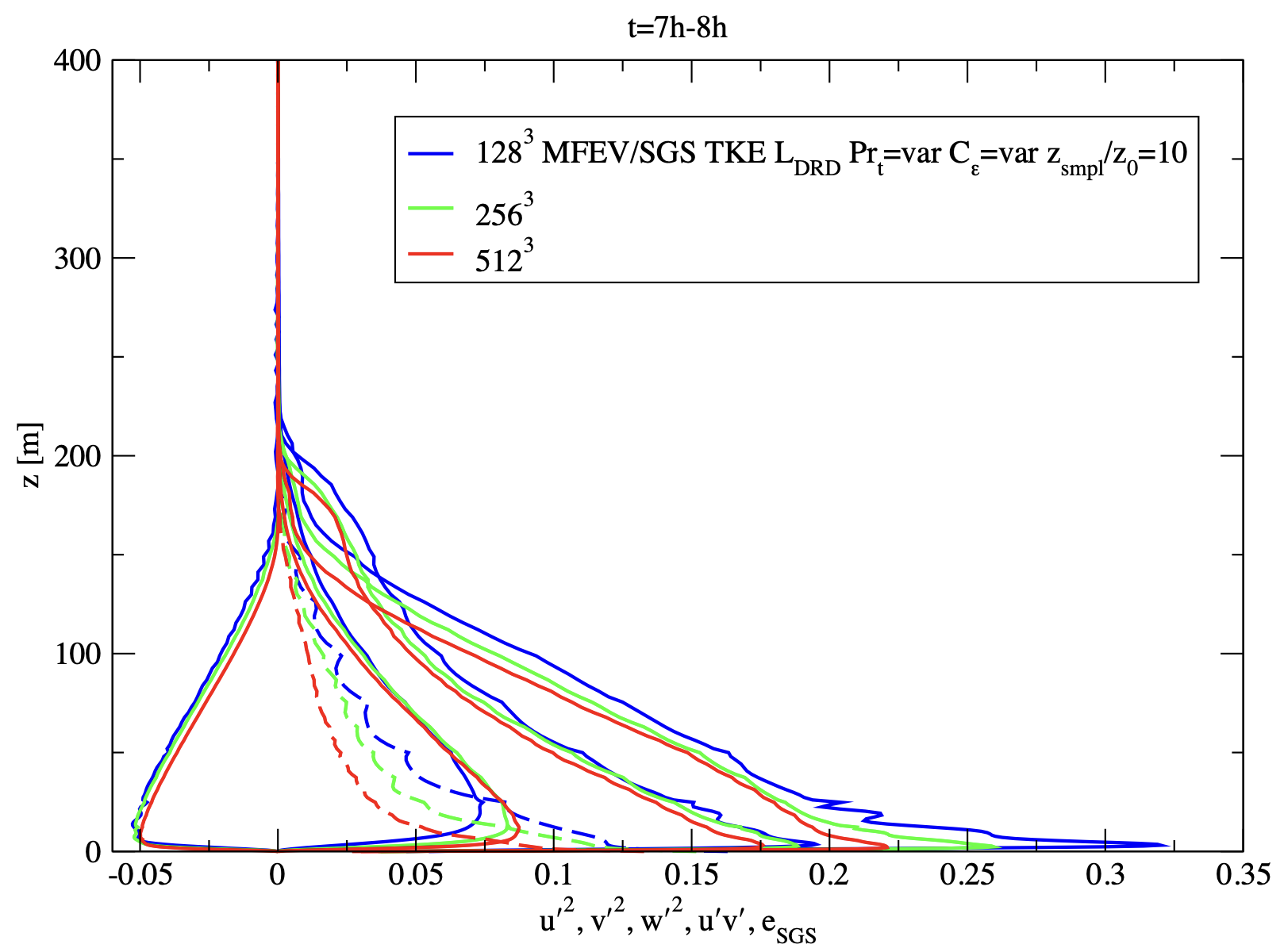}
    }
  \end{center}
  \caption{\label{fig.fluc_sgstke_vs_deard_z10}
   Nek5000/RS, Horizontally averaged streamwise, spanwise and normal fluctuation
   profiles at $t=7h$ for $z_{smpl}/z_0=10$ using (a) MFEV/SGS-TKE with $L_{SMG}$
   and (b) MFEV/SGS-TKE with $L_{DRD}$ for resolutions $n=128^3$, $n=256^3$, and $n=512^3$.}
\end{figure*}

\vspace{-0.5em}
\section{Convergence and Verification Tests} 

Results obtained with both the MFEV/HPF approach and the MFEV/SMG approach  
lead to converged results with increasing resolution as well as to asymptotic
convergence with $Re$ and $z_1^+$. Moreover, convergence with resolution
seems to be faster with resolution using MFEV/SMG as compared with MFEV/HPF 
as can be observed in Fig.~\ref{fig.mfev_hpf_smg} (a) and (b), 
which show horizontally averaged streamwise and spanwise velocities at $t=7h$ 
using MFEV/SMG and MFEV/HPF, respectively, using traction boundary conditions.
The same behavior can be observed in Fig.~\ref{fig.mfev_hpf_smg} (c) and (d) 
which shows horizontally averaged profiles of potential temperature for the
same two appoaches.

Interestingly, the difference between the mean profiles obtained with MFEV/HPF and MFEV/SMG
is reduced with increasing resolution as can be observed in Fig.~\ref{fig.plot_MFEV_HPF_SMG}
(a)--(d).

Figure~\ref{fig.nek-amr-512} (a) shows the horizontally averaged streamwise and
spanwise velocities and Fig.~\ref{fig.nek-amr-512}(b) the horizontally averaged
potential temperature at $t=6h$ for the two highest resolutions $512^3$ and $1024^3$
for Nek5000/RS MFEV/HPF and MFEV/SMG and for $512^3$ for AMR-Wind, respectively.
As can be observed, Nek5000/RS converges to the same profiles as resolution
is increased; they also agree well with the AMR-Wind obtained profiles at $512^3$.

Figure~\ref{fig.nek-tke-smg} compares horizontally averaged streamwise and spanwise velocity
profiles at $t=7h$, using MFEV/SMG and MFEV/SGS-TKE with $L_{SMG}$ for resolutions $n=128^3$, $n=256^3$ and $n=512^3$.
For completeness, the MFEV/HPF profiles for $n=1024^3$ are also shown. As can be observed, the difference between
MFEV/SMG and MFEV/SGS-TKE using $L_{SMG}$ is negligible for all resolutions.

In addition to the above, sampling for the evaluation of the wall momentum and heat fluxes was
extended to include specified $z-$locations away from the lower wall. Specifically, as explained in~\cite{ANL23_report},
it was possible to sample tangential velocities and potential temperature at specified $z-$locations away from the lower
wall in order to evaluate $u_{\tau}$ and $\theta_{\tau}$, i.e. the wall momentum and heat fluxes.
The evaluation of $u_{\tau}$ and $\theta_{\tau}$
is performed using the system of equations and method presented in section 4 of~\cite{ANL22_report}. To investigate the effect of the
sampling location $z_{smpl}$ on the results, simulations using MFEV/SGS-TKE with $L_{SMG}$ were performed
for various values of $z_{smpl}/z_0$ ranging between 10 and 50, for various resolutions. Figure~\ref{fig.nek-pr1_z10_z50}
shows horizontally averaged streamwise and spanwise velocity profiles at $t=7h$ using $z_{smpl}/z_0$ values between 10
and 50 for resolutions (a) $n=128^3$, (b) $n=256^3$, and (c) $n=512^3$. An important conclusion from this study was that
the effect of the sampling location diminishes with resolution and already at $n=512^3$ it is almost negligible.

The effect of $Pr_t$ was also investigated for the case MFEV/SGS-TKE with $L_{SMG}$ 
and the results are shown in Fig.~\ref{fig.nek-var-pr-conv-z10-z50} (a) and
(b) at resolutions $n=128^3$, and $n=256^3$, with $Pr_t$ taking values $1$ and $1/3$
for $z_{smpl}/z_0=10$ and $z_{smpl}/z_0=50$, respectively. As can be observed,
the effect of $Pr_t$ is almost negligible for both resolutions studied.


Figure~\ref{fig.mean_sgstke_vs_deard_z10_z50} shows a comparison of the horizontally averaged streamwise and spanwise velocity profiles at $t=7h$ 
for (a) $z_{smpl}/z_0=10$ and (b) $z_{smpl}/z_0=50$, between the MFEV/SGS-TKE with $L_{SMG}$ and the MFEV/SGS-TKE with $L_{DRD}$ for resolutions $n=128^3$, 
$n=256^3$, and $n=512^3$. For completeness, the MFEV/HPF profiles for $n=1024^3$ are also shown.
As resolution is increased, differences between profiles obtained by the two approaches, MFEV/SGS-TKE with $L_{SMG}$ and MFEV/SGS-TKE with $L_{DRD}$, 
become negligible.


Figure~\ref{fig.fluc_sgstke_vs_deard_z10} shows a comparison of the horizontally averaged streamwise, spanwise and normal fluctuation profiles at $t=7h$ 
for $z_{smpl}/z_0=10$ using (a) MFEV/SGS-TKE with $L_{SMG}$ and (b) MFEV/SGS-TKE with $L_{DRD}$ for resolutions $n=128^3$, $n=256^3$, and $n=512^3$.
Good convergence is also observed in second-order quantities in Fig.~\ref{fig.fluc_sgstke_vs_deard_z10} (a) and (b),
for MFEV/SGS-TKE with (a) $L_{SMG}$ and with (b) $L_{DRD}$. This is especially the case for resolutions $n=256^3$ and $n=512^3$ and for 
$L_{SMG}$. The resolved fluctuations obtained by the two approaches MFEV/SGS-TKE+with $L_{SMG}$ and $L_{DRD}$ compare reasonably well
for the same effective resolution.

\vspace{-0.5em}
\section{GABLS results from Nek5000/NekRS and AMR-Wind}

Table 1 provides a summary of the simulation details and bulk boundary layer values from simulations of the GABLS case using five grids with increasing resolution,
N = $(128^3$, $256^3$, $512^3, 1024^3, 2048^3)$. The variables in Table~\ref{tab.global_quant} are the case, the effective resolution,
the approximate total number of timesteps, the mesh spacing, the average surface friction velocity $u_\tau$, the average surface kinematic temperature 
flux $Q^{\star}$, boundary-layer depth $z_i$, Monin-Obukhov stability length $L_{MO}=-u_{\tau}^3/\kappa\beta Q^{\star}$
with von Karman constant $\kappa = 0.4$, and boundary-layer stability parameter $z_i/L_{MO}$. The height of the low-level jet (LLJ) or wind maximum defined
as the vertical location $z_j$, where the horizontal velocity reaches a maximum is also included in the last column. In contrast to~\cite{Sullivan2016} 
the ABL depth $z_i$ is defined as the height where the vertical gradient of the horizontal velocity drops to negligible values
and is not based on the maximum vertical gradient of the mean potential temperature. 
The ABL height $z_i$ as well as $u_\tau, Q^{\star}, L_{MO}$, $z_i/L_{MO}$, and $z_j / z_i$ appear to be converging with grid resolution. 
In contrast, in~\cite{Sullivan2016} and~\cite{Sullivan2023} these parameters were found to vary with the grid resolution primarily because of 
the variability in $z_i$. 
Statistics, denoted by angle brackets are obtained by averaging in the $x$-$y$ planes and over the time period $8 < t < 9 h$. 
A turbulent fluctuation from a horizontal mean is denoted by a superscript prime $( )^{\prime}$.

\begin{table} 
\caption{Global quantities}
\label{tab.global_quant}
\begin{tabular}{|c|c|c|c|c|c|c|c|c|c|c|}
\hline \text { Run } & \text { Pts } & $N_{\text {steps }}$ & $\Delta(\mathrm{m})$ & $z_i(\mathrm{~m})$ & $u_{\tau}\left(\mathrm{~m} \mathrm{~s}^{-1}\right)$ & $Q^{\star}\times 10^3\left(\mathrm{Km} \mathrm{s}^{-1}\right)$ & $L_{MO}(\mathrm{~m})$ & $z_i / L_{MO}$ & $z_j / z_i$ \\
\hline $\mathrm{A}$ & $128^3$ & 140000 & 3.125 & 223.8& 0.266 & -10.24 & 122.978 & 1.82 & 0.74  \\ 
\hline $\mathrm{B}$ & $256^3$ & 266000 & 1.56 & 217.9 & 0.264 & -9.89 & 124.385 & 1.752 & 0.74  \\ 
\hline $\mathrm{C}$ & $512^3$ & 540000 & 0.78 & 212.9 &  0.257 & -9.41 & 121.745 & 1.749 & 0.72 \\ 
\hline $\mathrm{D}$ & $1024^3$ & 1400000 & 0.39 & 215.7 & 0.257  & -9.44 & 120.747& 1.786 & 0.71 \\ 
\hline $\mathrm{E}$ & $2048^3$ & 3200000 & 0.19 & 216.4 & 0.259 & -9.68 & 120.747& 1.796 & 0.72 \\ 
\hline
\end{tabular}
\end{table}

\begin{figure*}
  \begin{center}
   \subfloat[]
    {
    \includegraphics[width=0.44\textwidth]{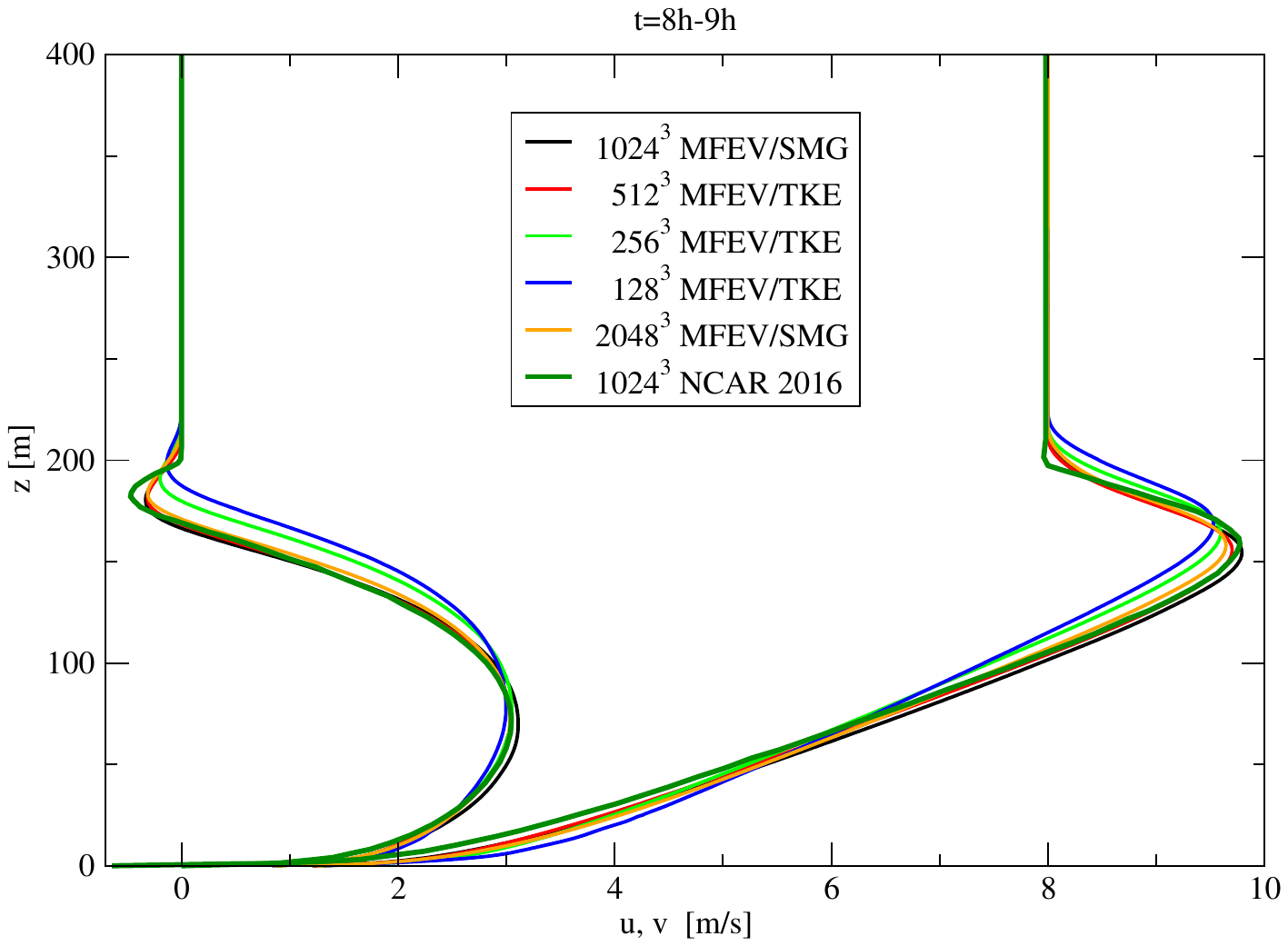}
    }
   \subfloat[]
    {
    \includegraphics[width=0.44\textwidth]{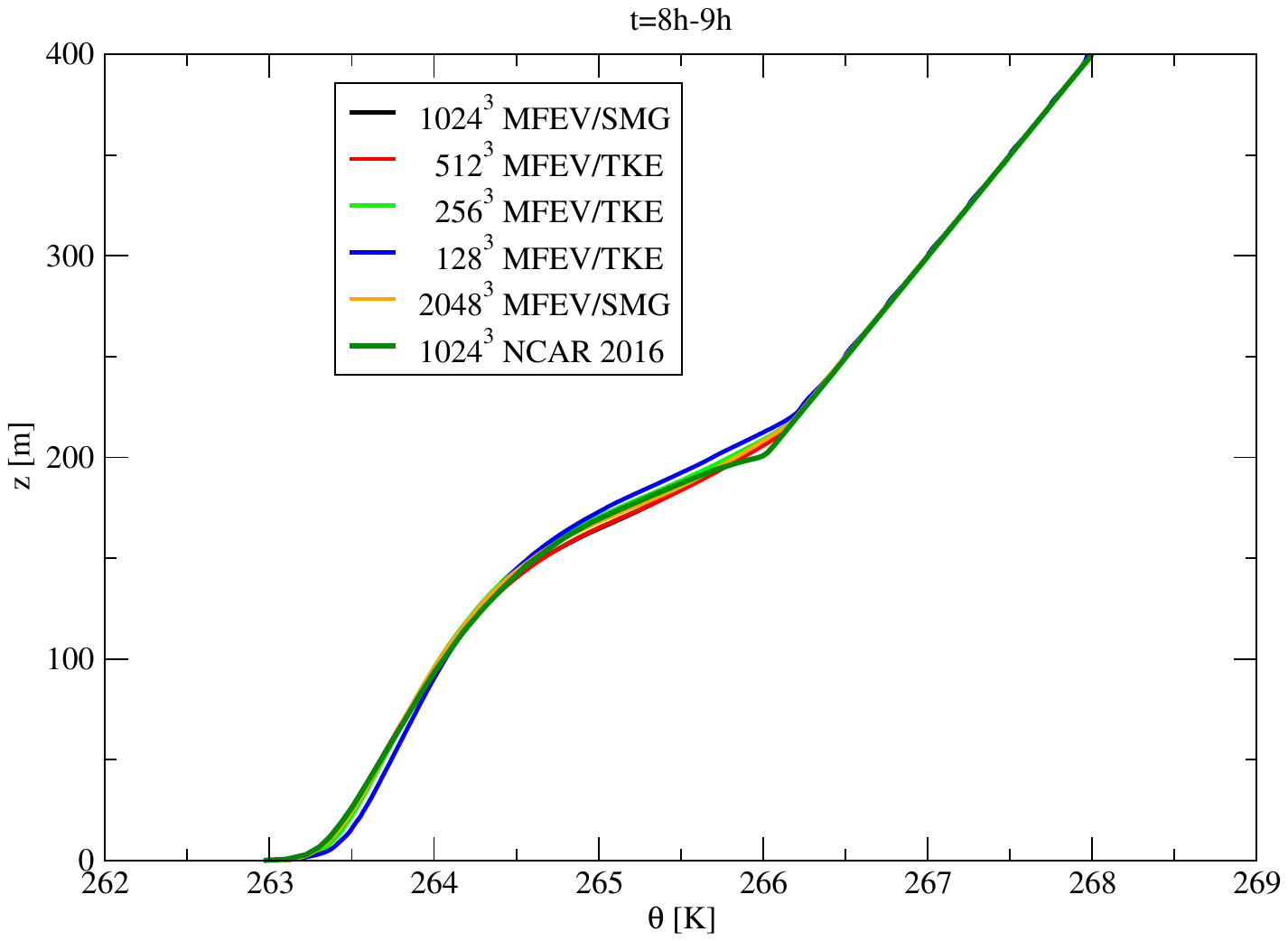}
    }
  \end{center}
  \caption{\label{fig.plot_mean_128_256_512_1024_mfev_smag_8h_9h}
   Planar and time averaged profiles of streamwise/spanwise velocity (left) and potential temperature (right)
  for $8 < t < 9 h$, using MFEV/TKE for resolutions $N=128^3$, $N=256^3$, $N=512^3$, $N=1024^3$ and $N=2048^3$.}
\end{figure*}

In addition to the bulk quantities in Table 1, our analysis of the stable ABL includes computation of vertical profiles 
of low-order moments, namely, means, variances, and momentum and temperature fluxes. 
Figures~\ref{fig.plot_mean_128_256_512_1024_mfev_smag_8h_9h}, left and right, compare 
horizontally and time-averaged profiles of streamwise/spanwise velocity and potential temperature, respectively 
for $8 < t < 9 h$, using MFEV/TKE for resolutions $N=128^3$, $N=256^3$, $N=512^3$, $N=1024^3$, and $N=2048^3$.
In these figures the vertical coordinate was not normalized with the boundary-layer depth, and as can be observed, the 
profiles above a resolution of $N=512^3$ do not appreciably change; the same holds for the boundary-layer depth. 
The same profiles from the work of~\cite{Sullivan2016} at resolution of $1024^3$ are shown in the same
figures for comparison.

\begin{figure*}
  \begin{center}
   \subfloat[]
    {
    \includegraphics[width=0.44\textwidth]{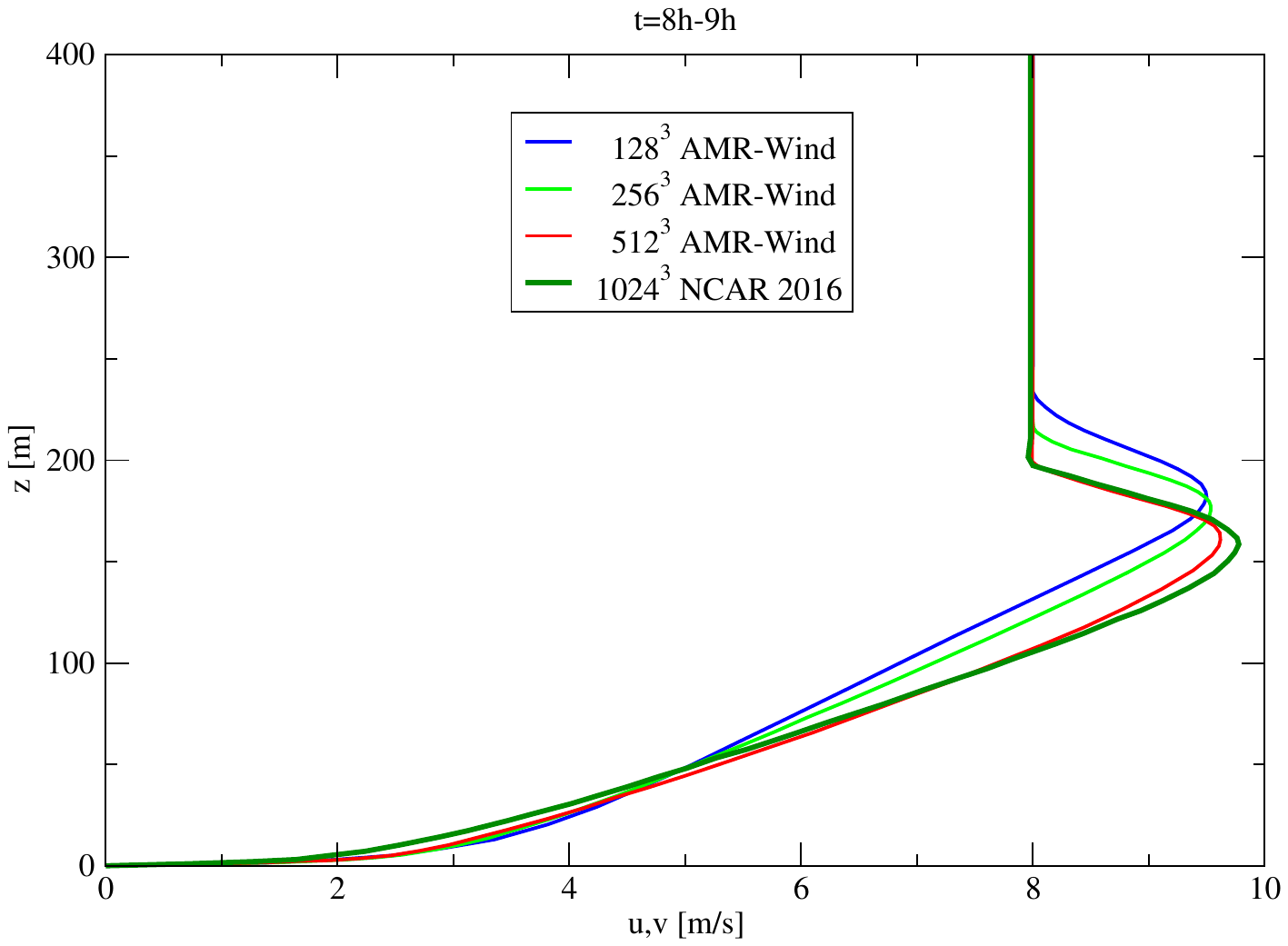}
    }
   \subfloat[]
    {
    \includegraphics[width=0.44\textwidth]{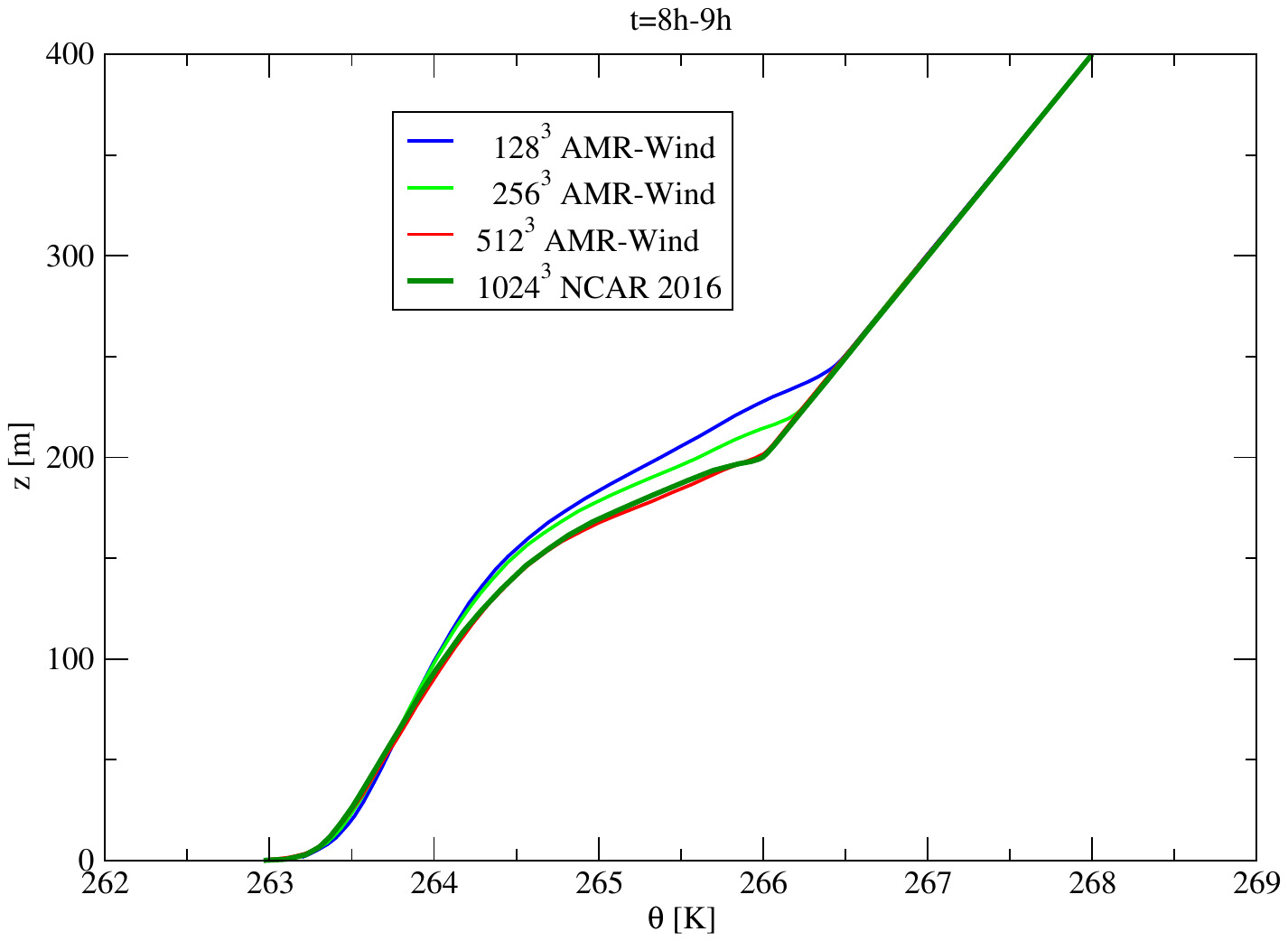}
    }
  \end{center}
  \caption{\label{fig.amr_u_v_temp}
 Vertical profiles of planar averaged $u$ and $v$-velocity (left) and of planar averaged potential temperature from AMR-Wind.}  
\end{figure*}

For AMR-Wind, the planar and time averaged profiles of streamwise and spanwise velocities 
are shown in Fig.~\ref{fig.amr_u_v_temp}, left, and of potential temperature Fig.~\ref{fig.amr_u_v_temp}, right.
Again the same profiles from the work of~\cite{Sullivan2016} at resolution of $1024^3$ are shown in the same
figures for comparison.

%
%
%
%

\begin{figure*}
  \begin{center}
   \subfloat[]
    {
    \includegraphics[width=0.44\textwidth]{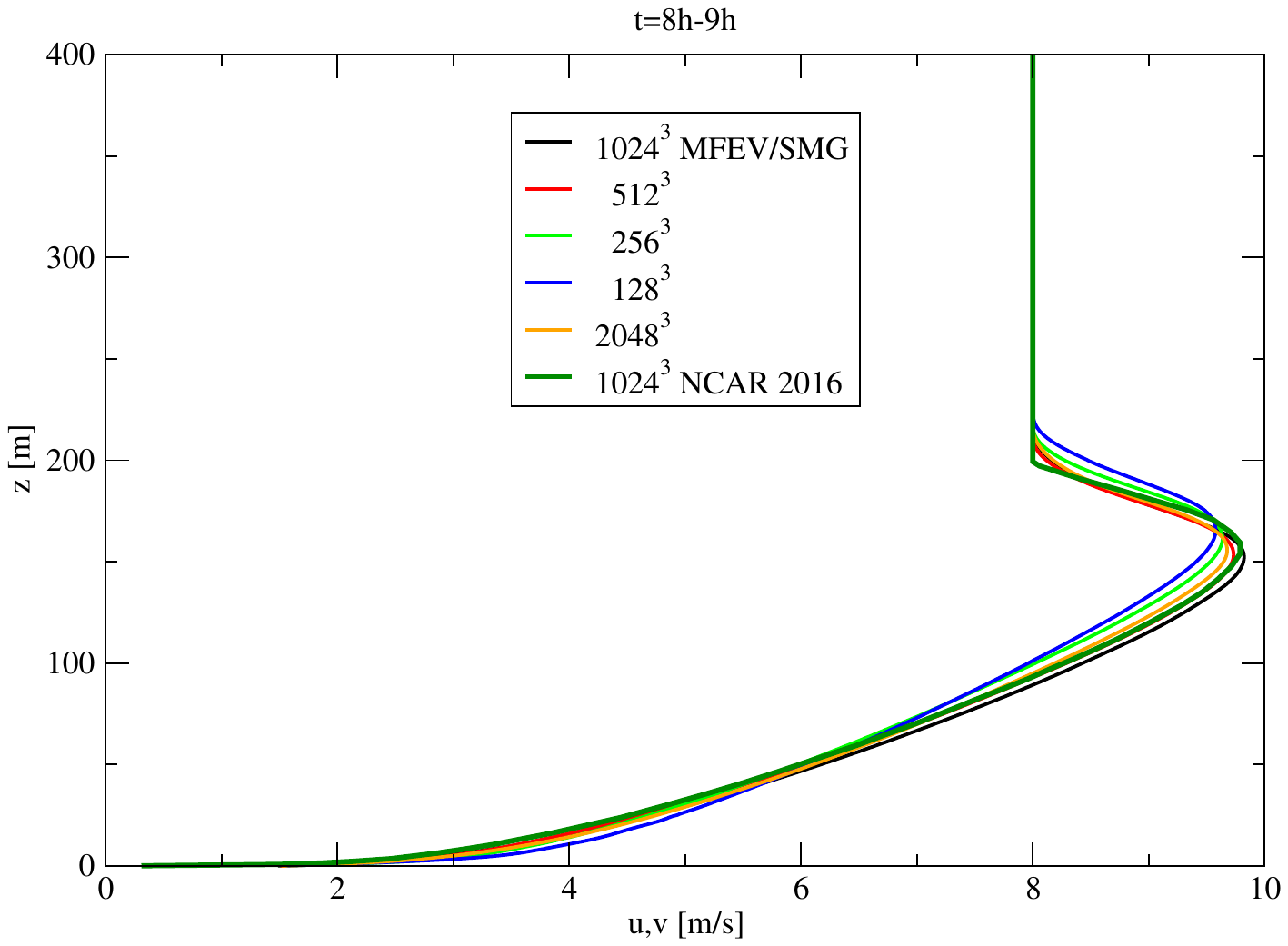}
    }
   \subfloat[]
    {
    \includegraphics[width=0.44\textwidth]{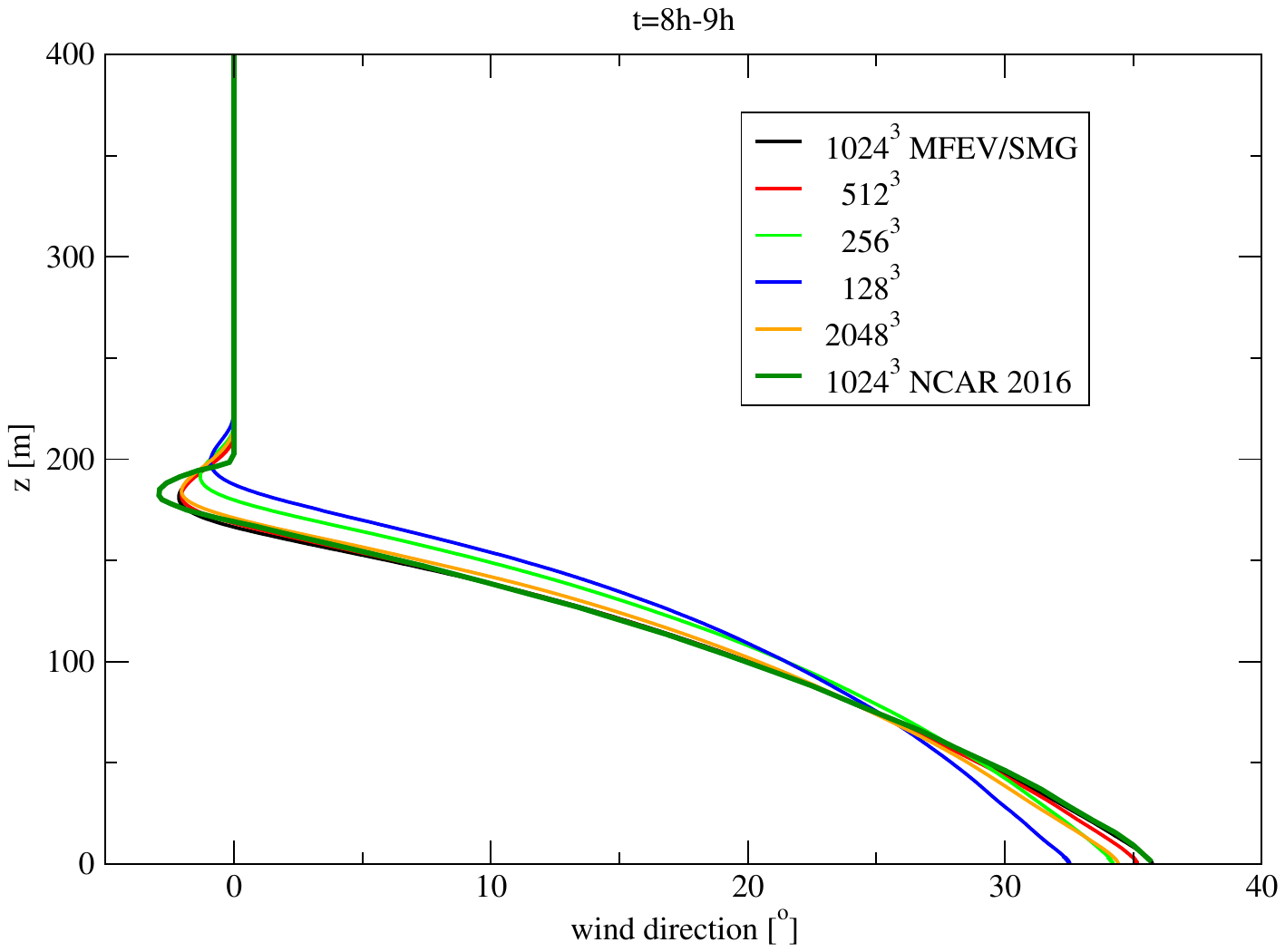}
    }
  \end{center}
  \caption{\label{fig.plot_horvel_angle_128_256_512_1024_mfev_smag_8h_9h}
   Nek5000/RS planar and time averaged for $8 < t < 9 h$ horizontal velocity (left) and wind direction (right),
   using MFEV/TKE for resolutions $N=128^3$, $N=256^3$, $N=512^3$ and $N=1024^3$}
\end{figure*}

The horizontally and time averaged profiles of the magnitude of the horizontal velocity and wind direction
for Nek5000/RS are shown in Fig.~\ref{fig.plot_horvel_angle_128_256_512_1024_mfev_smag_8h_9h}, left and right,
respectively. The same figures from the work of~\cite{Sullivan2016} at resolution of $1024^3$ are shown in the same
figures for comparison.

\begin{figure*}
  \begin{center}
   \subfloat[]
    {
    \includegraphics[width=0.44\textwidth]{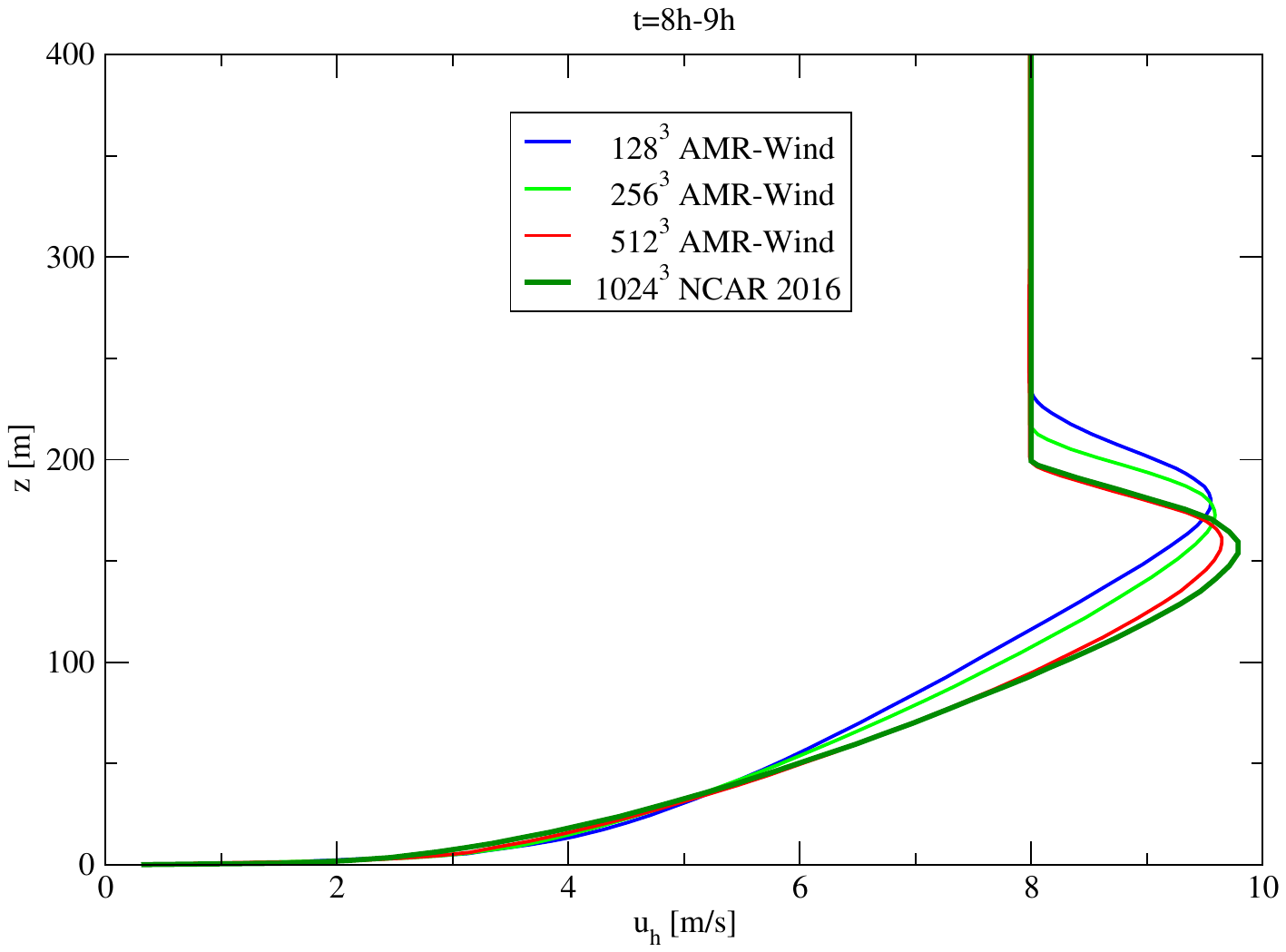}
    }
   \subfloat[]
    {
    \includegraphics[width=0.44\textwidth]{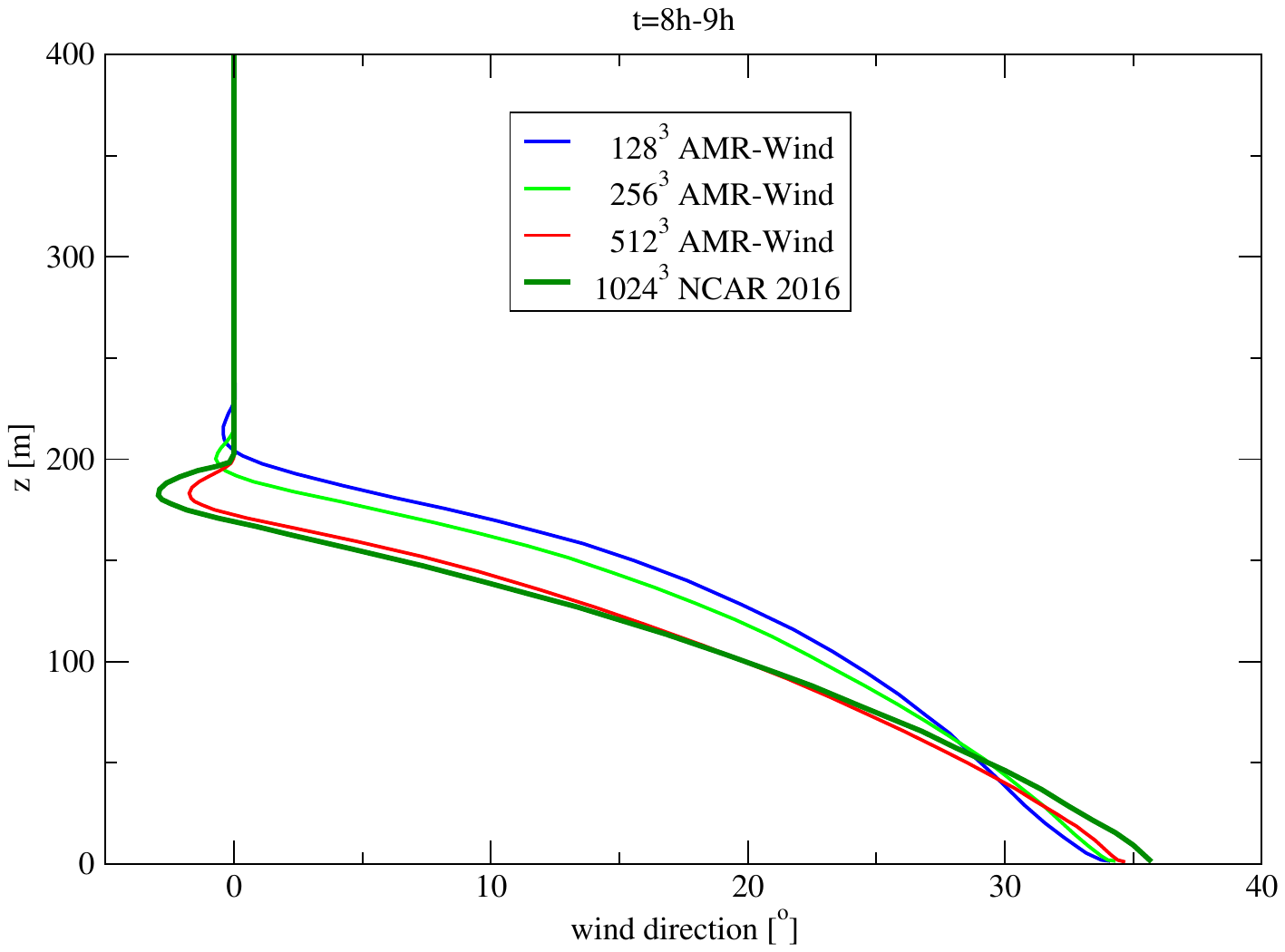}
    }
  \end{center}
  \caption{\label{fig.Uh_angle}
 Vertical profiles of planar averaged averaged horizontal velocity (left) and wind direction (right) AMR-Wind
 for resolutions $N=128^3$, $N=256^3$, $N=512^3$.}  
\end{figure*}

The vertical profiles of the planar and time averaged horizontal velocity and wind direction
for AMR-Wind are shown in Fig.~\ref{fig.Uh_angle}, left and right, respectively, together with 
same profiles from~\cite{Sullivan2016} at resolution of $1024^3$.

%
%

%

\begin{figure*}
  \begin{center}
   \subfloat[]
    {
    \includegraphics[width=0.44\textwidth]{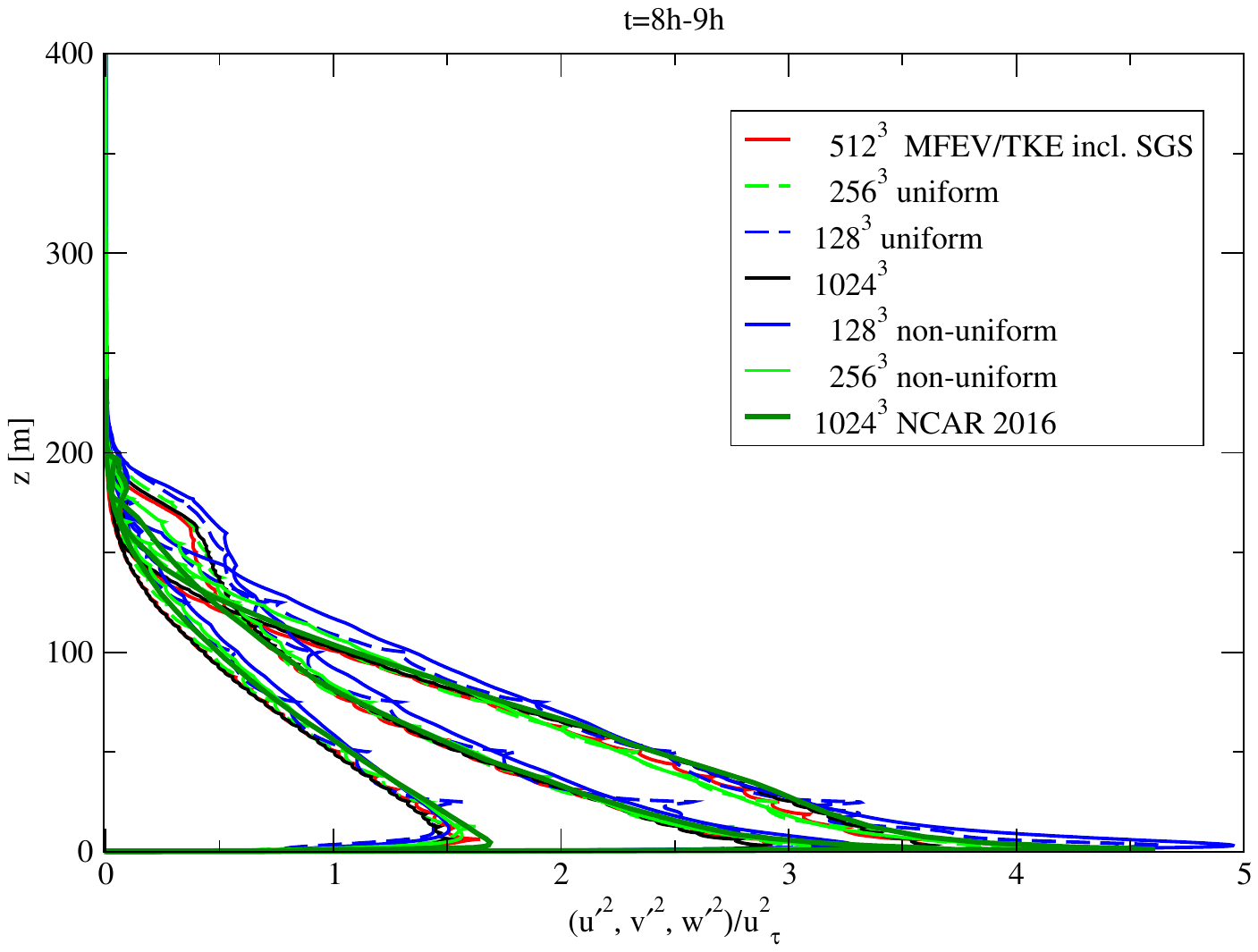}
    }
   \subfloat[]
    {
    \includegraphics[width=0.44\textwidth]{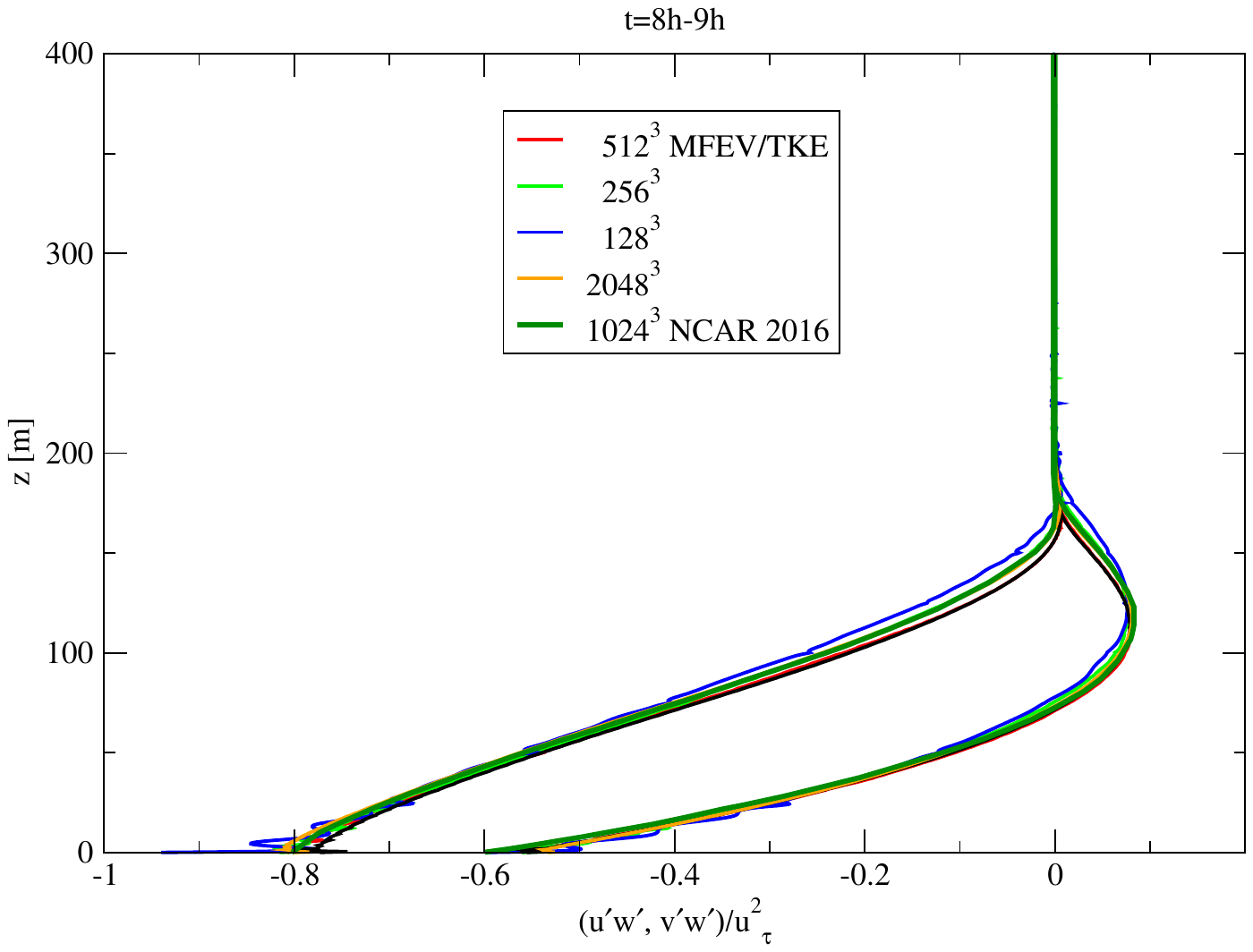}
    }
  \end{center}
  \caption{\label{fig.plot_fluc_esgs_u2_v2_w2_NCAR} 
   Nek5000/RS, MFEV/TKE fluctuation velocity profiles incl. SGS (left) and variance and covariance profiles incl. SGS (right) with resolution.}
\end{figure*}

A similar behavior is observed in Fig.~\ref{fig.plot_fluc_esgs_u2_v2_w2_NCAR}, left, for horizontally and time averaged fluctuations
between $8 < t < 9 h$. This figure compares fluctuation profiles obtained using MFEV/TKE normalized with $(u^\tau)^2$ for resolutions
$N=128^3$, $N=256^3$, $N=512^3$ and $N=1024^3$. SGS contributions are included to all quantities in the figure and as can be seen
observed, the profiles do not appreciably change above $N=256^3$. 
In this figure, the lower resolution simulations at $N=128^3$, $N=256^3$ were repeated with non-uniform resolution in the
vertical direction to reduce spikes at interelemental boudaries. This is a common feature of spectral element simulations, which
however, does not affect convergence of these quantities. Except for the spikes, the profiles are almost identical 
to the profiles obtained using uniform resolution (dashed).
As can be observed, the velocity variances from all simulations, which include SGS contributions collapse quite well for the 
four higher mesh resolutions considered. 

The same holds for the streamwise and. spanwise vertical momentum fluxes $\langle u^{\prime}w^{\prime} \rangle$ and
$\langle u^{\prime}v^{\prime} \rangle$ that include both the resolved and SGS contributions, which are shown in 
Fig.~\ref{fig.plot_fluc_esgs_u2_v2_w2_NCAR}, right, and which are in close agreement as the mesh spacing varies for all
resolutions.

\begin{figure}[hp!]
        \centering
            \includegraphics[width=3.in]{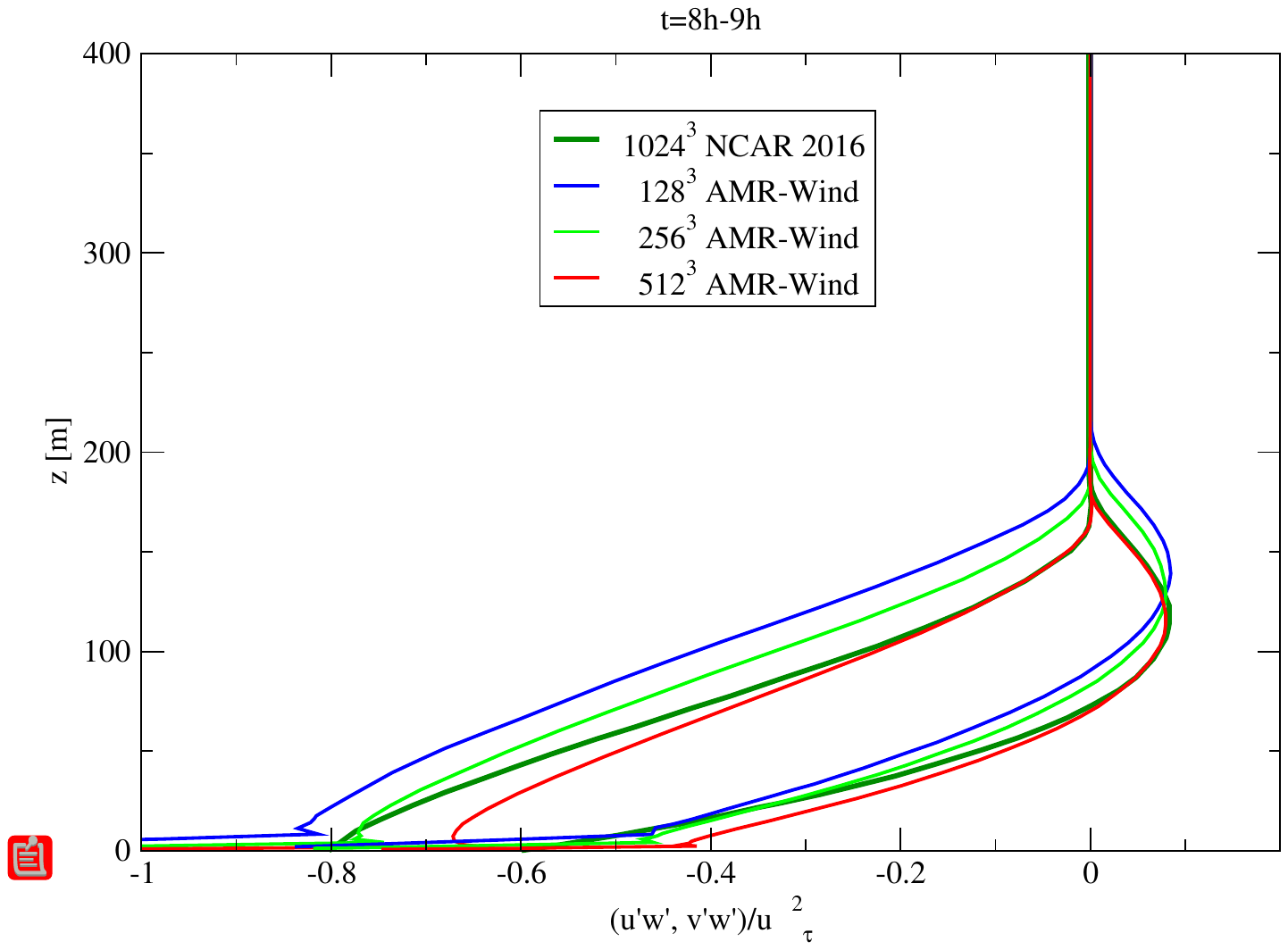}\
        \caption{\label{fig:var}Vertical profiles of velocity variance from AMR-Wind.}
\end{figure}

The streamwise and spanwise vertical momentum fluxes $\langle u^{\prime}w^{\prime} \rangle$ and
$\langle u^{\prime}v^{\prime} \rangle $ that include both the resolved and SGS contributions for 
AMR-Wind are shown in~\ref{fig:var} together with the same profiles from~\cite{Sullivan2016}.

\begin{figure*}
  \begin{center}
   \subfloat[]
    {
    \includegraphics[width=0.44\textwidth]{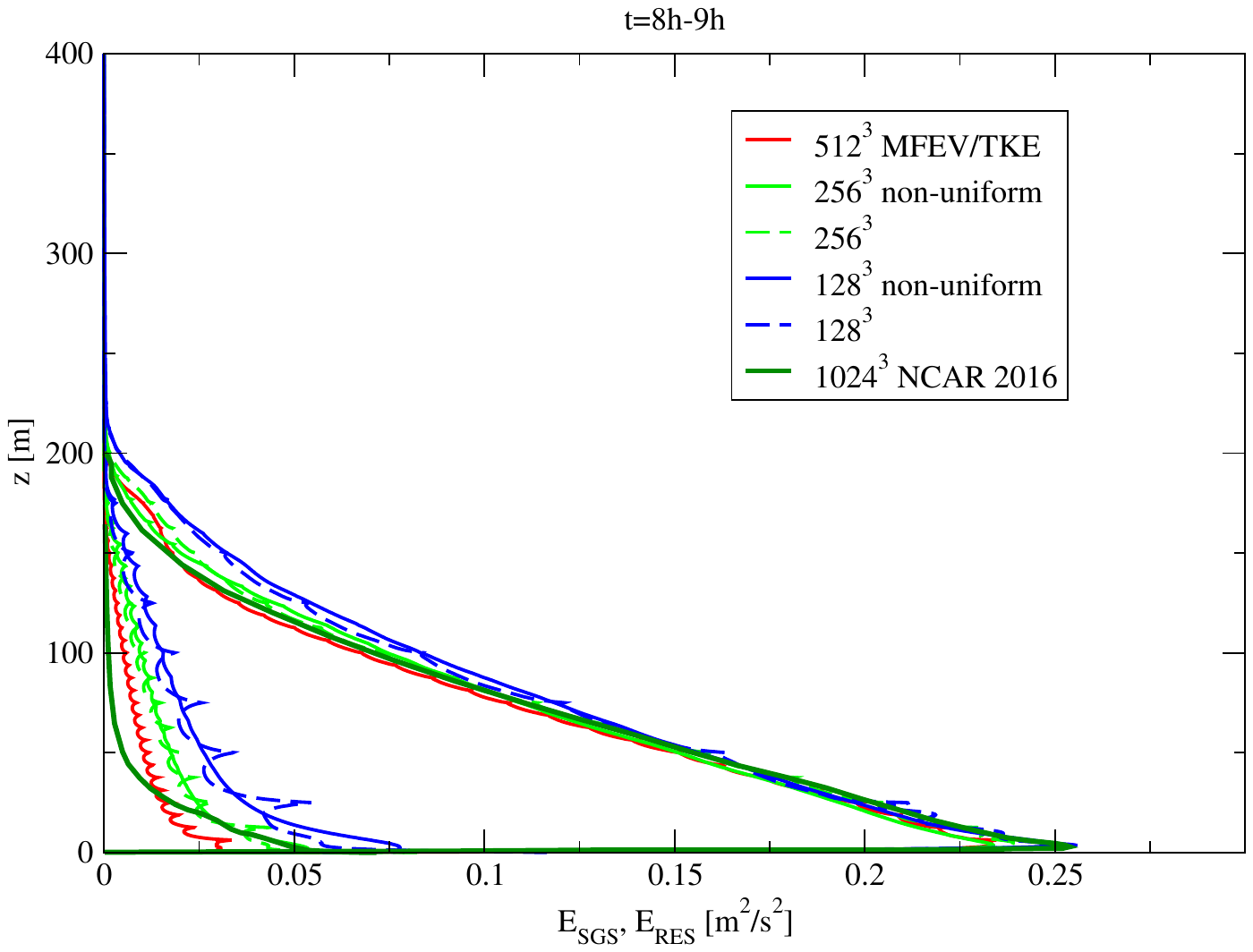}
    }
   \subfloat[]
    {
    \includegraphics[width=0.44\textwidth]{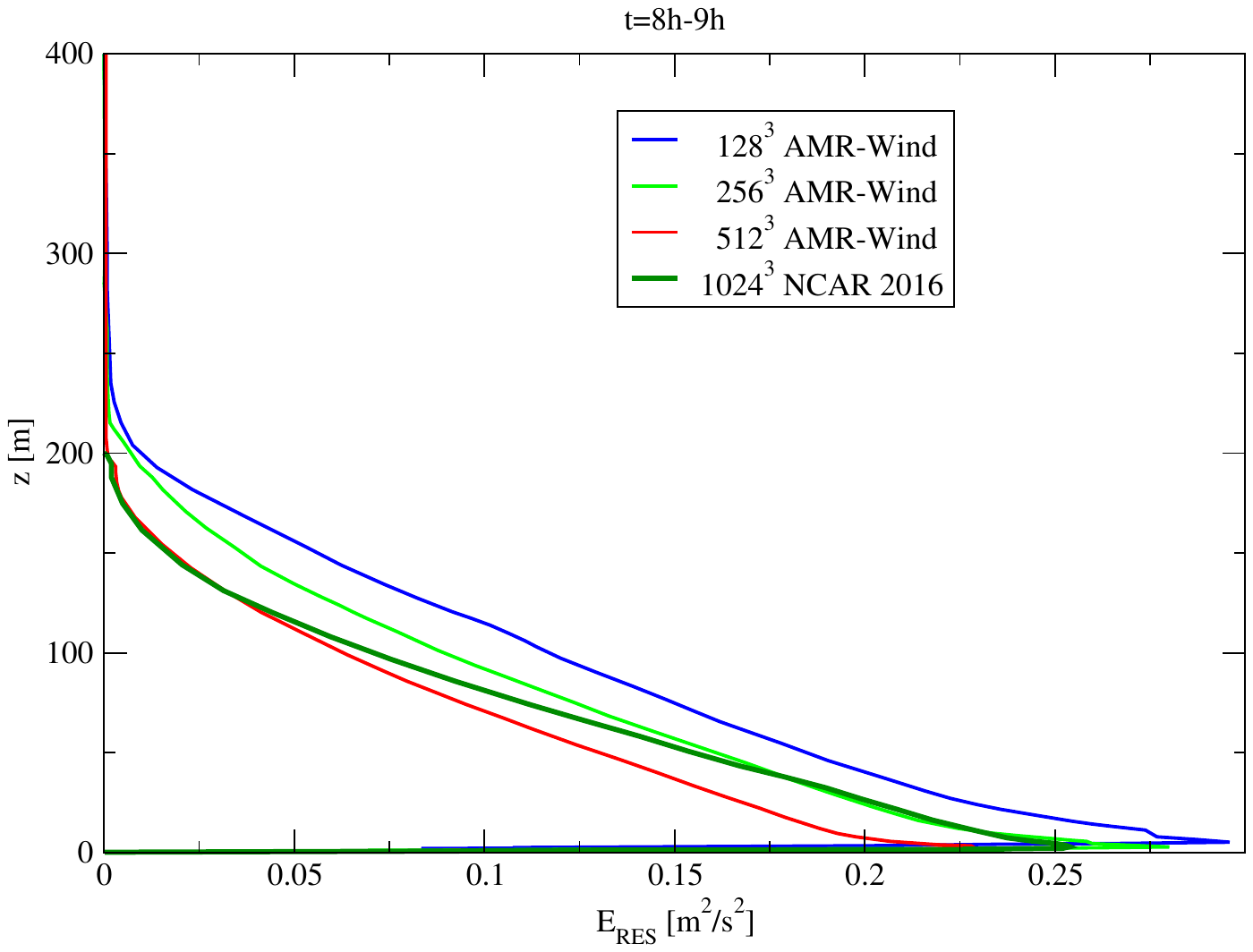}
    }
  \end{center}
  \caption{\label{fig.plot_esgs_eres}
  Vertical profiles of resolved and SGS turbulent kinetic energy from Nek5000/RS (left) and of resolved turbulent kinetic energy from AMR-Wind (right).}
\end{figure*}

A measure of the resolved nature of the flow fields is provided in Fig.~\ref{fig.plot_esgs_eres}. 
The vertical profiles of resolved and SGS turbulent kinetic energy from Nek5000/RS are shown in Fig.~\ref{fig.plot_esgs_eres},
left, and profiles of resolved turbulent kinetic energy from AMR-Wind are shown in~\ref{fig.plot_esgs_eres}, right.
As can be observed, the SGS energy computed near the surface is less than $20\%$ of the total for the coarsest resolution $N=128^3$, and reduces 
to values below $10\%$ at a resolution of $512^3$. This ratio is reduced to even smaller values at the highest resolution of
$1024^3$. The profile of SGS energy shown in Fig.~\ref{fig.plot_esgs_eres} shows a systematic decrease with resolution over 
the whole stable ABL. In fact, as noted in~\cite{Sullivan2023} the SGS TKE scales with $\Delta^{2/3}$ which corresponds to an
approximate reduction of $40\%$ with a mesh size reduction by a factor of 2. 

\begin{figure*}
  \begin{center}
    \subfloat[]
    {
    \includegraphics[width=0.44\textwidth]{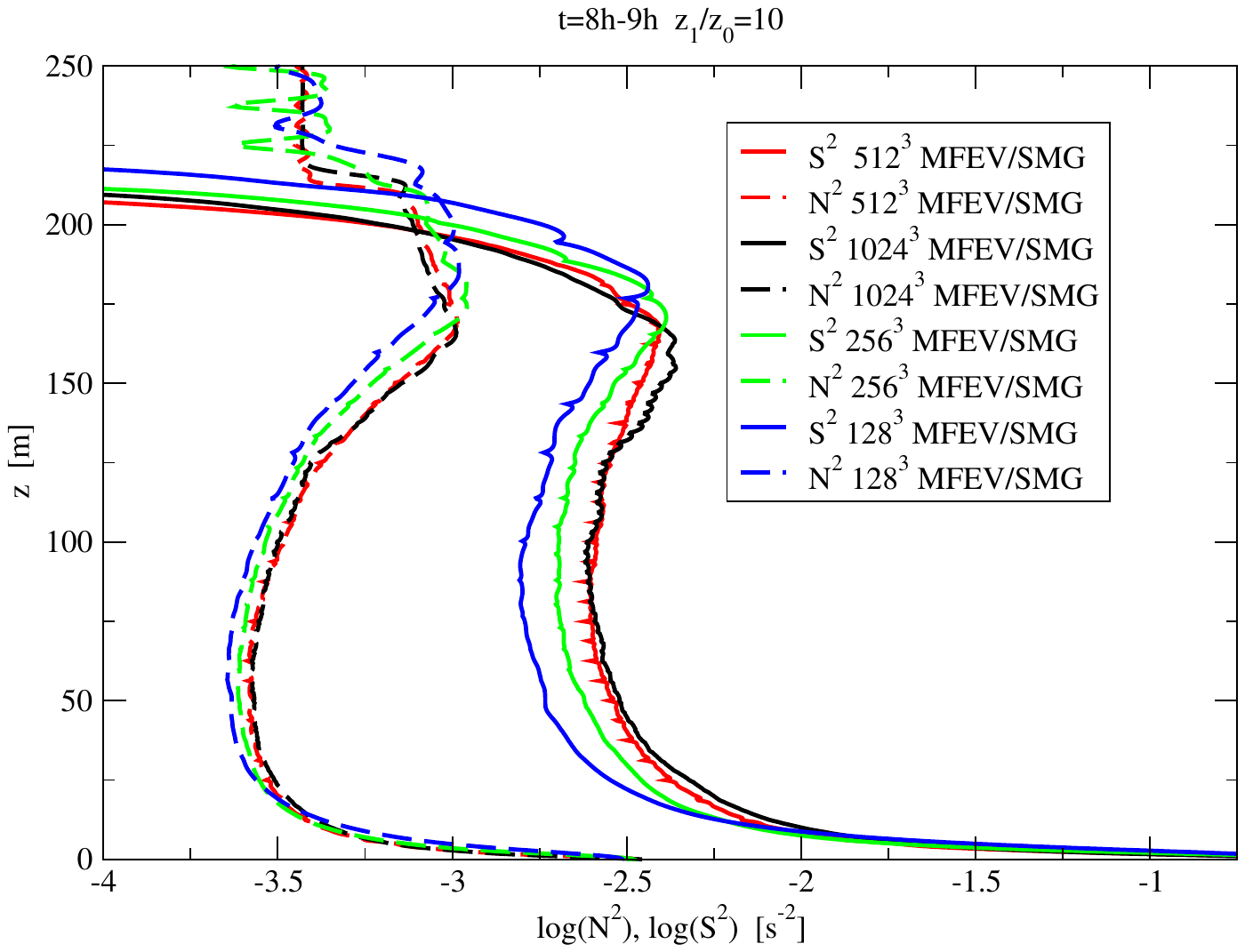}
    }
    \subfloat[]  
    {
    \includegraphics[width=0.44\textwidth]{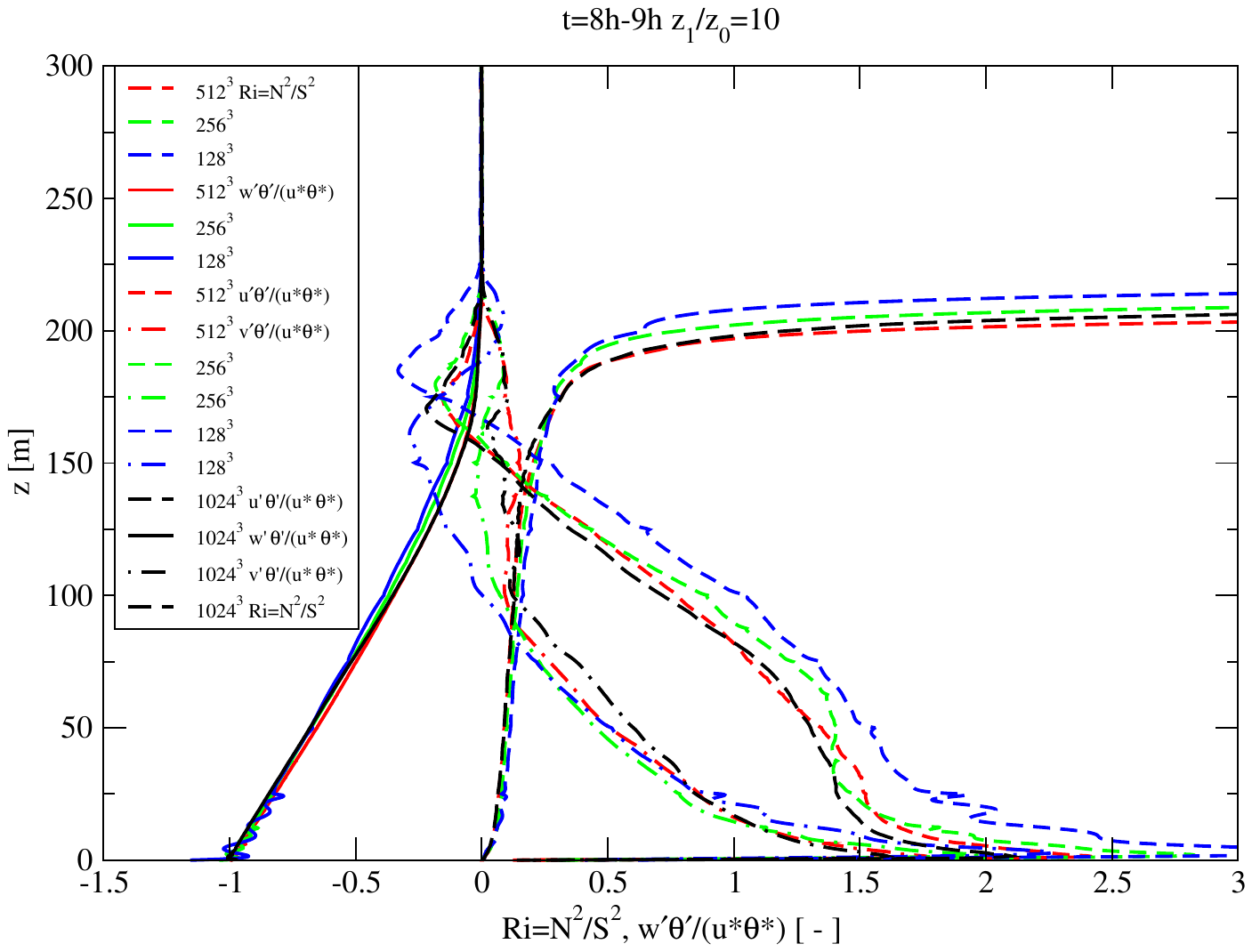}
    }
  \end{center}
  \caption{\label{fig.plot_test_all_N2_S2_new} 
   (a)
   MFEV/SMG $N^2$ and $S^2$ profiles with resolution and
   (b) 
   MFEV/SMG heat flux and Ri profiles.}
\end{figure*}

The vertical profiles of the squared shear and buoyancy frequency ($S^2, N^2$) from Nek5000/RS simulations 
are shown in Fig.~\ref{fig.plot_test_all_N2_S2_new} (a) for the four resolutions considered. Here
\begin{equation}
N^2=\frac{g}{\theta_0}\frac{\partial\langle \theta \rangle}{\partial z},
\label{eqn.N2}
\end{equation}
and 
\begin{equation}
S^2=\left(\frac{\partial\langle \bu_h \rangle}{\partial z}\right),
\label{eqn.S2}
\end{equation}

%
where $\bu_h$ is the horizontal velocity. The Richardson number $Ri$, defined as
\begin{equation}
Ri(z)=\frac{N^2}{S^2},
\label{eqn.Ri}
\end{equation}
is shown together with the vertical and horizontal temperature fluxes, which include SGS contributions, 
in Fig.~\ref{fig.plot_test_all_N2_S2_new} (b). As can be observed, the profiles are in close agreement as the mesh 
spacing is reduced.
In agreement with~\cite{Sullivan2016}, above the LLJ, $z > z_j$, $S^2$, and $N^2$ both decrease but at rates sufficient to 
maintain a constant $Ri$ near or slightly below the critical value of $0.25$. We note that a value of $Ri(z) \approx 0.2$ 
shows the approximate validity of the simple RANS parameterization of a constant Richardson number above the Monin--Obukhov surface layer.
The profiles of the vertical, $\langle w^{\prime}\theta^{\prime}\rangle$, and horizontal temperature fluxes, $\langle w^{\prime}\theta^{\prime}\rangle$ 
and $\langle w^{\prime}\theta^{\prime}\rangle$ are normalized by the product of the surface values $u^{\tau} Q^{\star}$. As can be observed from
the figure, in the mid- to lower BL, the vertical temperature fluxes are near-linear functions of $z$ as expected~\cite{Sullivan2016}. 
In the upper region, the mean flux profile displays more curvature, and approaching $z_i$ the vertical flux nearly collapses because of the 
increasing stratification. The horizontal temperature fluxes are comparable in magnitude to the vertical flux throughout the bulk of the BL. 
Futhermore, although the vertical temperature flux converges very quickly with resolution, the horizontal temperature fluxes seem to converge 
slower with increasing resolution.

\begin{figure*}
  \begin{center}
    \subfloat[]
    {
    \includegraphics[width=0.48\textwidth]{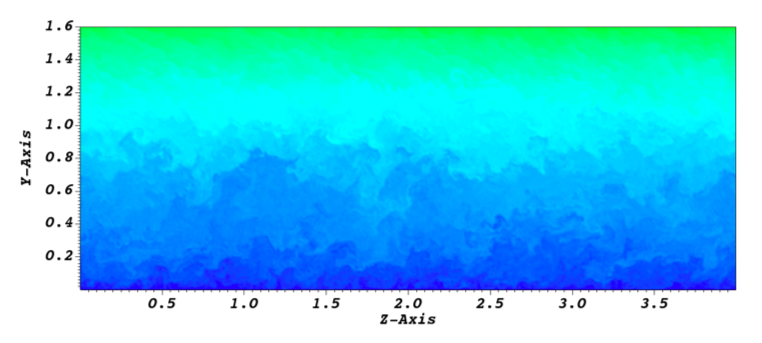}
    }
    \hspace{-1em}
    \subfloat[]
    {
    \includegraphics[width=0.48\textwidth]{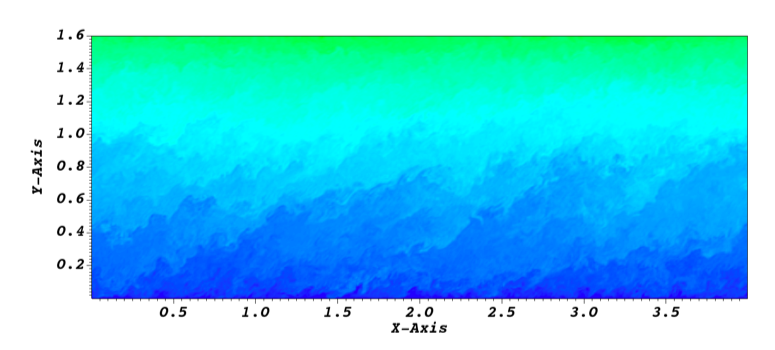}
    }
  \end{center}
  \caption{\label{fig.temp_isocontours} 
   Nek5000/RS Instantaneous potential temperature isocontours at $t=9h$ at 
   (a) $y=200$ m and (b) at $x=300$ m.}
\end{figure*}

Figures~\ref{fig.temp_isocontours} (a) and (b) show the instantaneous temperature isocontours at $t=9h$
in an $x–z$ plane at $y=200$ m and a $y-z$ plane at $x=300$ m, respecively. In agreement with~\cite{Sullivan2016},
inspection of these figures reveals that the temperature fronts are sharp warm - cold fronts tilted in the downstream direction, 
primarily a consequence of the sheared streamwise velocity. Near the low-level jet (z between 150-160m), the fronts 
are weaker with values of tilt angle.


\begin{figure*}
  \begin{center}
    \includegraphics[width=0.6\textwidth]{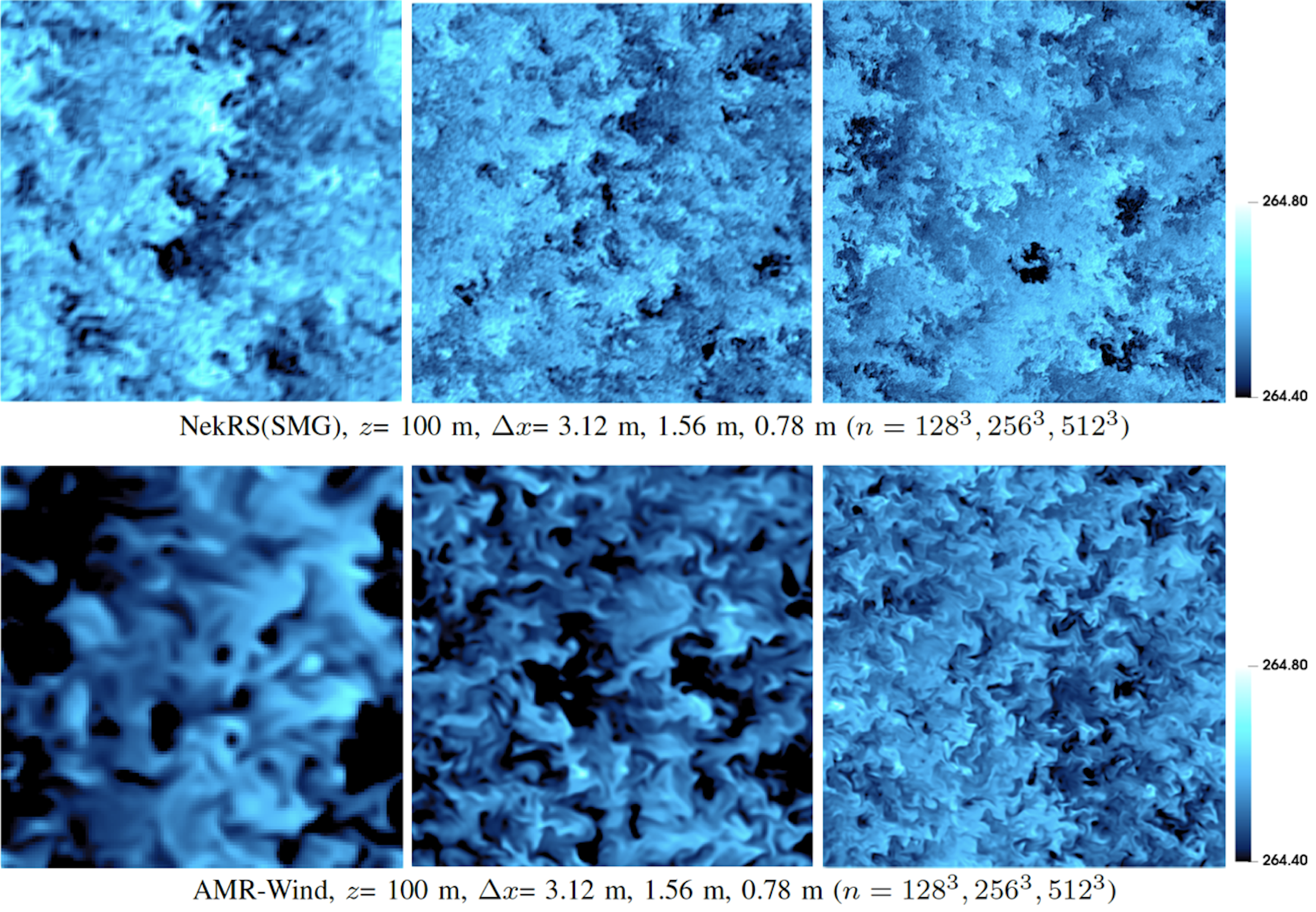}
  \end{center}
  \caption{\label{fig.temp_isocontours_compare_Nek_AMR} 
   Nek5000/RS MFEV/SMG and AMR-Wind at three grid refinement levels for 
   potential temperature at  $t=6h$.}
\end{figure*}

Figure~\ref{fig.temp_isocontours_compare_Nek_AMR} shows the instantaneous temperature isocontours
in an $x - z$ plane at $z=100$ m for the two codes Nek5000/RS and AMR-Wind and for three different
resolutions. As can be observed, finer scales do get resolved by both codes as resolution increases.
However, the same scales seem to be resolved by the high-order Nek5000/RS code using half the resolution 
of AMR-Wind. For example, the finest-resolved scales in the upper middle and lower right figures are very similar,
and they correspond to $256^3$ for Nek5000/RS and $512^3$ for AMR-Wind.

\begin{figure*}
  \begin{center}
    \includegraphics[width=0.6\textwidth]{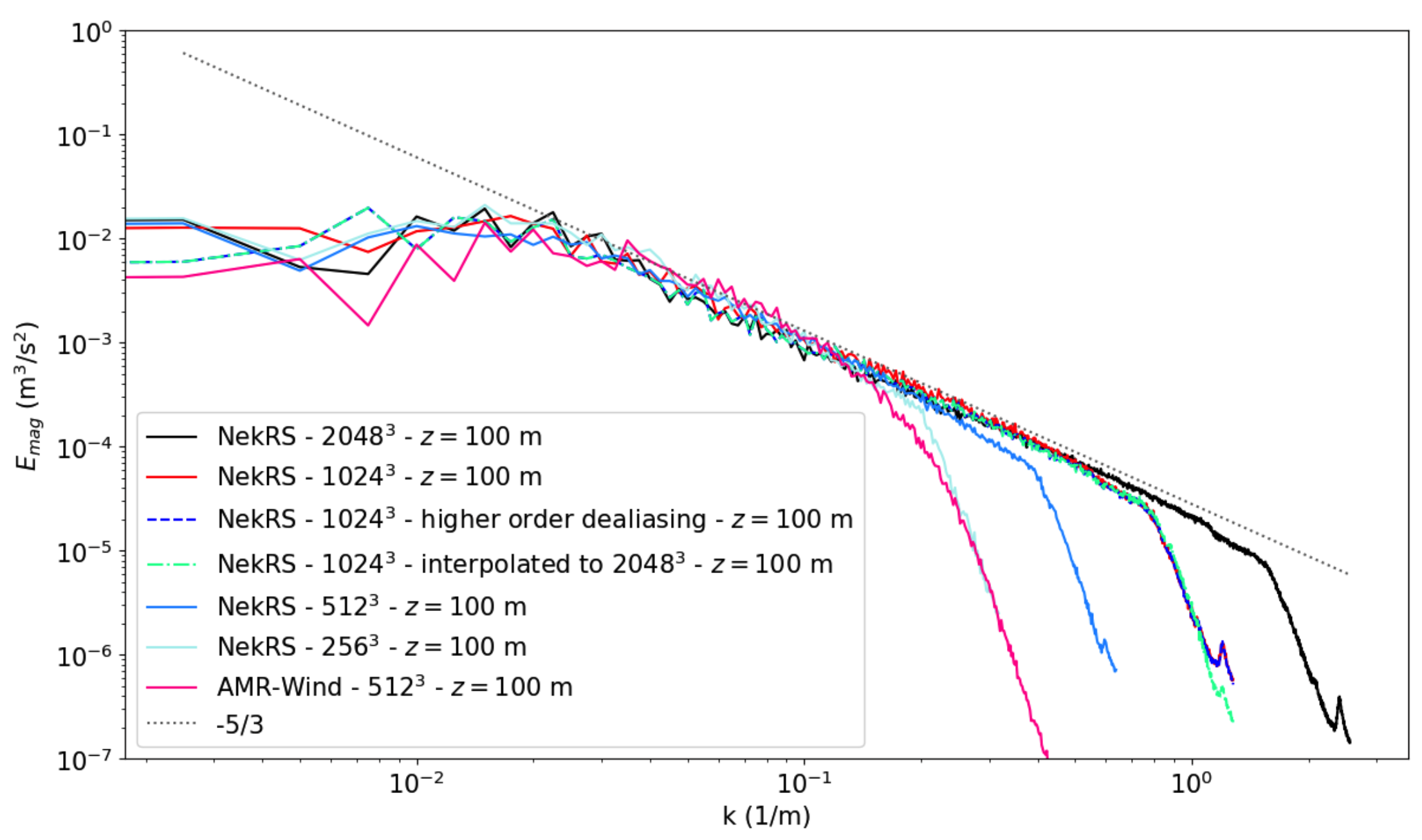}
  \end{center}
  \caption{\label{fig.power_spectrum_full} 
   Nek5000/RS and AMR-Wind: Spatial spectrum of the horizontal velocity at $z=100$m with resolution.}
\end{figure*}

Figure~\ref{fig.power_spectrum_full} demonstrates the spatial spectrum for
velocity magnitude at $t=8h$ and at $z=100$ m for $N=256^3$, $512^3$, $1024^3$, and $2048^3$ resolutions. 
We can observe that Nek is able to resolve out to $nx/\pi$, as would be expected given that the max
spacing for the collocation points is $(\pi/2) (L/nx)$, rather than ($L/nx$), for which Nyquist dictates
that one can resolve only to $nx/2$. Beyond ($nx/\pi$), we have a viscous-like decay, which
corresponds to SGS dissipation. On the other hand, the same spectrum obtained with AMR-Wind using a 
resolution of $512^3$ is almost identical with the one obtained by Nek using a resolution of $256^3$.
This is attributed to the second order spatial convergence of AMR-Wind compared with the hgih-order spectral
convergence of Nek.

\begin{figure*}
  \begin{center}
    \subfloat[]
    {
    \includegraphics[width=0.44\textwidth]{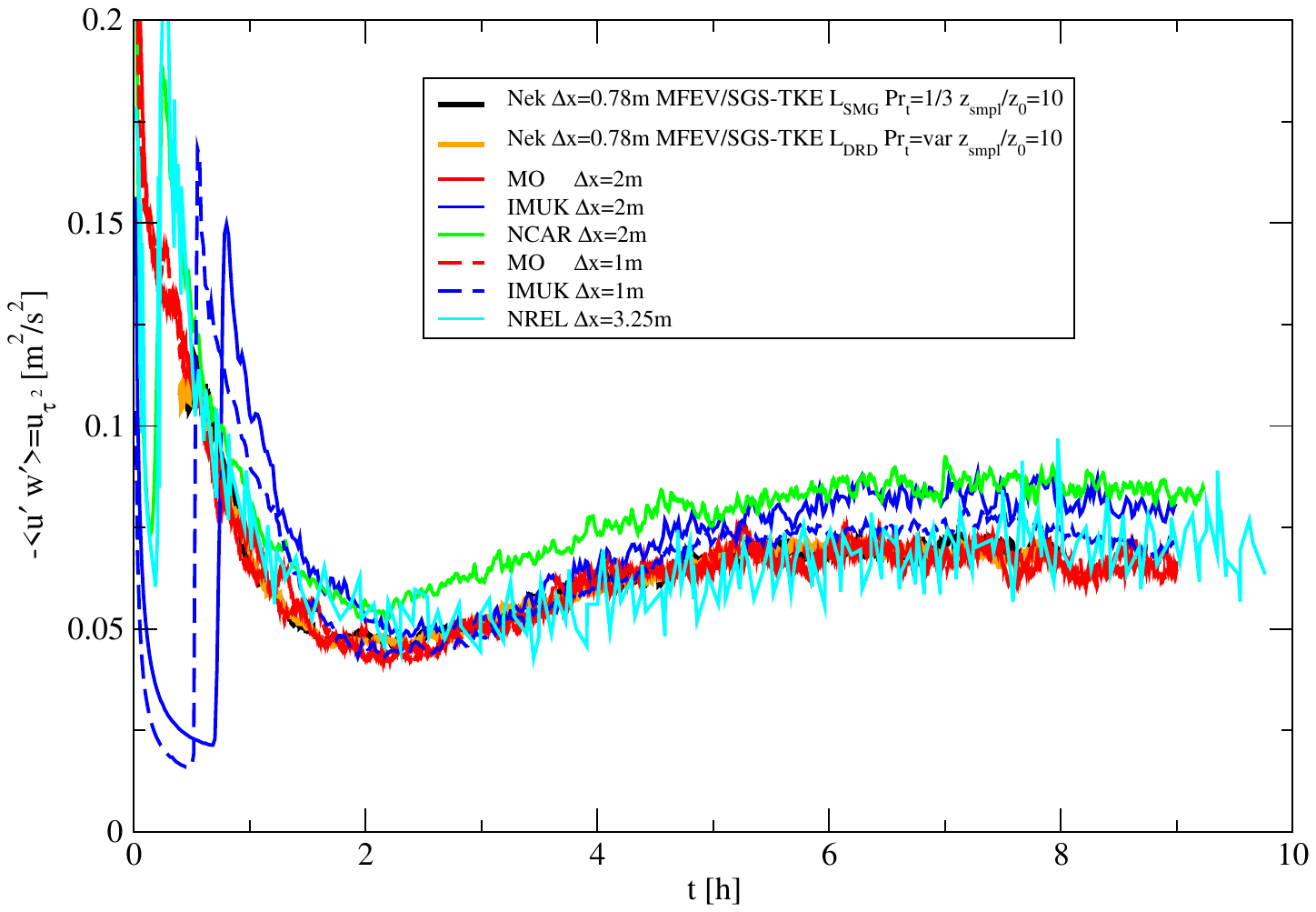}
    }
    \subfloat[]
    {
    \includegraphics[width=0.44\textwidth]{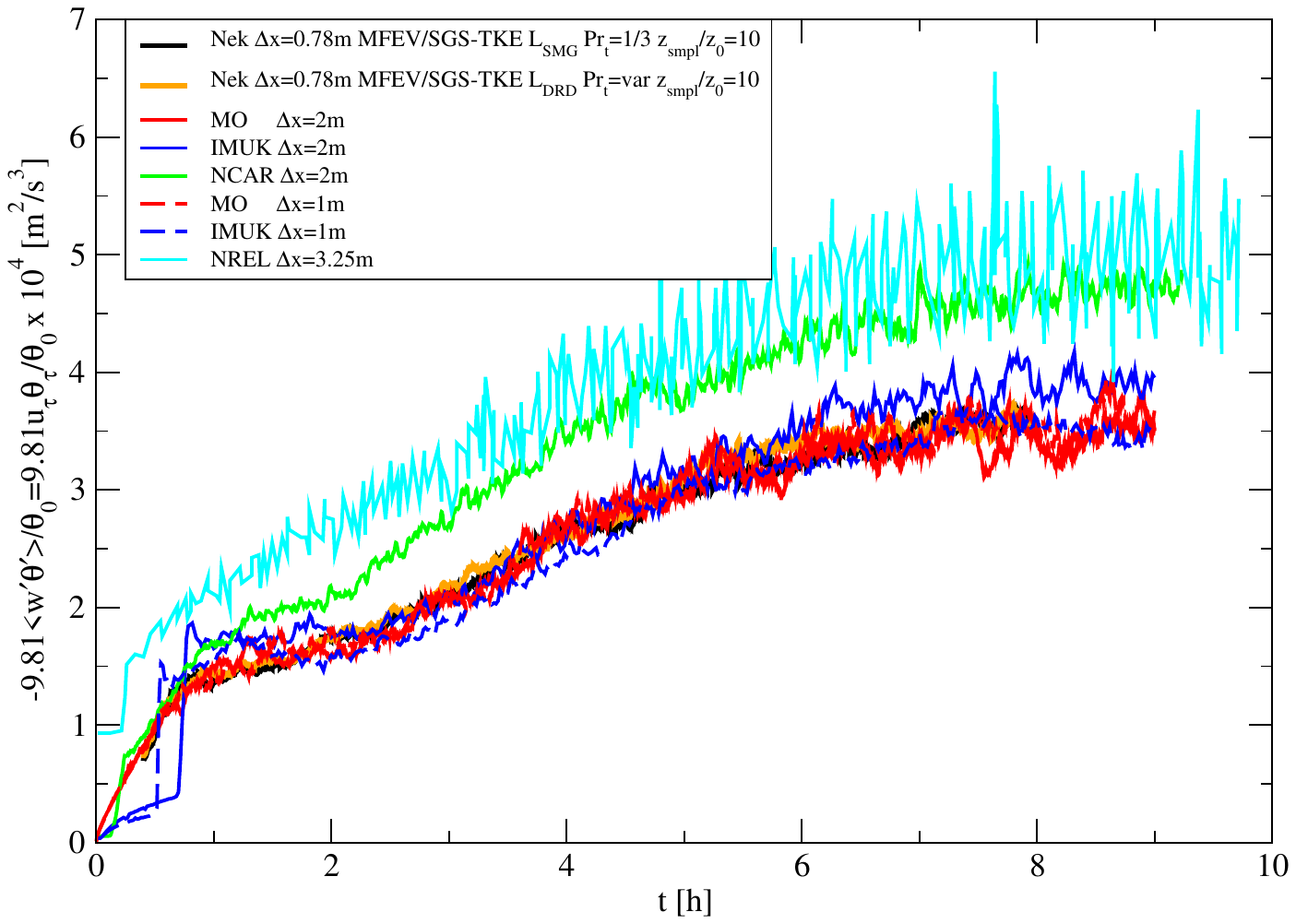}
    }
    \\
    \vspace{-1em}
    \subfloat[]
    {
    \includegraphics[width=0.44\textwidth]{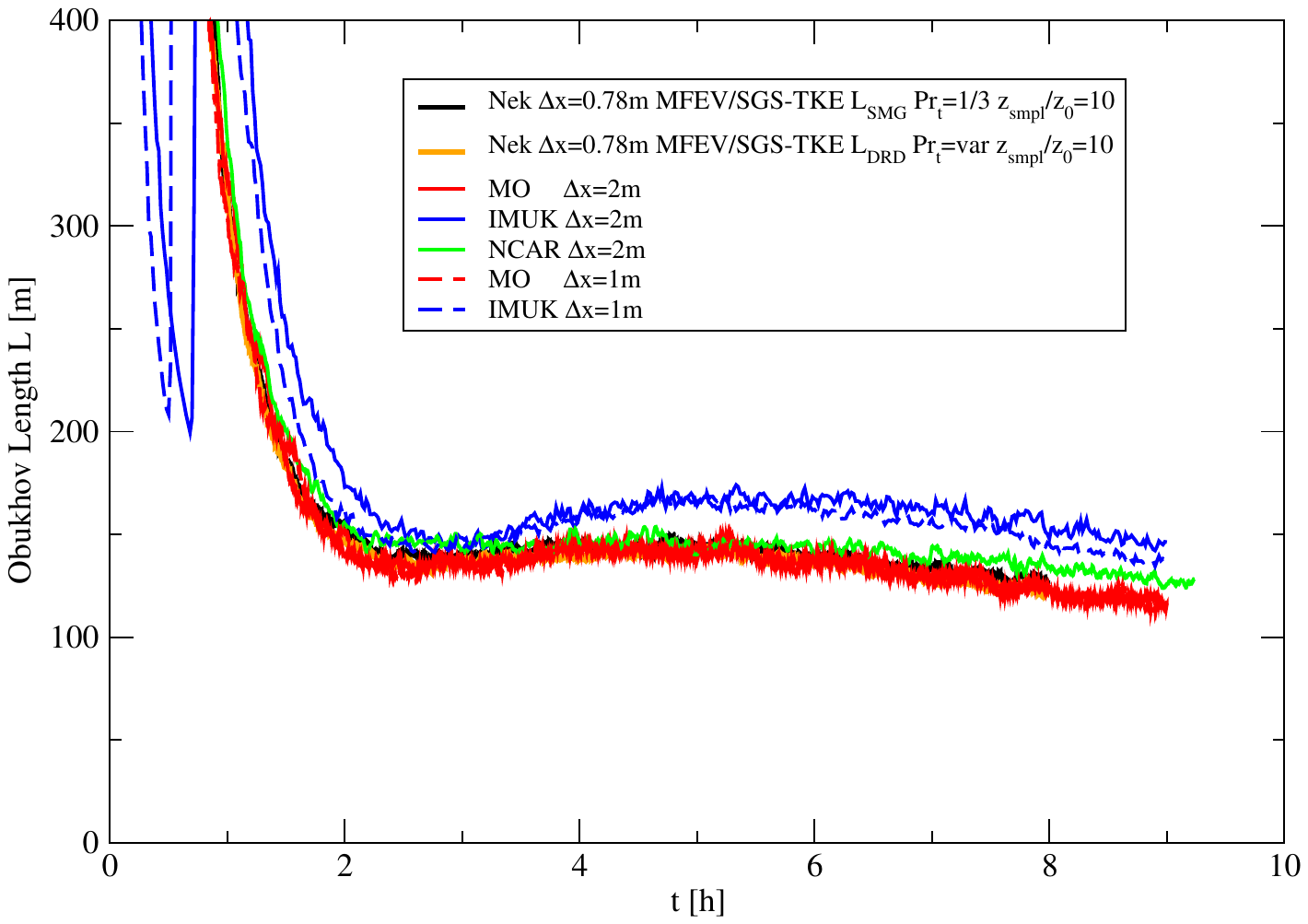}
    }
  \end{center}
  \caption{\label{fig.plot_surface_momflux_ncarBC_all} Comparison of Nek5000/RS with the NCAR, IMUK,
     and MO results for (a) the surface momentum flux, (b) the surface heat flux,
     and (c) the Monin-Obukhov length with resolution of $256^3$, $512^3$, $1024^3$, and $2048^3$.}
\end{figure*}

The remaining Fig.~\ref{fig.plot_surface_momflux_ncarBC_all} (a)--(c)
show the comparison of Nek5000/RS to the NCAR, IMUK,
and MO results for the surface momentum flux, the surface heat flux, and the Monin-Obukhov length, 
respectively.


%
%
      
\section{Conclusion}
We presented high-fidelity LES turbulence models for the atmospheric boundary
layer flows. We considered the GABLS1 benchmark problem and extended the range of our SGS modeling 
approaches in the context of the mean-field eddy viscosity provided with cross-verification 
and validation of two different codes, Nek5000/RS and AMR-Wind,
that are based on unstructured high-order and structured low-order order discretizations.

\section*{Acknowledgments}


This research was supported by the  U.S. Department of Energy, Office of Science, under contract DE-AC02-06CH11357  and
by the Exascale Computing Project (17-SC-20-SC), a joint project of the U.S.\ Department of Energy Office of Science and the National Nuclear Security Administration, responsible for delivering a capable exascale ecosystem, including software, applications, and hardware technology, to support the nation's exascale computing imperative. Funding was also provided by the U.S. Department of Energy, Office of Energy
Efficiency and Renewable Energy, Wind Energy Technologies Office. This work was authored in part by the National Renewable Energy Laboratory, operated by Alliance for Sustainable Energy, LLC, for the U.S.\ Department of Energy (DOE) under Contract No.\ DE-AC36-08GO28308.
The research used resources at the Oak Ridge Leadership
Computing Facility at Oak Ridge National Laboratory, which is supported by
the Office of Science of the U.S. Department of Energy under Contract DE-AC05-00OR22725.  The research also used computational resources sponsored by the DOE Office of Energy Efficiency and Renewable Energy and located at the National Renewable Energy Laboratory.
Part of this work has been supported by funding from the European
High-Performance Computing Joint Undertaking and Sweden, Germany, Spain, Greece and Denmark
under the CEEC project, Grant agreement No101093393. The highest resolution simulations of
$2048^3$ were performed at the Gauss Centre for Supercomputing, on the GCS Supercomputer
JUWELS at Julich Supercomputing Centre (JSC) under project LESABL.

%
%

%
%


 \bibliographystyle{sty/ametsocV6}
 \bibliography{bibs/emmd,bibs/ananias,bibs/abl,bibs/references}

\end{document}